\documentclass[showpacs,twocolumn,aps,preprintnumbers,letterpaper]{revtex4}
\usepackage{amsmath,amssymb}
\usepackage{epsfig}
\usepackage{graphicx}
\usepackage{amsmath}
\usepackage{slashed}
\usepackage{amsfonts}
\usepackage{epstopdf}
\usepackage{color}
\usepackage[caption=false]{subfig}
\usepackage[pdftex, pdfborder= 0 0 0, citecolor=blue, urlcolor=blue, linkcolor=blue, colorlinks=true, bookmarksopen=true]{hyperref}

\newcommand{\be}{\begin{equation}}
\newcommand{\ee}{\end{equation}}
\newcommand{\bear}{\begin{eqnarray}}
\newcommand{\eear}{\end{eqnarray}}
\newcommand{\ba}{\begin{array}}
\newcommand{\ea}{\end{array}}

\def\be{\begin{eqnarray}}
\def\ee{\end{eqnarray}}
\def\bea{\be}
\def\eea{\ee}

\def\roughly#1{\mathrel{\raise.3ex\hbox{$#1$\kern-.75em%
\lower1ex\hbox{$\sim$}}}}

\def\abs#1{{\left| #1 \right|}}

  \long\def\comment#1{ }

  \newcommand{\tr}{{\rm tr}}

  \newcommand{\beq}{\begin{eqnarray}}
  \newcommand{\eeq}{\end{eqnarray}}
  
 \def\simge{\mathrel{%
   \rlap{\raise 0.511ex \hbox{$>$}}{\lower 0.511ex \hbox{$\sim$}}}}
\def\simle{\mathrel{
   \rlap{\raise 0.511ex \hbox{$<$}}{\lower 0.511ex \hbox{$\sim$}}}}

\newcommand{\boldfX}{\mbox{${F}_{\rm X}$}}
\newcommand{\boldfXgZ}{\mbox{${F}_{\rm X}^{\gamma Z}$}}
\newcommand{\boldfXZ}{\mbox{${F}_{\rm X}^Z$}}
\newcommand{\boldftwo}{\mbox{$\tilde{F}_2$}}
\newcommand{\boldxft}{\mbox{$x\tilde{F}_3$}}
\newcommand{\boldfl}{\mbox{$\tilde{F}_{\rm L}$}}

\newcommand{\ncred}{\mbox{$ \sigma_{r,{\rm NC}}^{\pm}$}}

\newcommand{\ncdd}{\mbox{$\frac{\textstyle {\rm d^2} \sigma^{e^{\pm}p}_{{\rm NC}}}{\textstyle {\rm d}x_{\rm Bj}{\rm d} Q^2}$}}

\begin{document}

\title{Neutrino-nucleon DIS  from Holographic QCD:\\
PDFs of sea and valence quarks, form factors, and structure functions of the proton}

\author{Kiminad A. Mamo}
\email{kmamo@anl.gov}
\affiliation{Physics Division, Argonne National Laboratory, Argonne, Illinois 60439, USA
}
\author{Ismail Zahed}
\email{ismail.zahed@stonybrook.edu}
\affiliation{Center for Nuclear Theory, Department of Physics and Astronomy, Stony Brook University, Stony Brook, New York 11794-3800, USA}



\date{\today}
\begin{abstract}
We discuss unpolarized neutrino- and anti-neutrino-nucleon deep inelastic scattering (DIS)  using a chiral doublet of baryonic sources with explicit symmetry breaking,
in a slice of AdS$_5$ with both a hard and soft wall. We explicitly derive the direct and transition form factors for the vector and axial-vector currents for the holographic
dual of a proton and neutron. We use them to derive the s-channel structure functions for neutrino and anti-neutrino scattering  on a proton and neutron in bulk. The t-channel
contributions stemming from the Pomeron and Reggeon exchanges are also evaluated explicitly. The pertinent even and odd structure functions in the limit of large and 
small parton momentum fraction $x$ are given. The results allow for the extraction of the nonperterbative parton distribution functions 
carried by the sea and valence quarks both at large-x and small-x regimes. Our holographic PDF sets compare well with LHAPDF and CTEQ PDF sets in the large-x and small-x regimes in the intermediate range of $Q^2<10~\rm{GeV^2}$.
\end{abstract}


\maketitle

\setcounter{footnote}{0}


\section{Introduction}

At extremely low-x, the measured nucleon structure functions on unpolarized nucleon targets,
show a rapid growth of 
sea quarks and  gluons at low-x~\cite{H1,ZEUSS}. Phenomenological arguments suggest that
the growth saturates~\cite{GW}, a point supported by perturbative QCD arguments~\cite{BK}. 
A central question is then: what is the primary mechanism for the growth of the sea quarks
at low-x? 

DIS in holography at moderate-x is different from weak coupling as it involves hadronic  and
not partonic constituents~\cite{POL}. The large gauge coupling at the low renormalization point,
causes the color charges to rapidly deplete
their energy and momentum, making them visible to hard probes only through double trace operators.
However, because the holographic limit enjoys approximate conformal
symmetry,  the structure functions and form factors exhibit various scaling laws including the 
parton-counting rules~\cite{BF}. DIS scattering at low-x is partonic  and fully saturated~\cite{HATTA}.
Recently, we have extended the notion of DIS scattering to nuclei as extremal RN-AdS black holes
with an emerging Fermi surface~\cite{Mamo:2018eoy}.

In this paper we consider neutrino and anti-neutrino DIS scattering on an unpolarized 
nucleon described by a  Dirac  chiral doublet  using  holographic QCD.
The scattering is chiefly due to neutral and charge exchanges. In QCD the charge exchange currents
discriminate between partons and anti-partons. When combined with anti-neutrino DIS scattering, it allows for
the separation between sea and valence partons. In holography, the partonic description holds only for very large
$x$ and very small $x$ in the Regge limit. The purpose of the present study is to construct the even and odd parity
leptonic structure functions for neutrino and anti-neutrino scattering, and use them to extract the sea parton distribution
of the nucleon as a Dirac fermion in bulk.

The organization of the paper is as follows: in section II we briefly review the setting for the 
model with bulk chiral gauge fields and a doublet of Dirac fermions in a slice of AdS$_5$,
for both the soft and hard wall. The explicit breaking of chiral symmetry is enforced by the
boundary value of a scalar bulk field in the bi-fundamental representation. In section III
we derive the direct and transition form factors for the vector and axial-vector form factors 
for both the proton and neutron. 
In section IV, we briefly review the essentials of neutrino and anti-neutrino 
DIS scattering on unpolarized nucleon targets, and  show how the valence and sea up and down 
parton distribution functions are related to the structure functions of $W^\pm$ exchanges. 
The s-channel contributions to these structure functions are derived using the direct and
transition form factors in holographic QCD. The t-channel contributions are also derived
using  the Pomeron and Reggeon exchanges in bulk. In section V,  all the results are
combined to extract the valence and sea partonic distributions at large and small x, with a detailed comparison to 
the LHAPDF and CTEQ PDFs. 
 Our conclusions are in section V.  A number of Appendices are added to support some 
 of the derivations.

 \section{holographic model}

AdS/CFT duality maps a conformal and strongly coupled  gauge theory in 1+3 dimensions at the boundary to a weakly coupled type-II supergravity 
in 1+9 dimensions in bulk, with  a dilaton, an  antisymmetric tensor field and additional  odd  (IIA) or  even  (IIB) forms.  AdS$_5\times$S$^5$ geometry emerges 
from solitonic  and BPS charged D-brane supergravity solutions in bulk with the graviton and dilaton excitations dual to glueballs. The dual of the flavor excitations 
at the boundary, are obtained through probe Dp-branes and described by DBI and Chern-Simons effective actions. 

A simple way to capture AdS/CFT duality in the non-conformal limit is  to model it using a slice of AdS$_5$ with various bulk fields with assigned 
anomalous dimensions and pertinent boundary values, in the so-called bottom-up approach which we will follow here. 
 We consider AdS$_5$ both with a soft and hard wall, with a background metric  $g_{MN}=(\eta_{\mu\nu},-1)R^2/z^2$ with the flat metric  $\eta_{\mu\nu}=(1,-1,-1,-1)$
 at the boundary.
 Confinement will be described by  a background dilaton $\phi={\kappa}^2z^2$ for mesons and 
 $\phi={\tilde\kappa}^2z^2$ for nucleons  in the soft wall model. In the hard wall model, $\phi=0$ and confinement  is enforced  at $z=z_0$.

\begin{widetext}

\subsection{Bulk vector mesons}

The vector mesons fields in bulk  are denoted by $L_\mu, R_\mu$. They  are U(2) valued and described by the effective action~\cite{HIRN,GRIGORYAN}

\begin{eqnarray}
\label{01}
S_{M}=&&-\frac 1{4g_5^2}\int d^5x e^{-\phi(z)}\sqrt{g}\,g^{MP}g^{NQ}{\rm Tr}\bigg({\cal F}^L_{MN}{\cal F}^L_{PQ}+{\cal F}^R_{MN}{\cal F}^R_{PQ}\bigg)
+\int d^5x\,\bigg(\omega^L_5({\cal A})-\omega^R_5({\cal A})\bigg)
\end{eqnarray}
with ${\cal F}=d{\cal A}-i{\cal A}^2$ in form notations and ${\cal A}={\cal A}^aT^a$  with $T^0=\frac{3}{2}{\bf 1_2}$ and $T^i=\frac{1}{2}\tau^i$. 
Note that the generators $T^a$ are  fixed in such a way that the electromagnetic charge of the proton is 1,  while that of the  neutron is 0. The
bulk U(2) vector field $V=(R+L)/2$ and axial-vector  field $A=(R-L)/2$  reduce to the QCD flavor
source fields at the boundary. The Chern-Simons contributions in (\ref{01}) are

\be
\omega_5({\cal A})=
\frac{N_c}{24\pi^2}\int d^5x \,{\rm Tr}\left({\cal AF}^2+\frac 12 {\cal A}^3{\cal F}-\frac 1{10}{\cal A}^5\right)=
\frac{N_c}{24\pi^2}\int d^5x \,{\rm Tr}\left({\cal A}(d{\cal A})^2-\frac {3i}2{\cal A}^3d{\cal A}-\frac 35{\cal A}^5\right)
\ee
The  equation of motions for the bulk gauge fields  follow by variation
(${\cal D}=d-i{\cal A}$)

\be
\frac{1}{\sqrt{g}}\left[{\cal D}_M,\big(\sqrt{g}e^{-\phi}{\cal A}^{MN}\big)\right]=0
\ee
The coupling $g_5$ in (\ref{01})  is fixed by the brane embeddings in bulk, with 
${1}/{g_{5}^2}\equiv{3N_c}/(12\pi^2)$ (D7-branes),
and ${1}/{g_{5}^2}\equiv ({3\sqrt{\lambda}}/{2^{5/2}\pi}){N_c}/({12\pi^2})$ (D9-branes).
When ignoring these embedding, the standard assignement is
${1}/{g_{5}^2}\equiv{N_c}/(12\pi^2)$ to match the  vector 2-point correlation functions of QCD in the UV~\cite{CHERMAN}.

 \subsection{Bulk fermionic doublet}

The bulk Dirac fermion action in curved AdS$_5$ with minimal coupling to the {left}  U(2) gauge fields  is

\be
S_F=\frac 1{2g_5^2}\int d^{5} x \,e^{-\phi(z)}\,\sqrt{g}\,\Big(\mathcal{L}_{F1}+\mathcal{L}_{F2}\Big)+\frac 1{2g_5^2}\int d^4 x \sqrt{-g^{(4)}}\,\Big(\mathcal{L}_{UV1}+\mathcal{L}_{UV2}\Big)\,,\nonumber\\
\label{Action}
\ee
with  $\mathcal{L}_{UV1,2}=(\overline{\Psi}_{1,2}\Psi_{1,2})_{z=\varepsilon}$.  This extra UV contribution  is needed to maintain  the bulk to boundary correspondence.
The Dirac and Pauli fermionic contributions 
$ {\mathcal L}_{F1,2}=\mathcal{L}_{\rm Dirac1,2}+\mathcal{L}_{\rm Pauli1,2}$  are explicitly

\bea
\label{fermionAction}
\mathcal{L}_{\rm Dirac1,2}&=&\bigg( \frac{i}{2} \overline{\Psi}_{1,2} e^N_A \Gamma^A\big(\overrightarrow{D}_N^{L,R}-\overleftarrow{D}_N^{L,R}\big)\Psi_{1,2}-(\pm M+V(z))\bar{\Psi}_{1,2}\Psi_{1,2}\bigg)\,,\nonumber\\
\mathcal{L}_{\rm Pauli1,2}&=&\pm 2g_5^2\times\eta\, \bar{\Psi}_{1,2} e^M_Ae^N_B\sigma^{AB}{\cal F}^{L,R}_{MN}\Psi_{1,2}\,,\nonumber\\
\eea
with $\sigma^{AB}=\frac i2  [\Gamma^A,\Gamma^B]$, and the left and right covariant derivatives 
\bea
\overrightarrow{D}_N^{L}=&&\overrightarrow{\partial}_N +\frac{1}{8}\omega_{NAB}[\Gamma^A,\Gamma^B]-iL_N^aT^a\equiv \overrightarrow{\mathcal{D}}_N-iL_N^aT^a \nonumber\\
\overleftarrow{D}_N^{L}=&&\overleftarrow{\partial}_N +\frac{1}{8}\omega_{NAB}[\Gamma^A,\Gamma^B]+iL_N^aT^a\equiv \overleftarrow{\mathcal{D}}_N+iL_N^aT^a \nonumber\\
\overrightarrow{D}_N^{R}=&&\overrightarrow{\partial}_N +\frac{1}{8}\omega_{NAB}[\Gamma^A,\Gamma^B]-iR_N^aT^a\equiv \overrightarrow{\mathcal{D}}_N-iR_N^aT^a \nonumber\\
\overleftarrow{D}_N^{R}=&&\overleftarrow{\partial}_N +\frac{1}{8}\omega_{NAB}[\Gamma^A,\Gamma^B]+iR_N^aT^a\equiv \overleftarrow{\mathcal{D}}_N+iR_N^aT^a 
\eea
The left and right flavor  field strengths are set to ${\cal A}^{L}_{MN}=L_{MN}$, ${\cal A}^{R}_{MN}=R_{MN}$. The Dirac or minimal fermionic coupling is standard, 
but the Pauli coupling is not. It can be shown to follow from the SUGRA action by reduction from the top-down approach. Without it, the neutron form factor vanishes. 
The nucleon doublet  refers to 

\be
\Psi_{1,2}\equiv 
\begin{pmatrix} 
  \Psi_{p1,2}\\ 
  \Psi_{n1,2}
\end{pmatrix}\,.
\ee
The nucleon fields in bulk form an iso-doublet $p,n$ with $1,2$ referring to their  $\pm=R,L$ chirality at the {\it boundary}~\cite{YEE}.  They are dual to a boundary chiral doublet of   baryonic sources
 with $\Psi_{p1,2}\leftrightarrow {\cal O}_{p,\pm}$  and $\Psi_{n1,2}\leftrightarrow {\cal O}_{n,\pm}$ with anomalous dimensions
$\pm M=\pm (\Delta-2)=\pm (\tau-3/2)$. They map onto  a non-normalizable solution
or source,  plus  a normalizable or expectation value of the corresponding  source at the boundary. This doubling is best seen in the top-down approach using a left and a right bulk filling brane.

The fermionic potential $V(z)={\tilde \kappa}^2z^2$  will be  used for both the soft and hard wall. Here  $e^N_A=z \delta^N_A$ is the inverse vielbein.
The components of the spin connection are $\omega_{\mu z\nu}=-\omega_{\mu\nu z}=\frac{1}{z}\eta_{\mu\nu}$, the Dirac gamma matrices  
$\Gamma^A=(\gamma^\mu, -i\gamma^5)$ are chosen in the chiral representation, and satisfy the flat
anti-commutation relation $\{\Gamma^A,\Gamma^B\}=2\eta^{AB}$. The  equation of motions for the bulk Dirac chiral doublet  follows by variation

\bea
\label{EOM12}
&&\bigg(i e^N_A \Gamma^A D_N^{L,R} -\frac{i}{2}(\partial_N\phi)\, e^N_A \Gamma^A- (\pm M+V(z))\bigg)\Psi_{1,2}=0\,,
\eea
with $1,2=R,L=\pm$.
The inclusion of a fermionic potential which is set to the dilaton profile for simplicity, breaks conformal symmetry and garentees the Reggeization of the nucleon spectrum in bulk.
(\ref{EOM12}) is supplemented with confining  boundary conditions for the hard wall, and vanishing fields asymptotically for the soft wall. The solutions to (\ref{EOM12}) are briefly
discussed in Appendix~\ref{FERMIONFIELDS}. In short, for the soft wall and in the absence of a tachyon coupling in bulk, 
the spectrum Reggeizes with $M_n^2=4\tilde\kappa^2(n+\tau-1)$ and the ground state proton, neutron 
states with $n=0$  are degenerate. They follow from the mixed representation

\bea
\Psi_1(p,z)&=&\psi_R(z)\Psi^0_{R}(p)+ \psi_L(z)\Psi^0_{L}(p)
\eea
for $n=0$, with $\psi_R(z\approx 0)\approx z^{\tau+1/2}$ and $\psi_L(z\approx 0)\approx z^{\tau+3/2}$ for the positive parity states $1\equiv +$ at the boundary. Similar relations
follow for the negative parity states $2\equiv -$ at the boundary through the substitution $\psi_{R,L}\leftrightarrow \mp \psi_{L,R}$ by parity.

Note that the canonical dimension for the QCD baryonic sources is $\Delta=9/2$ which would suggest a twist $\tau=4$. However, we expect non-vanishing anomalous dimensions to develop
at strong coupling. We do not know of any reliable calculational scheme to assess them. We will assume $\Delta$ and thus $\tau$ as a parameter with a twist
$\tau=3$ to recover the hard scattering rules. The inclusion of additional twist contributions is discussed  in~\cite{VEGA}.

\subsection{Fermionic currents}

For later use, it will be useful to define the bulk U(2) Dirac 1-form currents

\begin{eqnarray}
J^{aN}_L=&&\frac{\partial\mathcal{L}_{\rm Dirac1}}{\partial L^a_N}=\overline{\Psi}_1 e^N_A \Gamma^A T^a\Psi_1\,,\nonumber\\
J^{aN}_R=&&\frac{\partial\mathcal{L}_{\rm Dirac2}}{\partial R^a_N}=\overline{\Psi}_2 e^N_A \Gamma^A T^a\Psi_2\,,
\end{eqnarray}
and the bulk U(2) Pauli 2-form currents

\begin{eqnarray}
J^{aMN}_L=&&\frac{\partial\mathcal{L}_{\rm Pauli1}}{\partial L^a_{MN}}=+2g_5^2\times\eta\overline{\Psi}_1 e^M_Ae^N_B \sigma^{AB}T^a\Psi_1\,,\nonumber\\
J^{aMN}_R=&&\frac{\partial\mathcal{L}_{\rm Pauli2}}{\partial R^a_{MN}}=-2g_5^2\times\eta\overline{\Psi}_2 e^M_Ae^N_B\sigma^{AB}T^a\Psi_2\,,
\end{eqnarray}
in terms of which the pertinent 1- and 2-form charged currents read respectively

\bea
J_{L,R}^{+N}=&&J_{L,R}^{1N}-iJ_{L,R}^{2N}=\overline{\Psi}_{n1,2} e^N_A \Gamma^A\Psi_{p1,2}\,,\nonumber\\
J_{L,R}^{-N}=&&J_{L,R}^{1N}+iJ_{L,R}^{2N}=\overline{\Psi}_{p1,2} e^N_A \Gamma^A\Psi_{n1,2}\,.
\eea
and

\bea
J_{L}^{+MN}=&&J_{L}^{1MN}-iJ_{L}^{2MN}=+2g_5^2\times\eta\overline{\Psi}_{n1} e^M_Ae^N_B \sigma^{AB}\Psi_{p1}\,,\nonumber\\
J_{L}^{-MN}=&&J_{L}^{1MN}+iJ_{L}^{2MN}=+2g_5^2\times\eta\overline{\Psi}_{p1} e^M_Ae^N_B \sigma^{AB}\Psi_{n1}\,,\nonumber\\
J_{R}^{+MN}=&&J_{R}^{1MN}-iJ_{R}^{2MN}=-2g_5^2\times\eta\overline{\Psi}_{n2} e^M_Ae^N_B \sigma^{AB}\Psi_{p2}\,,\nonumber\\
J_{R}^{-MN}=&&J_{R}^{1MN}+iJ_{R}^{2MN}=-2g_5^2\times\eta\overline{\Psi}_{p2} e^M_Ae^N_B \sigma^{AB}\Psi_{n2}\,.\nonumber\\
\eea

\subsection{Bulk effective action}

In terms of the charged and left 1-form gauge fields  $L_{N}^{\pm}\equiv \frac{1}{\sqrt{2}}(L_{N}^1\mp iL_{N}^2)$, the bulk meson effective action 
(\ref{01})  can be recast in the following form

\bea
\label{012}
S_M\supset && \frac 1{g_5^2}\int d^{5} x e^{-\phi(z)}\sqrt{g}
\,\bigg(-\partial^{M}L^{-N}\partial_{M}L_{N}^{+}+\partial^{M}L^{-N}\partial_{N}L_{M}^{+}
-\frac{1}{4}L^{0MN}L^{0}_{MN}-\frac{1}{4}L^{3MN}L^{3}_{MN}+L\rightarrow R\bigg)\nonumber\\
&&+\frac{N_c}{48\pi^2}\int d^{5} x \epsilon^{\mu\nu\rho\sigma}\bigg(\mathcal{L}^{CS}_{\mu\nu\rho\sigma}(L)-\mathcal{L}^{CS}_{\mu\nu\rho\sigma}(R)\bigg)
\nonumber\\
\ee
with  the Chern-Simons contribution restriced to the charged left-currents through a neutral

\bea
\label{CHERNLLL}
\mathcal{L}^{CS}_{\mu\nu\rho\sigma}(L)=
\partial_z L_{\mu}^{+}\partial_{\nu}L_{\rho}^{0}L_{\sigma}^{-}+\partial_z L_{\mu}^{-}\partial_{\nu}L_{\rho}^{0}L_{\sigma}^{+}
-\partial_z L_{\mu}^{+}L_{\rho}^{0}\partial_{\nu}L_{\sigma}^{-}-\partial_z L_{\mu}^{-}L_{\rho}^{0}\partial_{\nu}L_{\sigma}^{+}
+\partial_z L_{\mu}^{0}\partial_{\nu}L_{\rho}^{-}L_{\sigma}^{+}+\partial_z L_{\mu}^{0}\partial_{\nu}L_{\rho}^{+}L_{\sigma}^{-}\,,\nonumber\\
\eea
where we made use of the  Abelian field strengths
\bea
L_{MN}^{\pm}=&&\partial_ML^{\pm}_N-\partial_NL^{\pm}_M\,, \nonumber\\
L_{MN}^{0,3}=&&\partial_ML^{0,3}_N-\partial_NL^{0,3}_M\,.
\eea
In terms of the 1- and 2-form currents, the  bulk fermion effective action (\ref{fermionAction})   now reads

\bea
\label{fermionAction2}
\mathcal{L}_{F1}+\mathcal{L}_{F2}\supset &&
\frac{i}{2} \bar{\Psi}_1 e^N_A \Gamma^A\big(\overrightarrow{\mathcal{D}}_N-\overleftarrow{\mathcal{D}}_N\big)\Psi_1-(M+V(z))\bar{\Psi}_1\Psi_1
+\frac{1}{\sqrt{2}}L_{N}^{+}J_{L}^{-N}+\frac{1}{\sqrt{2}}L_{N}^{-}J_{L}^{+N}+L_{N}^{0}J_{L}^{0N}+L_{N}^{3}J_{L}^{3N}\nonumber\\
&&+L^{+}_{MN}J_{L}^{-MN}
+L^{-}_{MN}J_{L}^{+MN}+L^0_{MN}J_{L}^{0MN}
+L^3_{MN}J_{L}^{3MN}
+(1\leftrightarrow 2, L\leftrightarrow R, M\leftrightarrow -M)\nonumber\\\nonumber\\
=&&\frac{i}{2} \bar{\Psi}_1 e^N_A \Gamma^A\big(\overrightarrow{\mathcal{D}}_N-\overleftarrow{\mathcal{D}}_N\big)\Psi_1-(M+V(z))\bar{\Psi}_1\Psi_1 +\frac{i}{2} \bar{\Psi}_2 e^N_A \Gamma^A\big(\overrightarrow{\mathcal{D}}_N-\overleftarrow{\mathcal{D}}_N\big)\Psi_2-(-M+V(z))\bar{\Psi}_2\Psi_2\nonumber\\
&&+\frac{1}{\sqrt{2}}V_{N}^{+}J_{V}^{-N}+\frac{1}{\sqrt{2}}A_{N}^{+}J_{A}^{-N}+\frac{1}{\sqrt{2}}V_{N}^{-}J_{V}^{+N}+\frac{1}{\sqrt{2}}A_{N}^{-}J_{A}^{+N}+V_{N}^{0}J_{V}^{0N}+A_{N}^{0}J_{A}^{0N}+V_{N}^{3}J_{V}^{3N}+A_{N}^{3}J_{A}^{3N}\nonumber\\
&&+\frac{1}{\sqrt{2}}V^{+}_{MN}J_{V}^{-MN}+\frac{1}{\sqrt{2}}A^{+}_{MN}J_{A}^{-MN}+\frac{1}{\sqrt{2}}V^{-}_{MN}J_{V}^{+MN}+\frac{1}{\sqrt{2}}A^{-}_{MN}J_{A}^{+MN}\nonumber\\
&&+V^{0}_{MN}J_{V}^{0MN}+A^{0}_{MN}J_{A}^{0MN}+V^{3}_{MN}J_{V}^{3MN}+A^{3}_{MN}J_{A}^{3MN}\,,
\eea
where we defined

\bea
J_{V,A}^{\bar{\alpha} N}=J_{L}^{\bar{\alpha}N}\pm J_{R}^{\bar{\alpha}N}\,,\,\,\,\,J_{V,A}^{\bar{\alpha} MN}=J_{L}^{\bar{\alpha}MN}\mp J_{R}^{\bar{\alpha}MN}\,,\nonumber\\
\eea
with $\bar{\alpha}\equiv -,+,0,3$, and $V_N,A_N\equiv ({L_N\pm R_N})/{2}$. 
\end{widetext}

\subsection{Hard and soft wall models}

We now detail the specifics of the hard and soft wall model with chiral symmetry breaking through the use
of a flavor scalar in the bi-fundamental representation in a slab of AdS. These well motivated bottom-up
models capture the essentials of the top-down approaches and make more transparent the essential
aspects of holography. Their parameters are fixed by the brane embeddings and reduction in higher
dimensions.

\subsubsection{Hard wall with bi-fundamentals}

To distinguish vector and axial-vector spectra and introduce a scalar
$\tilde X$ in the bi-fundamental representation of $U(N_f)_L\times U(N_f)_R$ with a scaling dimension $\Delta=3$ and bulk action

\be\label{xaction}
{\cal S}[\tilde X]=\frac{1}{2g_5^2}\int\,d^5x\,\sqrt{g}\,{\rm Tr}\bigg(|D\tilde X|^2+3|\tilde X|^2\bigg)
\ee
using the 1-form $D\tilde X=d\tilde X-i{\cal A}^L\tilde X+i\tilde X{\cal A}^R$.  Note by replacing $X\rightarrow \sqrt{2g_5^2}\times \tilde{X}$, 
we can rewrite (\ref{xaction}), in a form similar to~\cite{KKSS}, as

\be\label{xaction2}
{\cal S}[X]=\int\,d^5x\,\sqrt{g}\,{\rm Tr}\bigg(|DX|^2+3|X|^2\bigg)\,.
\ee

The equation of motion derived from the above action (\ref{xaction2}), for ${\cal A}=0$, has a     background solution for the bi-fundamental scalar near the UV boundary ($z\rightarrow0$) given by

\be
\label{X0Z}
X_0(x,z\rightarrow 0)\approx \frac{1}{\sqrt{2g_5^2}}\Bigg(\frac{1}{2}M_qz+\frac{1}{2}\Sigma z^3\Bigg)\equiv \frac{1}{\sqrt{2g_5^2}}v(z)\nonumber\\
\ee
Both the current mass matrix and the quark bilinear are diagonal,  $M_q=m_q\textbf{1}$ and 
 $\Sigma=\sigma\textbf{1}$, with $\Sigma^{ij}=2g_5^2\left<\bar q_{R}^iq_{L}^j\right>$ where $i,j=1,2...N_f$ are the flavor indices. For the hard wall, the
boundary condition for the $U(N_f)$ gauge fields is ${\cal F}^{R,L}(x,z_0)=0$. 
\\
\\
{\bf Vectors:}
\\
\\
The bulk vector field $V_M=(V_\mu, V_5)$ splits into a $\mu$-transverse, $\mu$-longitudinal and 5-contribution. 
The 5-longitudinal components of the vector field,
mix through a Higgs-type effect. They can be decoupled by a pertinent choice of gauge.
The transverse part of the vector field decouples, and its 
mode decomposition   in terms of  $\psi_n(z)$ yields the bulk equation

\be
\label{VDM1}
\partial_z\bigg(\frac 1z \partial_z\psi_n(z)\bigg)+\frac 1zm_n^2\,\psi_n(z)=0
\ee
subject to the confining conditions $\psi_n(0)=\psi_n^\prime(z_0)=0$. The solutions are  readily found as
$\psi_n(z)\sim zJ_1(m_n z)$ with the vector spectrum $m_n$ fixed by the zeros $\gamma_{0,n}$
of the Bessel function  $J_0(\gamma_{0,n})=0$, with normalization

\be
\label{NORM}
\int_0^{z_0}\, dz \,\sqrt{g_{xx}}\,\psi_n(z)\psi_m(z)=\delta_{nm}
\ee
and  the completeness relation

\be
\label{COMP}
\sum_{n=1}^\infty\psi_n(z)\psi_n(z^\prime)=\sqrt{g^{xx}}\delta(z-z^\prime)
\ee
The first zero of $J_0$ or $\gamma_{0,1}=2.40483$, is identified with the rho meson state $m_{1=\rho}=\gamma_{0,1}/z_0=0.775$ GeV
which fixes the IR scale $z_0=3.103/{\rm GeV}$. Note that asymptotically $\gamma_{0,n}\approx n\pi$ with $m_n\approx n\pi/z_0$.


The bulk-to-boundary vector propagator follows from (\ref{VDM1}) through the substitution $m_n^2\rightarrow -Q^2$ 
with the boundary conditions ${\cal V}(Q,0)=1$ and $\partial_z{\cal V}(Q,z_0)=0$ (confining). The solution
can be obtained in closed form 

\be
{\cal V}(Q,z)=&&\sum_n\frac{g_5F_n\psi_n(z)}{Q^2+m_n^2}\nonumber\\
=&&Qz\bigg(K_1(Qz)+\frac{K_0(Qz_0)}{I_0(Qz_0)}\,I_1(Qz)\bigg)\nonumber\\
\ee
where we have also shown its mode decomposition with the decay constants
$g_5F_n=(-\partial_z\psi_n(z)/z)_0$.
\\
\\
{\bf Axials:}
\\
\\
Similarly the bulk axial-vector field $A_M=(A_\mu, A_5)$
 splits into a $\mu$-transverse, $\mu$-longitudinal and 5-contribution. 
The 5-longitudinal components of the axial-vector field,
mix through a Higgs-type effect. They can be decoupled by a pertinent choice of gauge,
at the expense of more coupling of the $A_5$ field with the tachyon field.  For instance in the R-gauge,
the $A_5$ field is identified with the pion field. These contributions will not
be followed except when discussing the pion contribution to the direct and transitional axial-vector form factors below. 

The $\mu$-transverse part of $A_\mu$ is always decoupled.
Its  mode decomposition  in terms of $\tilde\psi_n(z)$ yields the bulk equation
\be
\label{VDM21}
\partial_z\bigg(\frac 1z \partial_z\tilde\psi_n(z)\bigg)-\frac{1}{2}\frac{v^2(z)}{z^3}\tilde\psi_n(z)+\frac 1z\tilde m_n^2\,\tilde\psi_n(z)=0\,,\nonumber\\
\ee
Upon taking the chiral limit before the near-boundary limit  (i.e.,  $m_q\rightarrow 0$ before $z\rightarrow 0$), (\ref{VDM21}) reduces to
\be
\label{VDM22}
\partial_z\bigg(\frac 1z \partial_z\tilde\psi_n(z)\bigg)-\frac{1}{2}\sigma^2z^3\tilde\psi_n(z)+\frac 1z\tilde m_n^2\,\tilde\psi_n(z)=0\nonumber\\
\ee
with  the same confining boundary conditions $\tilde\psi_n(0)=\tilde\psi_n^\prime(z_0)=0$ and a similar normalization
(\ref{NORM}) and completeness relation (\ref{COMP}). Note that (\ref{VDM22}) can only be solved numerically. 

However, we can find an equation that can be solved analytically, if we take the near-boundary limit first without taking 
the chiral limit  (i.e.,   $z\rightarrow 0$ with $m_q \neq 0$) of (\ref{VDM21}) which reduces  to     
\be
\label{VDM23}
\partial_z\bigg(\frac 1z \partial_z\tilde\psi_n(z)\bigg)-\frac{1}{2z}m_q^2\,\tilde\psi_n(z)+\frac 1z\tilde m_n^2\,\tilde\psi_n(z)=0\,.\nonumber\\
\ee
Defining $\tilde{m}_n^2=m_n^2+\frac{1}{2}m_q^2$, we can recast  (\ref{VDM23})  in the form
\be
\label{VDM24}
\partial_z\bigg(\frac 1z \partial_z\tilde\psi_n(z)\bigg)+\frac 1zm_n^2\,\tilde\psi_n(z)=0\,,\nonumber\\
\ee
which maps onto the vector equation (\ref{VDM1}) with the replacement of $\tilde\psi_n(z)\leftrightarrow\psi_n(z)$, and $\tilde m_n\leftrightarrow m_n$.

In the near boundary or UV limit, the differences between the axial-vector and vector masses are  only due to the explicit symmetry breaking effect with $m_q\neq 0$,
and vanishes in the chiral limit. In general however, the difference is largely due 
to the spontaneous breaking of chiral symmetry through $\Sigma=\sigma{\bf 1}$, as  a numerical solution to  (\ref{VDM22}) shows. 
For   $\sigma z_0^3\ll 1$,  a  simple parametric  estimate can be obtained using 
first order perturbation theory  in (\ref{VDM22})

\be
\label{SUB0}
\tilde{m}_n^2\approx m_n^2+\frac{1}{2}\int_0^{z_0}dz \sqrt{g_{xx}}\left|\psi_n(z)\right|^2(\sigma z^2+m_q)^2\nonumber\\
\ee
which reduces to the  near boundary or UV limit result for $\sigma z_0^3\rightarrow 0$.  In the chiral limit  (\ref{SUB0})
yields a chiral splitting between  the  axials and vectors 

\be
\label{SUB01}
\tilde{m}_{n}^2\approx m^2_{n} +0.38\left(\frac 12 {\sigma^2 z_0^4}\right)
\ee
solely due to the chiral condensate. Note that asymptotically,  the {\it linear} mass splitting in the hard wall model vanishes, i.e.
$\tilde m_n-m_n\approx \sigma^2 z_0^5/n\rightarrow 0$ for $n\gg 1$. 
Below and for simplicity, we will  carry the analytical analysis using the near boundary limit,  using  the substitution

\be
\label{SUB02}
\frac 12 m_q^2\rightarrow \tilde{m}_n^2-m_n^2
\ee

The bulk-to-boundary axial-vector propagator follows from (\ref{VDM24}) through the substitution $m_n^2\rightarrow -\tilde Q^2=-(Q^2+\frac{1}{2}m_q^2)$ 
with the boundary conditions ${\cal A}(Q,0)=1$ and $\partial_z{\cal A}(Q,z_0)=0$ (confining). Note that ${\cal A}(Q,0)\neq{\cal A}(0,z)$ since ${\cal A}(0,z)\neq 1$. The solution
can be obtained in closed form as 

\be
\label{VQZ-approx0}
\tilde{\cal V}(Q,z)\approx &&\sum_n\frac{g_5\tilde F_n\tilde\psi_n(z)}{\tilde Q^2+m_n^2}=\sum_n\frac{g_5\tilde F_n\tilde\psi_n(z)}{Q^2+\tilde m_n^2}\nonumber\\
=&&\tilde Qz\bigg(K_1(\tilde Qz)+\frac{K_0(\tilde Qz_0)}{I_0(\tilde Qz_0)}\,I_1(\tilde Qz)\bigg)\nonumber\\
\ee
where we have also shown its mode decomposition with the decay constants
$g_5\tilde F_n=(-\partial_z\tilde\psi_n(z)/z)_{z=0}$.

\subsubsection{Hard wall without bi-fundamentals}

Both the bulk-to-boundary vector and axial-vector propagators can be obtained in closed form in 
a variant  of the Sakai-Sugimoto construction~\cite{SS}  using a hard wall model without the use of the
bi-fundamental scalar field in bulk but with modified boundary conditions. 
\\
\\
{\bf Vectors:}
\\
\\
The vector fields are still given 
with the same hard wall boundary conditions $\psi_n(0)=\psi_n^\prime(z_0)=0$, but the axial-vector fields
satisfy $\tilde\psi_n(0)=\tilde\psi_n(z_0)=0$ and $\tilde{\psi}_n^\prime(z_0)\neq 0$. 
The solutions are again readily found in the form $\psi_n(z)\sim zJ_1(m_n z)$ with $m_n=\gamma_{0,n}/z_0$.
\\
\\
{\bf Axials:}
\\
\\
Similarly, the axial vector spectrum follows  with $\tilde{\psi}_n(z)\sim zJ_1(\tilde m_nz)$ with  $\tilde m_n=\gamma_{1,n}/z_0$.
 If $z_0$ is fixed by the rho mass then the ratio of the axial-to-rho meson
mass is $\tilde m_1/M_1=\gamma_{1,1}/\gamma_{0,1}=1.593$ which is consistent with
the empirical ratio $m_A/m_\rho=1.587$ as in the Sakai-Sugimoto construction. The higher excited modes fare 
less better empirically in both formulations. 

The bulk-to-boundary axial-vector propagator follows  a similar reasoning as the
vector analogue, through the substitution $\tilde m_n^2\rightarrow -Q^2$ 
with the boundary conditions $\tilde{\cal V}(Q,0)=\tilde{\cal V}(0,z)=1$ and $\tilde{\cal V}(Q,z_0)=0$ (confining). The solution
follows  as

\be
\tilde{\cal V}(Q,z)\approx &&\sum_n\frac{g_5\tilde F_n\tilde\psi_n(z)}{Q^2+\tilde m_n^2}\nonumber\\
=&&\tilde Qz\bigg(K_1(\tilde Qz)-\frac{K_1(\tilde Qz_0)}{I_1(\tilde Qz_0)}\,I_1(\tilde Qz)\bigg)\nonumber\\
\ee

\subsubsection{Soft wall with bi-fundamentals}

In the soft wall model with scalar bi-fundamentals, we replace (\ref{xaction2}) by~\cite{KKSS}

\be\label{xactions}
{\cal S}[X]=\,\int\,d^5x\,e^{-\phi(z)}\sqrt{g}\,{\rm Tr}\bigg(|DX|^2+3|X|^2\bigg)\nonumber\\
\ee
with again the same background solution and boundary identification for the bi-fundamentals. 
\\
\\
{\bf Vectors:}
\\
\\
In this model, the bulk vector gauge field  in terms of  $\psi_n(z)$ yields the bulk equation

\be
\label{VDM1s}
\partial_z\bigg(\frac{e^{-\phi(z)}}{z} \partial_z\psi_n(z)\bigg)+\frac{e^{-\phi(z)}}{z}m_n^2\,\psi_n(z)=0\,.
\ee
The solutions are readily found as
$\psi_n(z)=c_{n}\tilde{\kappa}^2z^2 L_n^1( \tilde{\kappa}^2z^2)$, and $m_{n}^2=\,\,4\tilde\kappa^2(n+1)$ for $n=0,1,...$  with normalization coefficients $c_{n}=\sqrt{{2}/{n+1}}$ determined from the normalization condition (for the soft wall model with background dilaton $\phi=\tilde{\kappa}^2z^2$)
\be\label{NORMs}
\int dz\,\sqrt{g}e^{-\phi}\,(g^{xx})^2\,\psi_n(z)\psi_m(z)=\delta_{nm}\,.
\ee

For the soft wall model, the bulk-to-boundary vector propagator follows from (\ref{VDM1s}) through the substitution $m_n^2\rightarrow -Q^2$ 
with the boundary conditions ${\cal V}(Q,0)=1$. The solution
can be obtained in closed form as 

\bea
{\cal V}(Q,z)&=&\sum_n\frac{g_5F_n\psi_n(z)}{Q^2+m_n^2}\nonumber\\
&=&\tilde{\kappa}^2 z^2 \,\,\Gamma (1+\frac{Q^2}{4\tilde{\kappa}^2} )\,\,{\cal U} (1+\frac{Q^2}{4 \tilde{\kappa}^2} ; 2 ; \tilde{\kappa}^2 z^2 )\,,\nonumber\\
&=&\tilde{\kappa}^2z^2\int_{0}^{1}\frac{dx}{(1-x)^2}x^a{\rm exp}\Big[-\frac{x}{1-x}\tilde{\kappa}^2z^2\Big]\nonumber\\
\ee
where we have also shown its mode decomposition with the decay constants
$g_5F_n=(-\partial_z\psi_n(z)/z)_0$, and defined $a\equiv {Q^2}/{4\tilde{\kappa}^2}$.
\\
\\
{\bf Axials:}
\\
\\
For the soft wall model, the mode decomposition of the axial-vector gauge field in terms of $\tilde\psi_n(z)$ yields the bulk equation
\bea
\label{VDM21s}
&&\partial_z\bigg(\frac{e^{-\phi(z)}}{z}\partial_z\tilde\psi_n(z)\bigg)-\frac{e^{-\phi(z)}\frac{1}{2}v(z)^2}{z^3}\tilde\psi_n(z)\nonumber\\
&+&\frac{e^{-\phi(z)}}{z}\tilde m_n^2\,\tilde\psi_n(z)=0\,,
\eea
which, upon taking the chiral limit before the near-the-boundary one (i.e., taking $m_q\rightarrow 0$ before $z\rightarrow 0$), reduces to
\bea
\label{VDM22s}
&&\partial_z\bigg(\frac{e^{-\phi(z)}}{z} \partial_z\tilde\psi_n(z)\bigg)- e^{-\phi(z)}\frac{1}{2}\sigma^2z^3\tilde\psi_n(z)\nonumber\\
&+&\frac{e^{-\phi(z)}}{z}\tilde m_n^2\,\tilde\psi_n(z)=0\,,\nonumber\\
\eea
with a similar normalization
(\ref{NORMs}). Again, note that (\ref{VDM22s}) can only be solved numerically. 

But, similar to the hard wall case, we can find an equation that can be solved analytically, for the soft wall model, if we take the near-boundary limit first without taking the chiral one (i.e., $z\rightarrow 0$ with $m_q \neq 0$) of (\ref{VDM21s}) which reduces   to     
\be
\label{VDM23s}
&&\partial_z\bigg(\frac{e^{-\phi(z)}}{z}\partial_z\tilde\psi_n(z)\bigg)-\frac{e^{-\phi(z)}}{z}\frac{1}{2}m_q^2\,\tilde\psi_n(z)\nonumber\\
&+&\frac{e^{-\phi(z)}}{z}\tilde m_n^2\,\tilde\psi_n(z)=0\,.\nonumber\\
\ee
Defining $m_n^2=\tilde m_n^2-\frac{1}{2}m_q^2$, we can rewrite (\ref{VDM23s}) as 
\be
\label{VDM24s}
\partial_z\bigg(\frac{e^{-\phi(z)}}{z}\partial_z\tilde\psi_n(z)\bigg)+\frac{e^{-\phi(z)}}{z}m_n^2\,\tilde\psi_n(z)=0\,,\nonumber\\
\ee
which is essentially the same equation as the vector one (\ref{VDM1s}) with the replacement of $\tilde\psi_n(z)\leftrightarrow\psi_n(z)$, and $M_n\leftrightarrow m_n$.


For the soft wall model, the bulk-to-boundary axial-vector propagator follows from (\ref{VDM24s}) through the substitution $m_n^2\rightarrow -\tilde Q^2=-(Q^2+\frac{1}{2}m_q^2)$ 
with the boundary conditions $\tilde{\cal V}(Q,0)=1$. Note that $\tilde{\cal V}(Q,0)\neq\tilde{\cal V}(0,z)$ since $\tilde{\cal V}(0,z)\neq 1$. The approximate solution
follows as 

\begin{widetext}

\bea
\label{VQZ-approx}
\tilde{\cal V}(Q,z)\approx &&\sum_n\frac{g_5\tilde F_n\tilde\psi_n(z)}{\tilde Q^2+m_n^2}=\sum_n\frac{g_5\tilde F_n\tilde\psi_n(z)}{Q^2+\tilde m_n^2}\nonumber\\
=&&\tilde{\kappa}^2 z^2 \,\,\Gamma \bigg(1+\frac{\tilde Q^2}{4\tilde{\kappa}^2} \bigg)\,\,{\cal U} \bigg(1+\frac{\tilde Q^2}{4 \tilde{\kappa}^2} ; 2 ; \tilde{\kappa}^2 z^2\bigg)
=\tilde{\kappa}^2z^2\int_{0}^{1}\frac{dx}{(1-x)^2}x^{\tilde a}{\rm exp}\Big[-\frac{x}{1-x}\tilde{\kappa}^2z^2\Big]\nonumber\\
\eea
where we have also shown its mode decomposition with the decay constants
$g_5\tilde F_n=(-\partial_z\tilde\psi_n(z)/z)_0$, and defined $\tilde a\equiv {\tilde Q^2}/{4\tilde{\kappa}^2}$.

We emphasize that the exact form of  (\ref{VQZ-approx}) requires solving numerically (\ref{VDM21s}) for the normalizable modes
and using the mode decomposition (first line). Alternatively,  one  can  solve (\ref{VDM21s})  also numerically for the non-normalizable modes,
after the substitution $\tilde m_n^2\rightarrow -Q^2$. However, for DIS scattering  which is the main thrust of this paper,
this is not needed. Indeed,  in the DIS regime with $Qz_0\gg 1$,  the near-boundary approximation
giving (\ref{VQZ-approx}) (second line) is  sufficient. This will be assumed throughout.


\section{Direct and transition  form factors}

The vector and axial-vector couplings to the Dirac fermion in bulk follow from the Witten diagrams. 
They involve both the Dirac and Pauli form factors. We note that neutrino scattering through the 
charged currents involve solely the charged left-currents. Here we construct both the direct and
transition form factors needed for the vector and axial-vector currents and compare the direct ones 
 to the  most current data on the proton and neutron. We will use these form factors to construct
 the s-channel contributions for neutrino DIS scattering on nucleon targets.

\subsection{Direct vector and axial form factors}

The direct vector and axial form factors for the proton and neutron $V,A+N(p)\rightarrow N(p^\prime)$, can be extracted from the boundary to bulk
three point functions with pertinent LSZ reduction using

\begin{eqnarray}
{\cal C}^{(0,3)}_{NN}(p,p^\prime, q)=\lim_{p^{\prime 2},p^2\to m_N^2}(p^{\prime 2}-m_{N}^ 2)(p^2-m_N^2)\,\int d^4xd^4y\,e^{i(p^\prime\cdot x-p\cdot y-q\cdot z)}\,
\left<0\bigg|T\bigg({\cal O}_{N}(x)\,\tilde{J}_{V,A}^{(0,3)\mu}(z)\,{\cal O}_N(y)\bigg)\bigg|0\right>\nonumber\\
\end{eqnarray}
through the ratio

\be
\label{DIRECT}
{W}_{V,A}^{(0,3)\mu}(Q^2)=\left<N(p')\left|\tilde{J}_{V,A}^{(0,3)\mu}(0)\right|N(p)\right>=\frac{{\cal C}^{(0,3)}_{NN}(p,p^\prime, q)}{F_N(p')F_N(p)}
\nonumber\\
\ee
for the chargeless form factor, and similarly for the charged currents

\begin{eqnarray}
\label{DIREC2}
{W}_{V,A}^{\pm\mu}(Q^2)= \left< N(p')\left|\frac{1}{\sqrt{2}}\tilde{J}_{V,A}^{\pm\mu}(0)\,\right|N(p)\right>\,\nonumber\\
\end{eqnarray}
and the electomagnetic currents

\bea
\label{DIRECTEM}
{W}_{EM}^\mu(Q^2)=\left<N(p')\left|\tilde{J}_{EM}^\mu(0)\right|N(p)\right>
=\left< N(p')\left|\frac{1}{3}\tilde{J}_{V}^{0\mu}(0)+\tilde{J}_{V}^{3\mu}(0)\right|N(p)\right>\,
\eea
with $q^2=(p'-p)^2=-Q^2$. 
Here $N(p)$ refers to the U(2) proton-neutron doublet

\be
N(p)\equiv 
\begin{pmatrix} 
  N_{p}(p)\\ 
  N_{n}(p)
\end{pmatrix}\,,
\ee
and the  baryonic decay  constant  $F_N(p)$ is canonically defined as

\be
\left<0\left|{\cal O}_N(x)\right|N(p)\right>=F_N(p)e^{-ip\cdot x}
\ee
modulo the spin-flavor structure of the nucleon source ${\cal O}_N$. These definitions are commensurate with the lattice definitions for the three point functions and form factors
\cite{Alexandrou:2017hac}, with the baryonic decay constants defined and evaluated in~\cite{Tsutsui:2004qc} (and references therein). In our case they are  tied to the bulk  wave functions
and given in  (\ref{fermiondc}-\ref{SolutionFermions}).

The chargeless currents $\tilde{J}_{V,A}^{(0,3)\mu}$ at the boundary are identified with the quark (partonic)  currents, with the quarks in the fundamental representation of U$(N_f=2)$.  
They are sourced by the dual  bulk vector fields  $V_{\mu}^{0}(Q,z\rightarrow 0)$  and $V_{\mu}^{3}(Q,z\rightarrow 0)$ at the boundary, respectively. Similarly, the dual bulk axial vector
fields $A_{\mu}^{(0,3)}(Q,z\rightarrow 0)$ and  $\frac{1}{2}\times\frac{1}{\sqrt{2}} \times A_{\mu}^{\pm}(Q,z\rightarrow 0)$ at the boundary, are the dual of the quark currents 
 $\tilde{J}_{A}^{(0,3)\mu}(0)$ and  $2\tilde{J}_{A}^{\pm\mu}(0)$, respectively.  
 
 We now proceed to  evaluate the Abelian part,
 $U(1)_{V,A}^{\tilde{\alpha}}\subset U(2)$ of the Dirac and Pauli  contributions to the direct vector or axial form factors of the proton and neutron (\ref{DIRECT}). 
We  will give a  detailed account of the Dirac contribution to the direct parts of the ectromagnetic and axial-vector currents, setting up this way the various definitions
and normalizations. The Pauli contributions will follow a similar reasoning and will be only quoted.

\subsubsection{Direct vector form factor: Dirac}

The Dirac contribution to the direct part of the electromagnetic current can be extracted from the bulk Dirac part of the action  in (\ref{fermionAction2})
in the soft wall model

\begin{eqnarray}
\label{InteractionActionEM0}
S_{Dirac}^{EM} [i,X]=\frac{1}{2g_5^2}\, \int dz d^{4}y \sqrt{g} e^{- \tilde\kappa^2 z^2}\Big(\frac{1}{3}V_{N}^{0}J_{V}^{0N}+V_{N}^{3}J_{V}^{3N}\Big) 
\end{eqnarray}
or more explicitly

\begin{eqnarray}
\label{InteractionActionEM}
S_{Dirac}^{EM} [i,X]=&& \frac{1}{2g_5^2}\, \int dz d^{4}y \sqrt{g} e^{- \tilde\kappa^2 z^2} \,\frac{z}{R}
\nonumber\\
&&\times \bigg(\bar\Psi_{1X}  \gamma^N  \Big(\frac{1}{3}V_N^{0}T^0+ V_N^{3}T^3\Big)\Psi_{1i} 
+\bar{\Psi}_{2X} \, \gamma^N  \Big(\frac{1}{3}V_N^{0}T^0+ V_N^{3}T^3\Big)\Psi_{2i} \bigg) \,,\nonumber\\
=&&(2\pi )^4 \delta^4 ( P_X -p - q )\times F_X(P_X)\times F_N(p)\nonumber\\
&&\times\,\frac{1}{2g_5^2}\times 2g_5^2\times e_{nucleon}\times\bar{u}_{s_{X}}(P_X) \slashed\epsilon(q) u_{s_{i}}(p)\times\frac{1}{2}\Big[ {\cal I}_L (n_X, Q^2 )  
+ {\cal I}_R (n_X, Q^2 ) \Big]\,,\nonumber\\ 
\end{eqnarray}
following the conventions for the interaction action used in~\cite{BRAGA}. 
Here  $\epsilon^\mu(q)$ is the polarization of the EM probe.
In the last equality in (\ref{InteractionActionEM}) we substituted the bulk gauge fields by (\ref{Gauge}), and the bulk fermionic currents in terms of the fermionic fields (\ref{SolutionFermions}).
The charge assignments are $e_{nucleon}=1$ for the proton, and  $e_{nucleon}=0$ for the neutron. We have also defined

\be
 {\cal I}_L (n_X, Q^2 ) \equiv {\cal I}(M+5/2,n_X,  Q^2)\qquad {\cal I}_R (n_x , Q^2)\equiv {\cal I}(M+3/2, n_X, Q^2) 
 \ee
 with ($w=\tilde\kappa^2z^2$)

\bea
\label{generalintegral}
{\cal I} ( {\bar m}, n_X , Q^2)  = && C (  {\bar m}, n_X  ) \, 
\Gamma\bigg( 1 + \frac{Q^2}{4 \tilde\kappa^2} \bigg) \, \int_0^\infty dw w^{{\bar m} -1} e^{-w} {\cal U} (1+\frac{Q^2}{4\tilde\kappa^2} ; 2 ; w ) \,  L_{n_X}^{{\bar m} - 2}( w )\nonumber\\
=&&\frac{\Gamma \left(\bar{m}\right) 
\bigg({\frac{\Gamma \left(\bar{m}-1\right) \Gamma \left(n_X +\bar{m}-1\right)}{\Gamma (n_X +1)}}\bigg)^{\frac 12}}
{ \Gamma \left(\bar{m}-1\right) }
\frac{\frac {Q^2}{4\tilde\kappa^2}{\Gamma \left(\frac{Q^2}{4 \tilde\kappa ^2}+n_X\right)}}
{\Gamma \left(\frac{Q^2}{4 \tilde\kappa ^2}+n_X+\bar{m}\right)}
\eea
and the normalization
 
\begin{equation} 
C ( {\bar m}, n_X )  =  \bigg({\frac{ \Gamma(n_X +1)}{\Gamma({\bar m} -1 )\Gamma(n_X + {\bar m} -1 )}}\bigg)^{\frac 12}\,, 
\end{equation} 
Here  we have set $Q^2=-q^2$ (space-like),  and used the final state mass shell condition 

\be
\label{nx1}
P_X^2=(p+q)^2=M_0^2+Q^2\bigg(\frac 1x-1\bigg)\equiv M_X^2=4{\tilde \kappa}^2\bigg(n_X+M+\frac 12\bigg)
\ee
to identify $n_X=Q^2(1/x-1)/4{\tilde \kappa}^2$, with  $p^2=M_0^2=4{\tilde \kappa}^2(M+\frac 12 )=8\tilde\kappa^2$ for the
initial nucleon state.

The Dirac part of the electromagnetic current (\ref{DIRECTEM}) can be extracted from (\ref{InteractionActionEM}) using

\bea
{W}_{EM(Dirac)}^\mu(Q^2)=\bar{u}_{s'}(p') \gamma^{\mu}u_{s}(p)\times F_1^{EM(Dirac)}(Q)\equiv \frac{1}{F_N(p')F_N(p)}\frac{\delta S_{Dirac}^{EM}}{\delta \epsilon_{\mu}(q)}+\mathcal{O}(N_c^{-2})
\eea
which amounts to the  Dirac or minimal  contribution to the electromagnetic form factor  ($Q^2<0$)

\bea
F_{1}^{EM(Dirac)}(Q^2)&=&\frac{1}{2g_5^2}\times 2g_5^2\times e_{nucleon}\times\frac{1}{2}\Big[{\cal I}_L (n_X=0, Q^2 )  + {\cal I}_R (n_X=0, Q^2 ) \Big]+\mathcal{O}(N_c^{-2}),
\eea
or more  explicitly  in the soft wall model

\bea
\label{FEMDIRAC}
F_{1}^{EM(Dirac)}(Q)&=&e_{nucleon}
\bigg(
\frac 1{(\frac {Q^2}{4\tilde\kappa^2}+1)(\frac {Q^2}{4\tilde\kappa^2}+2)}+
\frac 3{(\frac {Q^2}{4\tilde\kappa^2}+1)(\frac {Q^2}{4\tilde\kappa^2}+2)(\frac {Q^2}{4\tilde\kappa^2}+3)}
\bigg)+\mathcal{O}(N_c^{-2})\nonumber\\
&=&e_{nucleon}
\bigg(\frac{m_0^4}{(Q^2+m_0^2)(Q^2+m_1^2)}+\frac{3m_0^6}{(Q^2+m_0^2)(Q^2+m_1^2)(Q^2+m_2^2)}\bigg)+\mathcal{O}(N_c^{-2})\,.
\eea
using the soft wall rho meson trajectory $m_n^2=4\tilde\kappa^2(n+1)$.
Note that normalizing $F_{1}^{EM(Dirac)}(0)=1$ for  the proton, fixes $1+\mathcal{O}(N_c^{-2})=1$. In other words, the $1/N_c$ corrections to the EM form factors must vanish at $Q=0$ due to the conservation of the electromagnetic  current. Also note that (\ref{FEMDIRAC}) asymptotes a dipole  form.

\subsubsection{Direct axial form factor: Dirac}

The direct axial form factors are derived using the same reasoning, with 
${W}_{A}^{\tilde{\alpha}\mu}(Q^2)$ (\ref{DIRECT})  following from the pertinent variation of the bulk Dirac action

\begin{eqnarray}
\label{InteractionActionAxial03pm}
S_{Dirac}^{(0,3)Axial} [i,X]=&&\frac{1}{2g_5^2}\, \int dz d^{4}y \sqrt{g} e^{- \tilde\kappa^2 z^2}A_{N}^{(0,3)}J_{A}^{(0,3)N} \nonumber\\
S_{Dirac}^{\pm Axial} [i,X]=&&\frac{1}{2g_5^2}\frac{1}{\sqrt{2}}\int dz d^{4}y \sqrt{g} e^{- \tilde\kappa^2 z^2}A_{N}^{\pm}J_{A}^{\mp N} 
\end{eqnarray}
or more explicitly  for the chargeless component,

\begin{eqnarray}
\label{InteractionActionAxial03}
S_{Dirac}^{(0,3)Axial} [i,X]
=&& \frac{1}{2g_5^2}\, \int dz d^{4}y \sqrt{g} e^{- \tilde\kappa^2 z^2} \,\frac{z}{R}
 \bigg(\bar\Psi_{1X} \, \gamma^N\, A_{N}^{(0,3)}T^{(0,3)}\,\Psi_{1i} -\bar{\Psi}_{2X}  \gamma^N\, A_{N}^{(0,3)}T^{(0,3)}\,\Psi_{2i}\bigg)\nonumber\\
\approx &&(2\pi )^4 \delta^4 ( P_X -p - q )\,\times F_X(P_X)\times F_N(p)\nonumber\\
&&\times\,\frac{1}{2g_5^2}\times 2g_5^2\times  g_{Anucleon}^{(0,3)}\times\bar{u}_{s_{X}}(P_X) 
\slashed{\epsilon}(q)\gamma^5 u_{s_{i}}(p)\times\frac{1}{2}\Big[{\cal I}_R (n_X, \tilde Q^2 )-{\cal I}_L (n_X, \tilde Q^2 )\Big]\,,\nonumber\\
\end{eqnarray}
where the substitution (\ref{SUB02}) was used in the second line, i.e.

\bea
\label{QSUB}
\tilde Q^2=Q^2+\frac 12 m_q^2\rightarrow Q^2+\tilde m_0^2-m_0^2
\eea
Similarly, for the charged components

\begin{eqnarray}
\label{InteractionActionAxialpm}
S_{Dirac}^{\pm Axial} [i,X]\equiv && \frac{1}{2g_5^2}\times \frac{1}{\sqrt{2}}\int dz d^{4}y \sqrt{g} e^{- \kappa^2 z^2} \,\frac{z}{R}
\bigg(\bar\Psi^{p/n}_{1X} \, \gamma^N\, A_{N}^{\pm}\,\Psi^{n/p}_{1i} -\bar{ \Psi}^{p/n}_{2X} \gamma^N\, A_{N}^{\pm}\,\Psi^{n/p}_{2i} \bigg) \,,\nonumber\\
=&&(2\pi )^4 \delta^4 ( P_X -p - q )\times F_X(P_X)\times F_N(p)\nonumber\\
&&\times \frac{1}{2g_5^2}\times 2g_5^2\,\times\frac{1}{\sqrt{2}}\times e_{Wnucleon}^{\pm}\times\bar{u}_{s_{X}}(P_X) 
\slashed{\epsilon}^\pm(q)\gamma^5 u_{s_{i}}(p)\times\frac{1}{2}\Big[{\cal I}_R (n_X, \tilde Q^2 )-{\cal I}_L (n_X, \tilde Q^2 )\Big]\,,\nonumber\\
\end{eqnarray}
Here $\epsilon^\pm_\mu(q)$ is the polarization of the charged vectors.
Following the normalization of the flavor  generators $T^a$  in (\ref{01}),  the  axial and electroweak charges are

\begin{eqnarray}
\label{axialcharges}
&&g_{Anucleon}^{0}=g_{Aproton}^{0}=g_{Aneutron}^{0}=\frac{3}{2}\nonumber\\
&&g_{Anucleon}^{3}=g_{Aproton}^{3}=\frac{1}{2}\nonumber\\
&&g_{Anucleon}^{3}=g_{Aneutron}^{3}=-\frac{1}{2}\nonumber\\
&&e_{Wnucleon}^{\pm}=e_{Wproton}^{\pm}=e_{Wneutron}^{\pm}=1
\end{eqnarray}


Finally, the isoscalar axial form factor can be extracted from (\ref{InteractionActionAxial03}) using

\bea
\label{isoscalar}
{\cal W}_{A(Dirac)}^\mu(Q^2)\equiv {\cal W}_{A}^{0\mu}(Q^2)
=\bar{u}_{s'}(p') \gamma^{\mu}\gamma^5 u_{s}(p)\times F_1^{0A(Dirac)}(Q)=
\frac{1}{F_N(p')F_N(p)}\frac{\delta S_{Dirac}^{0Axial}}{\delta \epsilon_{\mu}(q)}+\mathcal{O}(N_c^{-2})
\eea
which yields the minimal Dirac  contribution to the isoscalar axial form factor as

\bea
\label{011}
F_{1}^{0A(Dirac)}(Q)&\approx &\frac{1}{2g_5^2}\times 2g_5^2\times g_{Anucleon}^{0}\times\frac{1}{2}\Big[{\cal I}_R (n_X=0, \tilde Q^2 )  
- {\cal I}_L (n_X=0, \tilde Q^2 ) \Big]+\mathcal{O}(N_c^{-2})\,.
\eea
More explicitly,   in the soft wall model  we have

\bea
\label{FADIRAC}
F_{1}^{0A(Dirac)}(Q)\approx g_{Anucleon}^{0}\bigg(
\frac 1{(\frac {Q^2+\tilde m_0^2}{4\tilde\kappa^2})(\frac {Q^2+\tilde m_0^2}{4\tilde\kappa^2}+1)}-
\frac 3{(\frac {Q^2+\tilde m_0^2}{4\tilde\kappa^2})(\frac {Q^2+\tilde m_0^2}{4\tilde\kappa^2}+1)(\frac {Q^2+\tilde m_0^2}{4\tilde\kappa^2}+2)}
\bigg)+\mathcal{O}(N_c^{-2})\nonumber\\
\eea
following (\ref{FEMDIRAC}). (\ref{FADIRAC}) reduces to a dipole form factor  asymptotically.  Recall that the substitution 
(\ref{QSUB}) in the bulk-to-boundary axial-vector propagator is justified for $Qz_0\gg1$, as we noted in (\ref{VQZ-approx}). 
For $Qz_0\rightarrow 0$, (\ref{011}-\ref{FADIRAC})  give

\bea
\label{022}
g_A^{3(Dirac)}=\frac 13 g_A^{0(Dirac)}  \approx \bigg[\frac{\tilde m_0^2}{m_0^2}-1\bigg]
\bigg[\bigg(\frac{\tilde m_0^2}{m_0^2}\bigg)\bigg(\frac{\tilde m_0^2}{m_0^2}+1\bigg)\bigg(\frac{\tilde m_0^2}{m_0^2}+2\bigg)\bigg]^{-1}
\rightarrow  \frac 16 \bigg[\frac{\tilde m_0^2}{m_0^2}-1\bigg]+{\cal O}((\sigma z^3_0)^3)
\eea
with the rightmost result corresponding to the leading perturbative correction as in (\ref{SUB0}) for $\sigma z^3_0\ll 1$. 
Using the soft wall parameters in~\cite{KKSS} (model A) yields $m_0=0.775$ GeV and $\tilde m_0=1.363$ GeV, 
resulting in a large Dirac leading perturbative contribution to the isovector axial charge of the nucleon $g_A^{3(Dirac)}\approx 1/3$.
The near-boundary approximation is not justified in this limit, with the exact but numerical  bulk-to-boundary axial-vector propagator required.

\subsubsection{Pion pole}

Finally, we note that the axial-vector form factor is characterized by  two invariant form factors $F_A(q^2)$ and $H_A(q^2)$ in general, which are 
defined as

\be
\label{FAHA}
\left<p_1\left|\tilde J_A^{\mu a}(0)\right|p_2\right>=\overline{u}(p_1)\left(\gamma^\mu\gamma^5F_A(q^2)+q^\mu\gamma^5 H_A(q^2)\right)T^a\,u(p_1)
\ee
with $q=p_2-p_1$. In the chiral limit, the two form factors are tied by the conservation of the axial-vector current
$H_A(q^2)=-2m_NF_A(q^2)/q^2$ with $H_A(q^2\approx 0)\approx -2m_Ng_A/q^2$ exhibiting the pion pole. The absence of this
contribution in (\ref{InteractionActionAxialpm}) can be traced back to the  $A_5$ field  which we have ignored.
 As we noted above, 
the $A_5$ mixes with the tachyon $X$.
When taken into proper consideration, this mixing after gauge fixing locks $A_5$ with the phase $\Pi$ of the tachyon,
i.e. $X=X_0e^{i\Pi}$, which is identified with the pion field. Careful considerations in bulk yield precisely $H_A(q^2)$ as expected
from current conservation in the chiral limit~\cite{ABIDINK}. Here we can just reinstate  this contribution  by inspection, with the full pion pole 
$q^2\rightarrow q^2-m_\pi^2$.

\end{widetext}

\subsection{Transition form factors}

In DIS scattering of neutrinos off nucleons, we will need the transition form factors of the left chargeless and charged currents. 
For that, we define the left vector transition form factor for the process   $V,L+p\rightarrow X$ 

\be
\label{TRAN}
{\cal W}_{V,L}^\mu(Q^2)= \left<X\left|\tilde J_{V,L}^\mu(0)\right|p\right>\nonumber\\
\ee
with $Q^2=(P_X-p)^2$. We first evaluate the $U(1)_L$  Dirac  and Pauli  contributions to the transition vector form factor
(\ref{TRAN}) and then generalize  them to the corresponding $U(1)^{EM}_{V}, U(1)^{\pm}_L\subset U(2)_L$ contributions.

\subsubsection{$U(1)_L$ contributions}

The minimal Dirac interaction term between the bulk $U(1)_L$ gauge field $L_{N}$ and the bulk fermionic field $\Psi_1$ in the action is

\begin{eqnarray}
\label{InteractionAction}
S_{Dirac}^{L} [i,X] =  \frac{1}{2g_5^2}\, \int dz d^{4}y \sqrt{g} e^{- \tilde\kappa^2 z^2} \,\frac{z}{R}
\, L_N \bar\Psi_{1X} \, \gamma^N \, \Psi_{1i} \,.\nonumber\\
\end{eqnarray}
following the canonical normalizations

\bea
\label{SUBX}
\Psi\rightarrow \sqrt{2g_5^2}\,\Psi\qquad L_{N}\rightarrow g_5L_{N}
\eea
which makes the couplings and power counting manifest in Witten diagrams. The explicit form of (\ref{InteractionAction}) in terms of the
bulk fermions is

\begin{widetext}
\begin{eqnarray}
\label{InteractionAction2}
\tilde{S}_{Dirac}^{L} [i,X] =&&(2\pi )^4 \delta^4 ( P_X -p - q )\times F_X(P_X)\times F_N(p)\nonumber\\
&&\times \frac 1{2g^2_5}\times 2g_5^2\,\Big[ \bar{u}_{s_{X}}(P_X) \slashed\epsilon(q)\Big(\frac{1 - \gamma^5}{2} \Big) u_{s_{i}}(p) {\cal I}_L (n_x, Q^2 )  +  \,\bar{u}_{s_{X}}(P_X) 
\slashed{\epsilon}(q) \Big(\frac{1 + \gamma^5}{2} \Big) u_{s_{i}}(p) {\cal I}_R (n_x, Q^2 ) \Big]   \,.\nonumber\\
\end{eqnarray}
The Dirac contribution to the $U(1)_L$ transition form factor reads explicitly

\bea
{\cal W}_{L,Dirac}^\mu(q^2)=&&\frac{1}{F_X(P_X)F_N(p)}\frac{\delta S_{Dirac}^{L}}{\delta \epsilon_{\mu}(q)}+\mathcal{O}(N_c^{-2})\nonumber\\
=&&\frac{1}{F_X(P_X)F_N(p)}\times\frac{1}{2g_5^2}\times\frac{1}{g_5}\times\frac{\delta \tilde S_{Dirac}^{L}}{\delta \epsilon_{\mu}(q)}+\mathcal{O}(N_c^{-2})\nonumber\\
=&&\frac{1}{2g_5^2}\times 2g_5^2\times\,\Big[ \bar{u}_{s_{X}}(P_X) \gamma^\mu \Big(\frac{1 - \gamma^5}{2} \Big) u_{s_{i}}(p) {\cal I}_L (n_x , Q^2)  +  \,\bar{u}_{s_{X}}(P_X) \gamma^\mu \Big(\frac{1 + \gamma^5}{2} \Big) u_{s_{i}}(p) {\cal I}_R (n_x , Q^2) \Big]+\mathcal{O}(N_c^{-2}) 
\nonumber\\
\eea

\subsubsection{$U(1)^{\pm}_L$ contributions}

For neutrino and anti-neuutrino scattering, the pertinent  left-handed hadronic transition form factor is

\bea
\label{TRANSITIONPM}
{\cal W}_{\pm}^\mu(Q^2)=&&\left<N(P_{X})\left|\tilde J^{\pm\mu}_{L}(0)\right|N(p)\right>
\eea
which can be evaluated using the Dirac and Pauli contributions to the fermionic action. To illustrate the normalizations, consider for simplicity
the contribution due to the Dirac part, which yields the interaction vertex

\begin{eqnarray}
\label{InteractionActionAxialpm}
&&S_{Dirac}^{\pm L} [i,X]=\frac{1}{2g_5^2}\times\frac{1}{\sqrt{2}}\int dz d^{4}y \sqrt{g} e^{- \tilde\kappa^2 z^2}L_{N}^{\pm}J_{L}^{\mp N}=\frac{1}{2g_5^2}\times\frac{1}{\sqrt{2}}\int dz d^{4}y \sqrt{g} e^{- \tilde\kappa^2 z^2} \,\frac{z}{R}
\bar\Psi_{1X}^{p/n} \, \gamma^N\, L_{N}^{\pm}\,\Psi_{1i}^{n/p}\nonumber\\
&&=(2\pi )^4 \delta^4 ( P_X -p - q )\times\,\frac{1}{2g_5^2}\times 2g_5^2\times F_X(P_X)F_{p/n}(p)\nonumber\\
&&\times \frac{1}{\sqrt{2}}\times e_{Wnucleon}^{\pm}\times
\Big[ \bar{u}_{s_{X}}(P_X) \slashed\epsilon^\pm(q)\Big(\frac{1 - \gamma^5}{2} \Big) u_{s_{i}}(p) {\cal I}_L (n_x, Q_L^2 )  +  \,\bar{u}_{s_{X}}(P_X) 
\slashed{\epsilon}^\pm(q) \Big(\frac{1 + \gamma^5}{2} \Big) u_{s_{i}}(p) {\cal I}_R (n_x,  Q_L^2 ) \Big]\,,\nonumber\\
\end{eqnarray}
with the corresponding transition form factor

\bea\label{TRANSITIONPM2}
&&{\cal W}_{L(Dirac)}^{\pm\mu}(Q^2)=\frac{1}{F_X(P_X)F_{p/n}(p)}\frac{1}{\frac{1}{2\sqrt{2}}}\frac{\delta S_{Dirac}^{\pm L}}{\delta\epsilon^\pm_{\mu}(q)}+\mathcal{O}(N_c^{-2})\nonumber\\
&&=\frac{1}{2g_5^2}\times 2g_5^2\times e_{Wnucleon}^{\pm}\nonumber\\
&&\times2\Big[\bar{u}_{s_{X}}(P_X) \gamma^{\mu}\Big(\frac{1 - \gamma^5}{2} \Big) u_{s_{i}}(p) {\cal I}_L (n_x,Q_L^2 )  +  \,\bar{u}_{s_{X}}(P_X) 
\gamma^{\mu} \Big(\frac{1 + \gamma^5}{2} \Big) u_{s_{i}}(p) {\cal I}_R (n_x, Q_L^2 ) \Big]+\mathcal{O}(N_c^{-2})\,\nonumber\\
&&=\frac{1}{2g_5^2}\times 2g_5^2 \times e_{Wnucleon}^{\pm}\nonumber\\
&&\times  \bigg(\bar{u}_{s_{X}}(P_X) \gamma^{\mu} u_{s_{i}}(p)\Big[{\cal I}_R (n_x, Q_L^2 )+ {\cal I}_L (n_x, Q_L^2 ) 
\Big]+\bar{u}_{s_{X}}(P_X) \gamma^{\mu}\gamma^5 u_{s_{i}}(p)\Big[{\cal I}_R (n_x, Q_L^2 ) - {\cal I}_L (n_x, Q_L^2 ) \Big]\bigg)+\mathcal{O}(N_c^{-2})\,.\nonumber\\
\eea

\subsubsection{Charged transition form factor: Pauli}

The minimal Dirac bulk interaction does not contribute to the neutron transition form factor. This contribution arises from the $U(1)_L$ part of the 
Pauli interaction in bulk~\cite{CARLSON}

\bea
\label{PAULI}
S_{Pauli}^{L,R} [i,X] = \frac { \eta_{} }{2g_5^2}\int dz d^{4}y \sqrt{g} e^{- \tilde\kappa^2 z^2} \,\frac{z^2}{R^2}
\bigg(\bar\Psi_{1X }\,\sigma^{MN}\,L_{MN}\,\Psi_{1i }\,-\bar\Psi_{2X} \,\sigma^{MN}\,R_{MN} \Psi_{2i }\bigg)
\eea
The inclusion of this interaction is straightforward but tedious, with the final result for the charged currents

\bea
\label{LEFTFULL}
{\cal W}_{L}^{\pm\mu}(Q^2) &&=e_{Wnucleon}^{\pm}\nonumber\\
&&\times  \bigg(\bar{u}_{s_{X}}(P_X) \gamma^{\mu} u_{s_{i}}(p)
\Big[{\cal I}_R (n_x, Q_L^2 )+ {\cal I}_L (n_x, Q_L^2 )\Big]\nonumber\\
&&\qquad+\bar{u}_{s_{X}}(P_X) \gamma^{\mu}\gamma^5 u_{s_{i}}(p)
\Big[{\cal I}_R (n_x, Q_L^2 )- {\cal I}_L (n_x, Q_L^2 \Big]\bigg)\nonumber\\
&&+\eta^\pm \bigg(\bar{u}_{s_{X}}(P_X) \gamma^{\mu} u_{s_{i}}(p)
\Big[{\cal J}_R (n_x, Q_L^2 )+ {\cal J}_L (n_x, Q_L^2 \Big]\nonumber\\
&&\qquad+\bar{u}_{s_{X}}(P_X) \gamma^{\mu}\gamma^5 u_{s_{i}}(p)
\Big[{\cal J}_R (n_x, Q_L^2 )-{\cal J}_L (n_x, Q_L^2 )\Big]\bigg)\nonumber\\
&&+  2\eta^\pm\,\bigg(\bar{u}_{s_{X}}(P_X) \sigma^{\mu\nu}iq_\nu u_{s_{i}}(p)
\Big[{\cal I}_{LR}(n_x, Q_L^2 )+ {\cal I}_{RL}(n_x, Q_L^2 ) \Big]\nonumber\\
&&\qquad +\bar{u}_{s_{X}}(P_X) \sigma^{\mu\nu}iq_\nu\gamma^5 u_{s_{i}}(p)
\Big[{\cal I}_{LR} (n_x, Q_L^2 )- {\cal I}_{RL}(n_x, Q_L^2 )\Big]\bigg)\nonumber\\
&&-\eta^\pm\,\bigg(\bar{u}_{s_{X}}(P_X)q^\mu\slashed q u_{s_{i}}(p)
\Big[{\cal K}_{R}(n_x, Q_L^2 )+ {\cal K}_{L}(n_x, Q_L^2 ) \Big]\nonumber\\
&&\qquad +\,\bar{u}_{s_{X}}(P_X) q^\mu\slashed q\gamma^5 u_{s_{i}}(p)
\Big[{\cal K}_{R} (n_x, Q_L^2 )- {\cal K}_{L}(n_x, Q_L^2 )\Big]\bigg)+\mathcal{O}(N_c^{-2})\,.
\eea
with


\be
\sigma^{\mu\nu}=2S^{\mu\nu}=\frac{i}{2}[\gamma^\mu,\gamma^\nu]=i(\gamma^\mu\gamma^\nu-\eta^{\mu\nu})\,,
\ee
and  manifest current conservation, i.e. $q_\mu{\cal W}^{\pm \mu\nu}=0$ for on-shell spinors.
The integrals 

\be
 {\cal J}_L,{\cal K}_L (n_X, Q^2 ) \equiv {\cal J,K}(M+5/2,n_X,  Q^2)\qquad {\cal J}_R, {\cal K}_R (n_X , Q^2)\equiv {\cal J,K}(M+3/2, n_X, Q^2) 
 \ee
are related to the general integrals of the type (\ref{generalintegral}), namely

\bea
\label{generalintegralX}
{\cal J} ( {\bar m}, n_X , Q^2)  &=&  C (  {\bar m}, n_X  ) \, 
\Gamma( 1 + a ) \, \int_0^\infty dw w^{{\bar m} -1} e^{-w} \bigg(w{\cal U} (1+a ; 2 ; w )\bigg)^\prime \,  L_{n_X}^{{\bar m} - 2}( w )\nonumber\\
&=&C (  {\bar m}, n_X  ) \, 
\Gamma( 1 + a) \, \int_0^\infty dw w^{{\bar m} -1} e^{-w} \bigg(-a\,{\cal U} (1+a ; 1 ; w )\bigg) \,  L_{n_X}^{{\bar m} - 2}( w )\nonumber\\
&=& C (  {\bar m}, n_X  )\frac{\Gamma ({\bar m})\Gamma ({\bar m}+n_X-1)(n_X-a({\bar m}-1))a\Gamma (a+n_X)}{\Gamma (n_X+1)\Gamma \left(a+{\bar m}+n_X+1\right)}\,\nonumber\\
&=& \frac{-a({\bar m}-1)+n_X}{a+{\bar m}+n_X}\times{\cal I} ( {\bar m}, n_X , Q^2)\,,\nonumber\\
{\cal K} ( {\bar m}, n_X , Q^2) &=& \frac{1}{\tilde{\kappa}^2}\times {\cal J} ( {\bar m}, n_X , Q^2)\,,\nonumber\\
\eea
with  $a= {Q^2}/{4 \tilde\kappa^2}$, and

\bea
\label{generalintegralXX}
{\cal I}_{LR}(n_X,Q^2) &=&\frac{1}{\tilde{\kappa}}\times C (m+5/2, n_X  ) \, 
\Gamma( 1 + a )\, \int_0^\infty dw w^{m+1} e^{-w} {\cal U} (1+\frac{Q^2}{4\tilde\kappa^2} ; 2 ; w ) \,  L_{n_X}^{m+\frac{1}{2}}( w )\nonumber\\
{\cal I}_{RL}(n_X,Q^2)&=& \frac{1}{\tilde{\kappa}}\times C (  m+3/2, n_X  ) \, 
\Gamma( 1 + a ) \, \int_0^\infty dw w^{m+2} e^{-w} {\cal U} (1+\frac{Q^2}{4\tilde\kappa^2} ; 2 ; w ) \,  L_{n_X}^{m - \frac{1}{2}}( w )\,.\nonumber\\
\eea
We have made use of the recursive relation between the  confluent hypergeometric functions 

\begin{equation}
w \;\partial_w \;{\cal U}(a,b,w) = (1-b)\; {\cal U}(a,b,w) - (1+a-b)\;{\cal U}(a,b-1,w)\,,
\end{equation}
together with their  integral representation 

\be
{\cal U}(a,b,w)=\frac{1}{\Gamma(a) }\int _0^{\infty } dt\,t^{a-1}(t+1)^{-a+b-1} e^{-tw}\,,
\ee
and the property of the Laguerre polynomials
\be
L_n^a(w)=\frac{\Gamma(a+1+n)}{\Gamma(a+1)\Gamma(n+1)}\,_1F_1\left(-n,a+1,w\right)\,,
\ee
to evaluate the integrals. 

\subsubsection{$U(1)^{EM}_V\subset U(2)_V$ contributions}

The electromagnetic  transition vector form factor  ${\cal W}_{EM}^\mu(Q^2)$ as defined through

\bea
\label{TRANSITIONEM}
{\cal W}_{EM}^\mu(Q^2)=\left<N(P_{X})\left|\tilde{J}_{EM}^\mu(0)\right|N(p)\right>=
\left<N(P_X)\left|\frac{1}{3}\tilde{J}_{V}^{0\mu}(0)+\tilde{J}_{V}^{3\mu}(0)\right|N(p)\right>
\eea
can be extracted from (\ref{InteractionActionEM}) with the inclusion of the Pauli contribution (\ref{PAULI}), using

\bea\label{TRANSITIONEM2}
{\cal W}_{EM(Dirac)}^\mu(Q^2)=&&\frac{1}{F_X(P_X)F_N(p)}\frac{\delta S_{Dirac+Pauli}^{EM}}{\delta \epsilon_{\mu}(q)}+\mathcal{O}(N_c^{-2})\nonumber\\
\eea
The result is

\bea
\label{EMFULL}
\epsilon_{\mu}{\cal W}_{EM}^\mu(Q^2)&& =\epsilon_{\mu}\bigg(e_{nucleon}\
\times  \bar{u}_{s_{X}}(P_X) \gamma^{\mu} u_{s_{i}}(p)\,\frac 12\,
\Big[{\cal I}_R (n_x, Q^2 )+ {\cal I}_L (n_x, Q^2)\Big]\nonumber\\
&&+\eta^{p/n}\, \bar{u}_{s_{X}}(P_X) \gamma^{\mu} u_{s_{i}}(p)\,\frac 12\,
\Big[ {\cal J}_R (n_x, Q^2)- {\cal J}_L (n_x, Q^2) \Big]\nonumber\\
&&+  \eta^{p/n}\,\bar{u}_{s_{X}}(P_X) \sigma^{\mu\nu}iq_\nu u_{s_{i}}(p)
\Big[{\cal I}_{LR}(n_x, Q^2 )- {\cal I}_{RL}(n_x, Q^2 ) \Big]\bigg)+\mathcal{O}(N_c^{-2})\,.
\eea

\subsubsection{$U(1)_A\subset U(2)_A$ contributions}

The chargeless axial transition form factor  ${\cal W}_{0,3}^\mu(Q^2)$ as defined through

\bea
\label{TRANSITIONAXIAL}
{\cal W}_A^{(0,3)\mu }(Q^2)=\left<N(P_{X})\left|\tilde{J}^{(0,3)\mu}_{A}(0)\right|N(p)\right>
\eea
follows the same reasoning, with the full result including the Dirac and Pauli contributions

\bea
\label{AXIALFULL}
\epsilon_\mu^{(0,3)}{\cal W}_A^{(0,3)\mu }(Q^2)&&=\epsilon_\mu^{(0,3)} \bigg(g^{(0,3)}_{Anucleon}
\times  \bar{u}_{s_{X}}(P_X) \gamma^{\mu}\gamma^5 u_{s_{i}}(p)\,\frac 12\,
\Big[{\cal I}_R (n_x, \tilde Q^2 )- {\cal I}_L (n_x, \tilde Q^2)\Big]\nonumber\\
&&+\eta^{(0,3)}\, \bar{u}_{s_{X}}(P_X) \gamma^{\mu}\gamma^5 u_{s_{i}}(p)\,\frac 12\,
\Big[{\cal J}_R (n_x, \tilde Q^2)+ {\cal J}_L (n_x, \tilde  Q^2)\Big]\nonumber\\
&&+  \eta^{(0,3)}\,\bar{u}_{s_{X}}(P_X) \sigma^{\mu\nu}iq_\nu \gamma^5u_{s_{i}}(p)
\Big[{\cal I}_{LR}(n_x, \tilde  Q^2 )+ {\cal I}_{RL}(n_x, \tilde  Q^2 ) \Big]\bigg)+\mathcal{O}(N_c^{-2})\,.
\eea

\section{Neutrino and anti-neutrino DIS scattering in QCD}

In QCD lepton nucleon scattering follows from the contraction of the leptonic tensor and hadronic tensor through the
exchange of neutral currents carried by $\gamma, Z$ and  charged currents  carried by $W^{\pm}$. 
Some useful insights on standard neutrino DIS scattering on  a nucleon can be found in~\cite{REVIEW}.
In this  section we briefly review the key definitions and characteristics of this scattering  as a prelude to the holographic 
analysis, which will make use of the transition form factors established above for the s-wave contributions to DIS.

\subsection{Structure functions for $\nu, \overline \nu$ scattering}

An overall review of neutrino DIS scattering can be found in~\cite{REVIEW}, so we will be brief in our presentation of the results
for our ensuing analysis. 
For unpolarized nucleons, the hadronic 
tensor for neutrino (antineutrino) scattering can be organized under the strictures of Lorentz symmetry, parity and current conservation in terms of three
invariant structure functions

\begin{eqnarray}
\label{1}
W^{\nu, \overline \nu}_{\mu\nu}=&&\frac{1}{4\pi}\int d^4x\,e^{iq\cdot x}\left<P\left|\Big[\tilde J^\mp_{\mu}(x),\tilde J^\pm_\nu(0)\Big]\right|P\right>\nonumber\\
=&&\left(-\eta_{\mu\nu}+\frac{q_\mu q_\nu}{q^2}\right)F^{W^\pm}_1(x,q^2)
+\left(P_\mu+\frac{q_\mu}{2x}\right)\left(P_\nu+\frac{q_\nu}{2x}\right)\,\frac{2x}{q^2}F^{W^\pm}_2(x, q^2)\mp i\epsilon_{\mu\nu\alpha\beta}q^\alpha P^\beta\,\frac{x}{q^2}F^{W^\pm}_3(x, q^2)\nonumber\\
\end{eqnarray}
$q_\mu, P_\mu$ are the 4-momenta of the virtual current and nucleon respectively. 
$x=-q^2/2P\cdot q$ is the Bjorken parameter which is kinematically bounded $0\leq x\leq 1$. 
$F_{1,2}$ refer to the symmetric structure functions,
while $F_3$ to the antisymmetric one. The formers are parity preserving, while the latter is not for neutrino and antineutrino probes. 
In the DIS limit with $Q^2=-q^2\gg P^2$ and $x$ fixed, the structure functions in QCD obey Bjorken scaling. In this limit, the parity even structure
functions satisfy the Callan-Gross relation $F_2=2x F_1$. 
We note that analyticity allows to relate the hadronic DIS tensor (\ref{1}) to the discontinuity of the forward Compton amplitude
of a lepton on a nucleon

\be
\label{2}
4\pi W^{\nu, \overline \nu}_{\mu\nu}=2\pi{\rm Im}\,T^{\nu, \overline \nu}_{\mu\nu}\equiv 2\pi{\rm Im}\,i\int d^4x e^{iq\cdot x} \left<P\right|T^*\left(\tilde J^\mp _\mu(x)\tilde J^\pm_\nu(0)\right)\left|P\right>
\ee

To calculate the hadronic tensor (\ref{1}) we use the completeness of the hadronic spectrum f
\be
\label{1X}
W^{\nu, \overline \nu}_{\mu\nu}=\frac{1}{2} \sum_{s,s_X}  \sum_{M_X} \,\delta \Big( M_X^2 - (P+q)^2 \, \Big)\,{\cal W}^{\nu, \overline \nu *}_\mu\,{\cal W}^{\nu, \overline \nu}_\nu\equiv W^{\nu, \overline \nu S}_{\mu\nu}+iW^{\nu, \overline \nu A}_{\mu\nu}\nonumber\\
\ee
with the transition current matrix element (\ref{TRAN}), 
for excited states of squared mass $M_X^2=Q^2(1/x-1)+m_N^2$, and $P^2=M_0^2=m_N^2$.
For neutrino scattring the explicit form of the QCD quark (partonic) currents in (\ref{2}) are given by

\bea
\label{3}
e\tilde J^\mu_{EM}=&&e\,\overline q Q\gamma^\mu q\nonumber\\
e_W\times 2\tilde J^{\pm\mu}_{L}=&&e_W\,\overline qT^\pm \gamma^\mu \frac 12 {(1-\gamma^5)} q\nonumber\\
e_W\tilde J^\mu_Z =&& e_W\,\tilde J^{\mu}_{L}-2e_W\,{\rm sin^2\theta_W}\tilde J^\mu_{EM}=e_W\,\overline q\gamma^\mu \frac 12 {(1-\gamma^5)}q-2e_W\,{\rm sin^2\theta_W}\,\overline qQ\gamma^\mu q\nonumber\\
\eea
with $e, e_W$ the electric and electroweak charges, and ${\rm sin\theta_W}=\frac{e}{e_W}$ with the Weinberg angle $\theta_W\approx \frac {\pi}6$.
\be
\label{4}
Q=\begin{pmatrix} 
  \frac 23  & 0\\ 
 0& -\frac 13
\end{pmatrix}\qquad 
T^+=(T^-)^\dagger=
\begin{pmatrix} 
  0& 2\\ 
  0 & 0
\end{pmatrix}
\ee
The imaginary part in (\ref{2}) receives contribution from the neutral $\gamma\gamma$, $ZZ$, $\gamma Z$ 
as they mix, and the charged $W^\mp  W^\pm$ as they are conjugate. 

\end{widetext}

\subsection{Unpolarized parton distributions  through charged currents}

Neutrino and anti-neutrino scattering on a hadron through the charged currents yield very
informative information on the parton content of a hadron in the DIS limit. At weak coupling,
the parton model gives a very simple descriptive of the structure functions in terms of the
partonic distribution functions of the hadron. Assuming two flavors for simplicity and isospin symmetry, the partonic
model for $\nu p\rightarrow l^-X$ through $W^+$ exchange gives

\be
\label{5}
F_1^{W^+p}(x,Q)=&&d(x,Q)+\overline u(x,Q)\nonumber\\
F_2^{W^+p}(x,Q)=&&2x(d(x,Q)+\overline u(x,Q))\nonumber\\
F_3^{W^+p}(x,Q)=&&2(d(x,Q)-\overline u(x,Q))\nonumber\\
\ee
while for $\overline\nu p\rightarrow l^+X$ through $W^-$ exchange it gives

\be
\label{6}
F_1^{W^-p}(x,Q)=&&u(x,Q)+\overline d(x,Q)\nonumber\\
F_2^{W^-p}(x,Q)=&&2x(u(x,Q)+\overline d(x,Q))\nonumber\\
F_3^{W^-p}(x,Q)=&&2(u(x,Q)-\overline d(x,Q))\nonumber\\
\ee
modulo $e_W^2$.
The corresponding neutron structure functions follow by isospin symmetry. 
(\ref{5}-\ref{6}) can be inverted to give the unpolarized valence and sea parton distributions in the proton.
We now proceed to evaluate these unpolarized partonic distributions
using the holographic dual of neutrino scattering.

\begin{widetext}

\section{Neutrino and anti-neutrino DIS in Holography}

We now consider neutrino  DIS scattering on a nucleon as a Dirac fermion in bulk using holography, 
in the double limit of a large number of colors and strong gauge coupling $\lambda=g^2N_c$. 
Anti-neutrino DIS scattering follows from  pertinent re-arrangements.
DIS scattering on a nucleon as a bulk dilatino using holography was first addressed in~\cite{POL} and later by 
 others~\cite{BRAGA,HATTAX,KOVENSKY,Mamo:2019mka}. At large-x, DIS scattering using $U(1)_V$ probes follows from the direct and crossed 
Witten diagrams in bulk, and at small-x it follows from Pomeron exchange in the form of a Reggeized and warped 
close string exchange in bulk. We now review the analysis for a $U(1)_V$ current and then extend it to the electromagnetic vector $U(1)^{EM}_V\subset U(2)_V$, and the left-handed $U(1)^{\pm}_L\subset U(2)_L$ currents.


\subsection{Structure functions: Baryonic exchange in the s-channel}

For unpolarized $U(1)_L$ scattering, the hadronic tensor is 
\begin{equation}
W_{L}^{\mu\nu} \, = \frac{1}{4 \pi} \sum_s \, \int d^4y\, e^{iq\cdot y} \langle P, s \vert \, \Big[\tilde J_{L}^\mu (y), \tilde J_{L}^\nu (0)\Big] 
\, \vert P, s  \rangle \,,
 \label{HadronicTensor}
\end{equation}
with the spectral decomposition for the s-channel contributions

\begin{eqnarray}
\label{hadronic2}
W_{L,s}^{\mu\nu} & = & \frac{1}{4 \pi} \sum_{s,s_X}  \sum_{M_X} \int \frac{d^4P_X}{(2\pi)^4} \theta (P^0_X )
(2\pi)\delta \Big( P^2_X -  M_X^2 \, \Big) (2\pi )^4 \delta^4 (P +q - P_X ) 
\, \langle P,s \vert \tilde J_{L}^\mu ( 0 )   \vert P_X , s_X  \rangle\,
\langle P_X ,\, s_X  \vert \tilde J_{L}^\nu ( 0 )   \vert P,\, s \rangle\,
\cr
& =& \frac{1}{2} \sum_{s,s_X}  \sum_{M_X} \,\delta \Big( M_X^2 - (P+q)^2 \, \Big) 
\langle P,s \vert \tilde J_{L}^\mu ( 0 )   \vert P + q, s_X  \rangle\,
\langle P + q,\, s_X  \vert \tilde J_{L}^\nu ( 0 )   \vert P,\, s \rangle\,
\cr
&=&\frac{1}{2} \sum_{s,s_X}  \sum_{M_X} \,\delta \Big( M_X^2 - (P+q)^2 \, \Big)\,{\cal W}_{L}^\mu\,{\cal W}_{L}^{\nu\,*}\,.\cr & &
\end{eqnarray}
The t-channel Reggeized contributions will be addressed below.

\subsection{Dirac contribution}

Using the relations

\begin{eqnarray}
\epsilon_\mu \langle P_X  \vert {\tilde J}_{L,Dirac}^\mu ( q )   \vert P  \rangle\,
&=&  (2 \pi)^4 \, \delta^4 ( P_X - P - q ) \,\epsilon_\mu \,  \langle P + q \vert \tilde J_{L,Dirac}^\mu ( 0 )  \vert P \rangle
\,= \, \mathcal{N}_{L}\times\frac{1}{g_5}\times \tilde{S}_{int,Dirac}^{L} [i,X]\, \cr \cr
\epsilon_\mu \langle P  \vert {\tilde J}_{L,Dirac}^\mu ( q )   \vert P_X  \rangle\,
&=&  (2 \pi)^4 \, \delta^4 ( P_X - P - q ) \,\epsilon_\mu \,  \langle P \vert \tilde J_{L,Dirac}^\mu ( 0 )  \vert P + q \rangle
\,= \, \mathcal{N}_{L}\times\frac{1}{g_5}\times \tilde{S}_{int,Dirac}^{L} [X,i]\,,\cr & & 
\label{Prescription}
\end{eqnarray}
we can make explicit the contracted hadronic tensor

\begin{eqnarray}
\label{Amplitudewithspinors}
& & W^{\mu\nu}_{L,s,Dirac} \, = \,\frac12 \,\sum_{M_X} \delta ( M_X^2 - (p+q)^2 ) \,\mathcal{N}_{L}^2\nonumber\\
&&\times   \sum_{ s_i } \sum_{s_X} \, \bigg({\cal I}_L {\cal I}_R
 \Big( \bar{u}_{s_{X}} \gamma^\mu P_{L} u_{s_{i}} \bar{u}_{s_{i}} \gamma^\nu P_{R} u_{s_{X}} + \bar{u}_{s_{X}} \gamma^\mu  P_{R} u_{s_{i}} \bar{u}_{s_{i}} \gamma^\nu P_{L} u_{s_{X}} \Big)
\nonumber\\&&\qquad\qquad \,\,\, +  {\cal I}^2_L  \bar{u}_{s_{X}} \gamma^\mu P_{L} u_{s_{i}} \bar{u}_{s_{i}} \gamma^\nu P_{L} u_{s_{X}}
\, + {\cal I}^2_R \bar{u}_{s_{X}}\gamma^\mu P_{R} u_{s_{i}} \bar{u}_{s_{i}} \gamma^\nu 
P_{R} u_{s_{X}}\bigg)
\end{eqnarray}
The additional normalization constant
$\mathcal{N}_{L}=({1}/{2g_5^2})\times 2g_5^2\times\tilde{\mathcal{N}}_{L}$ compensates  for the missing higher spin-j and $\mathcal{O}(g_5^0)$ corrections in the s-channel. 
Since   $\sum_s (u_s)(p)(\bar{u}_s) (p) = \not\!p+M$, then the contracted hadronic tensor contributions from the s-channel are

\begin{eqnarray}
\label{Amplitude}
&&\epsilon_\mu \epsilon_\nu W_{L,s,Dirac}^{\mu\nu}  =   \, \sum_{M_X} \delta ( M_X^2 - (p+q)^2 ) \,  \, \mathcal{N}_{L}^2\nonumber\\
&&\times 2\bigg({\cal I}_R (n_x ) {\cal I}_L (n_x ) M_X  M_0 \,\epsilon \cdot \epsilon
+ \Big(  {\cal I}^2_R (n_x ) + {\cal I}^2_L (n_x ) \Big) \Big(( p\cdot \epsilon )^2 - \frac{1}{2} ( p^2 + p \cdot q )\, 
\epsilon \cdot \epsilon \Big) +\frac{1}{2}\Big({\cal I}_R^2(n_x)-{\cal I}_L^2(n_x)\Big)\Big(i\epsilon^{\mu\nu\alpha\beta}P_{X\alpha}p_\beta\Big)\bigg)\nonumber\\
\end{eqnarray}
By approximating the sum by an integral in a continuous state, i.e.,
\be
\sum_{M_X} \delta ( M_X^2 - (p+q)^2 )\approx\frac{1}{4\tilde\kappa^2}\int dn\,\delta\Big(\frac{s}{4\tilde\kappa^2}-\frac{M_n^2}{4\tilde\kappa^2}\Big)=\frac{1}{4\tilde\kappa^2}\,,
\ee
we have

\begin{eqnarray}
\label{ApproximateAmplitude}
&&\epsilon_\mu \epsilon_\nu W_{L,s,Dirac}^{\mu\nu} \approx \frac{ \mathcal{N}_{L}^2}{4\tilde\kappa^2}\nonumber\\
&\times &\bigg(2{\cal I}_L (n_x ) {\cal I}_R (n_x ) M_X  M_0 \,\epsilon \cdot \epsilon
 + \Big(  {\cal I}^2_R (n_x ) + {\cal I}^2_L (n_x ) \Big) \Big( 2 ( p\cdot \epsilon )^2 - ( p^2 + p \cdot q )\, 
\epsilon \cdot \epsilon \Big) +\Big({\cal I}_R^2(n_x)-{\cal I}_L^2(n_x)\Big)\Big(i\epsilon^{\mu\nu\alpha\beta}P_{X\alpha}p_\beta\Big)\bigg)\nonumber\\
\end{eqnarray}
The same contraction applied to the canonical hadronic tensor decomposition (\ref{1}), for neutrino, with a transverse polarization $\epsilon \cdot q = 0$ (which we have already used in deriving (\ref{ApproximateAmplitude}), yields

\begin{equation}\label{Amplitude2} 
 \epsilon_\mu \epsilon_\nu W_{L}^{\mu\nu}  \,=\, -\epsilon^2  F_{1}^{L}  \,+\, \frac{2 x}{q^2 } 
  ( \epsilon \cdot p )^2 \, F_{2}^{L} -i\epsilon_\mu\epsilon_\nu\epsilon^{\mu\nu\alpha\beta}q_\alpha p_\beta\,\frac{x}{q^2}F^{L}_3(x, q^2)\,,
\end{equation} 
A comparison of (\ref{Amplitude2})  to (\ref{ApproximateAmplitude}),  allows for the extraction of the s-channel baryonic contributions to the DIS structure functions

\begin{eqnarray}
\label{Result1}
 F_{1s,Dirac}^{L}  &=&\frac{ \mathcal{N}_{L}^2}{2 \tilde\kappa^2  } 
\bigg( \Big(  {\cal I}^2_R (n_x ) + {\cal I}^2_L (n_x ) \Big) \Big( \frac{M_0^2}{2 } + 
\frac{Q^2}{4 x} \Big) - {\cal I}_R (n_x ) {\cal I}_L (n_x ) M_0
 \bigg({ M_0^2 + Q^2\left(\frac 1x -1\right) }\bigg)^{\frac 12} \bigg)\nonumber\\
 F_{2s,Dirac}^{L} & = &\frac{ \mathcal{N}_{L}^2}{4 \tilde\kappa^2 } 
 \Big(  {\cal I}^2_R (n_x ) + {\cal I}^2_L (n_x ) \Big) \frac{ Q^2 }{  x}\,, \nonumber\\
 F_{3s,Dirac}^L&=&\frac{ \mathcal{N}_{L}^2}{4 \tilde\kappa^2 } 
 \Big(  {\cal I}^2_R (n_x ) - {\cal I}^2_L (n_x ) \Big) \frac{ Q^2 }{  x}
\end{eqnarray}
with the DIS kinematics (\ref{nx1}) subsumed. 
Similarly, using the transition form factor (\ref{TRANSITIONEM2}),  we  find
\bea
F_{1s,Dirac}^{EM}&=&e_{nucleon}^2\times F_{1s,Dirac}^{L}\nonumber\\
F_{2s,Dirac}^{EM}&=&e_{nucleon}^2\times F_{2s,Dirac}^{L}\nonumber\\
F_{3s,Dirac}^{EM}&=&0 \,.\nonumber\\
\eea
while using the transition form factor (\ref{TRANSITIONPM2}), we find

\bea
F_{1s,Dirac}^{W^{\pm}}&=&4\big(e^{\pm}_{Wnucleon}\big)^2\times F_{1s,Dirac}^{L}=\frac{4\big(e^{\pm}_{Wnucleon}\big)^2}{e_{nucleon}^2}\times F_{1s,Dirac}^{EM}\nonumber\\
F_{2s,Dirac}^{W^{\pm}}&=&4\big(e^{\pm}_{Wnucleon}\big)^2\times F_{2s,Dirac}^{L}=\frac{4\big(e^{\pm}_{Wnucleon}\big)^2}{e_{nucleon}^2}\times F_{2s,Dirac}^{EM}\nonumber\\
F_{3s,Dirac}^{W^{\pm}}&=&4\big(e^{\pm}_{Wnucleon}\big)^2\times F_{3s,Dirac}^{L}\,.\nonumber\\
\eea


\subsection{Dirac plus Pauli contributions}

The full left current vertex with the Pauli contribution is given in (\ref{LEFTFULL}). A rerun of the preceding arguments yields the hadronic tensor for $\nu$ scattering in the
untraced form ( $\bar\nu$ scattering follows from $\nu\leftrightarrow \mu$)

\begin{eqnarray}
W_L^{+\mu\nu}=\frac 1{16\pi{\tilde\kappa}^2}&&\bigg[2M_0M_X\bigg(\alpha_R^2g^{\mu\nu}+(\beta_R^2+2\alpha_R\beta_R)\,q^\mu q^\nu+\lambda_R^2(g^{\mu\nu}q^2-q^\mu q^\nu)\bigg)\nonumber\\
&&+2M_0\lambda_R \bigg(\alpha_R g^{\mu\nu} q\cdot (P_X+p)+\beta_R q^\mu q^\nu q\cdot (P_X+p)-i\alpha_R\epsilon^{\nu\mu\alpha\beta}q_\alpha (P_X+p)_\beta\bigg)\nonumber\\
&&+2\bigg(\alpha_Rg^\mu_{\bar \mu}+\beta_Rq^\mu q_{\bar \mu}\bigg)\bigg(\alpha_Rg^\nu_{\bar \nu}+\beta_Rq^\nu q_{\bar \nu}\bigg)\, \bigg(p^{\bar \mu}P_X^{\bar \nu}+p^{\bar \nu}P_X^{\bar \mu}-g^{\bar \mu\bar \nu}p\cdot P_X-i\epsilon^{\bar \nu \alpha \bar \mu \beta}P_{X\alpha}p_\beta\bigg)\nonumber\\
&&-\frac 12 \lambda_R^2
\bigg(3g^{\mu\nu}(2q\cdot p q\cdot P_X-q^2p\cdot P_X)\nonumber\\
&&\qquad\qquad+4\bigg(q^2(p^\mu P_X^\nu+p^\nu P_X^\mu)+q^\mu q^\nu p\cdot P_X
-(q^\nu p^\mu+q^\mu p^\nu)q\cdot P_X-(q^\nu P_X^\mu+q^\mu P_X^\nu)q\cdot p\bigg)\bigg)\nonumber\\
&&+(R\rightarrow L\,,- i\rightarrow +i)\bigg]
\end{eqnarray}
with  $P_X=q+p$, $M_X^2=M_0^2+q^2(1-1/x)$ and  

\begin{eqnarray}
\alpha_R=2(e^+_W{\cal I}_R+\eta{\cal J}_R)\qquad \beta_R=4\eta{\cal I}_{LR}\qquad \lambda_R=-2\eta{\cal K}_R
\end{eqnarray}

\bea\label{WP}
&&\epsilon_{\mu}^{+}\epsilon_\nu^{+*}\,W_{L,s}^{+\mu\nu}=\frac{1}{2} \sum_{s,s_X}  \sum_{M_X} \,\delta \Big( M_X^2 - (P+q)^2 \, \Big)\,\epsilon_{\mu}^{+}{\cal W}_{L}^{+\mu}\,\epsilon_{\nu}^{+*}{\cal W}_{L}^{+*\nu}\,\nonumber\\
&&=\frac{\big(\mathcal{N}_L^+\big)^{2}}{8\tilde\kappa^2}\epsilon_{\mu}^{+}\epsilon_\nu^{+*}\times 8\Big[2\tilde{{\cal I}}_{LR}(n_x)M_0M_{X}\eta^{\mu\nu}+\tilde{{\cal I}}_{+}^2(n_x)\Big(p^\mu P_X^\nu +p^\nu P_X^\mu -\eta^{\mu\nu}p\cdot P_X\Big)+\tilde{{\cal I}}_{-}^2(n_x)\Big(i\epsilon^{\mu\nu\alpha\beta}P_{X\alpha}p_\beta\Big)\Big]
\nonumber\\
\eea
where we have defined

\bea
\tilde{{\cal I}}_{LR}(n_x)&\equiv & {\cal I}_{\eta L}(n_x){\cal I}_{\eta R}(n_x)+4\eta^{2}q^2{\cal I}_{RL}(n_x){\cal I}_{LR}(n_x)\nonumber\\
\tilde{{\cal I}}_{+}^2(n_x)&\equiv & {\cal I}_{\eta R}^2(n_x)+{\cal I}_{\eta L}^2(n_x)-4\eta^{2}q^2\Big({\cal I}_{LR}^2(n_x)+{\cal I}_{RL}^2(n_x)\Big)\nonumber\\
\tilde{{\cal I}}_{-}^2(n_x)&\equiv &{\cal I}_{\eta  R}^2(n_x)-{\cal I}_{\eta L}^2(n_x)\nonumber\\
{\cal I}_{\eta  R}(n_x)&\equiv &
e^{+}_{Wnucleon}{\cal I}_{R}(n_x)+\eta  {\cal J}_{R}(n_x)\nonumber\\
{\cal I}_{\eta  L}(n_x)&\equiv &
e^{+}_{Wnucleon}{\cal I}_{L}(n_x)+\eta  {\cal J}_{L}(n_x)\,,
\eea
and we have used $\epsilon^+\cdot q=0$, $\epsilon^{+*}\cdot q=0$, $q_{\tilde\nu}q_{\tilde\mu}\epsilon^{\tilde{\nu}\beta\tilde{\mu}\nu}=0$, and $q_{\tilde\nu}q_{\tilde\mu}\epsilon^{\tilde{\mu}\alpha\mu\tilde{\nu}}=0$.

Comparing (\ref{WP}) with the contraction of (\ref{1}), for neutrino $\nu$, i.e.,
\bea
\epsilon_{\mu}^{+}\epsilon_\nu^{+*}\,W_{L}^{+\mu\nu}=\epsilon_{\mu}^{+}\epsilon_\nu^{+*}\,W^{\nu\,,\mu\nu}=-\epsilon^{+}\cdot\epsilon^{+*}\,F_{1}^{W^{+}}  \,+\, \frac{2 x}{q^2 }\,\epsilon^{+}\cdot p\,\epsilon^{+*}\cdot p \, F_{2}^{W^{+}} -i\epsilon^{+}_\mu\epsilon^{+*}_\nu\epsilon^{\mu\nu\alpha\beta}q_\alpha p_\beta\,\frac{x}{q^2}F^{W^{+}}_3(x, q^2)\,,\nonumber\\
\eea
allows for the extraction of the structure functions

\begin{eqnarray}
\label{SFWP}
 F_{1s}^{W^{+}}  &=&\frac{ 2\big(\mathcal{N}_{L}^+\big)^2}{\tilde\kappa^2  } 
\bigg(\tilde{{\cal I}}_{+}^2(n_x)\Big( \frac{M_0^2}{2 } + 
\frac{Q^2}{4 x} \Big) - \tilde{{\cal I}}_{LR}(n_x) M_0
 \bigg({ M_0^2 + Q^2\left(\frac 1x -1\right) }\bigg)^{\frac 12} \bigg)\,,\nonumber\\
 F_{2s}^{W^{+}} & = &\frac{ \big(\mathcal{N}_{L}^+\big)^2}{\tilde\kappa^2  }\, 
 \tilde{{\cal I}}_{+}^2(n_x) \frac{ Q^2 }{  x}\,, \nonumber\\
 F_{3s}^{W^{+}}&=&\frac{ \big(\mathcal{N}_{L}^+\big)^2}{\tilde\kappa^2  }\, 
 \tilde{{\cal I}}_{-}^2(n_x)\frac{ Q^2 }{  x}\,,
\end{eqnarray}
where we have used $P_{X_{\alpha}}=p_\alpha+q_\alpha$, $p_{\alpha}p_{\beta}\epsilon^{\mu\nu\alpha\beta}=0$, 
$\epsilon^+\cdot q=0$, $\epsilon^{+*}\cdot q=0$, and $M_X^2=M_0^2+q^2(1-1/x)$. Strict bulk-to-boundary correspondence implies 
${\cal N}_L^+=1$ in the double limit of a large number of colors $N_c$ and strong gauge t Hooft gauge coupling.
  Here we follow~\cite{BRAGA} and assume proportionality between the bulk and boundary structure functions with
${\cal N}_L^+$ an overall parameter that captures parts of the finite corrections to the strict double limit. It will be fixed by a point
in the data.
Similarly, we can find the structure functions for anti-neutrino $\tilde\nu$ scattering through $W^-$ exchange as

\begin{eqnarray}
\label{SFWM}
 F_{1s}^{W^{-}}  &=& F_{1s}^{W^{+}}(e^{+}_{Wnucleon}\rightarrow e^{-}_{Wnucleon}\,;\,{\cal N}_L^+\rightarrow{\cal N}_L^-)\,,\nonumber\\
 F_{2s}^{W^{-}} & = & F_{2s}^{W^{+}}(e^{+}_{Wnucleon}\rightarrow e^{-}_{Wnucleon}\,;\,{\cal N}_L^+\rightarrow{\cal N}_L^-)\,, \nonumber\\
 F_{3s}^{W^{-}}&=& F_{3s}^{W^{+}}(e^{+}_{Wnucleon}\rightarrow e^{-}_{Wnucleon}\,;\,{\cal N}_L^+\rightarrow{\cal N}_L^-)\,.
\end{eqnarray}

\subsection{Structure functions: Pomeron exchange in the t-channel}

DIS scattering at small-x is dominated by Reggeon and Pomeron exchanges. In this section, we first consider DIS scattering
using $U(1)_L$ currents in the Pomeron regime and then generalize our results to the electromagnetic vector 
$U(1)^{EM}_V\subset U(2)_V$, and the left-handed $U(1)^{\pm}_L\subset U(2)_L$ currents. The Pomeron is dual
to a close string exchange or graviton in the t-channel~\cite{SIN,JANIK,POLX,HATTAX,KOVENSKY}. 
This is best obtained by recalling that the hadronic tentor ties to
the forward scattering amplitude of a $U(1)_L$ current through

\begin{equation}
W_{L}^{\mu\nu} \, = \frac{1}{4 \pi} \sum_s \, \int d^4y\, e^{iq\cdot y} \langle P, s \vert \, \Big[ \tilde J_{L}^\mu (y) , \tilde J_{L}^\nu (0) \Big] 
\, \vert P, s  \rangle \, = 2\pi{\rm Im}\,T_{L}^{\mu\nu}\,,
 \label{HadronicTensor}
\end{equation}
where the Compton scattering amplitude $T_{L}^{\mu\nu}$ is given by

\begin{equation}
\epsilon_{\mu}\epsilon_{\nu}T_{L,t}^{\mu\nu}\equiv{\cal A}^{h}_{L p\rightarrow L p} (s,t)\,,
\end{equation}
for massive graviton or glueball $h_{\mu\nu}$ exchange in the t-channel, with the explicit result  given in (\ref{pomeron3}). 
The Pomeron  contribution as a graviton exchange to the t-channel structure functions $F^{L}_{1t,2t}(x,Q)$ follow from

\begin{equation}
\epsilon_{\mu}\epsilon_{\nu}W_{L,t}^{\mu\nu}\, =\, \epsilon^2  F_{1t}^{L}  \,+\, \frac{2 x}{q^2 } 
  ( \epsilon \cdot p )^2 \, F_{2t}^{L}\,= \,2\pi{\rm Im}\,{\cal A}^{tot}_{L  p\rightarrow L p} (s,t=-K^2=0)\,,
 \label{HadronicTensor-j}
\end{equation}
for $\epsilon \cdot q=0$. Inserting (\ref{pomeron33}) into (\ref{HadronicTensor-j}) we obtain

\bea
\label{F1F2T}
2xF_{1t}^{L}&=&\frac{2\kappa^2}{g_5^2}\times\frac{\pi}{\sqrt{\lambda}}\times\Big(\frac{Q}{\tilde{\kappa}}\Big)^{2-2/\sqrt{\lambda}}\times\Big(\frac{1}{x}\Big)^{1-2/\sqrt{\lambda}}
\nonumber\\
&\times&{\rm exp}\bigg[-\frac{\tilde{\xi}^2}{2}\frac{\sqrt{\lambda}}{\log[Q^2/\tilde{\kappa}^2]+\log[1/x]}\bigg]\times\bigg(\frac{\sqrt{\lambda}}{\log[Q^2/\tilde{\kappa}^2]+\log[1/x]}\bigg)^{3/2}\nonumber\\  
&\times &(2\pi)^{1/2}\tilde{\xi}\left(1 + {\cal O}\bigg(\frac{\sqrt{\lambda}}{\log[Q^2/\tilde{\kappa}^2]+\log[1/x]}\bigg) \right)\times\mathcal{F}(j_0,K=0)\times I_{\xi}^T(j_0,Q=Q')\,,\nonumber\\
F_{2t}^{L}&=&\frac{2\kappa^2}{g_5^2}\times\frac{\pi}{\sqrt{\lambda}}\times\Big(\frac{Q}{\tilde{\kappa}}\Big)^{2-2/\sqrt{\lambda}}\times\Big(\frac{1}{x}\Big)^{1-2/\sqrt{\lambda}}
\nonumber\\
&\times&{\rm exp}\bigg[-\frac{\tilde{\xi}^2}{2}\frac{\sqrt{\lambda}}{\log[Q^2/\tilde{\kappa}^2]+\log[1/x]}\bigg]\times\bigg(\frac{\sqrt{\lambda}}{\log[Q^2/\tilde{\kappa}^2]+\log[1/x]}\bigg)^{3/2}\nonumber\\   
&\times &(2 \pi)^{1/2}\tilde{\xi}\left(1 + {\cal O}\bigg(\frac{\sqrt{\lambda}}{\log[Q^2/\tilde{\kappa}^2]+\log[1/x]}\bigg) \right)\times\mathcal{F}(j_0,K=0)\times \Big( I_{\xi}^T(j_0,Q=Q')+I_{\xi}^L(j_0,Q=Q')\Big)\,,\nonumber\\
\eea
with $\tilde\xi=\gamma+\pi/2$. 
The pre-exponents in (\ref{F1F2T}) are commensurate with the expected Pomeron behavior $s^{\alpha_{\mathbb P}(0)-1}$  with the intercept
$\alpha_{\mathbb P}(0)-1=1-2/\sqrt{\lambda}$ after the identification ${s}\sim Q^2/x$ in the DIS limit. The additional overall factor of $Q$ is
the left over (longitudinal) polarization from the overlaping incoming-outgoing $U(1)_L$ wavefunctions. The exponents reflect on the warped Gribov diffusion
in $D_\perp=3$,  for $2_\perp$ spatial dimensions and $1_\perp$ holographic dimension.
The result (\ref{F1F2T}) extends to the SU(2) currents by introducing a normalization factor $\mathcal{N}_L^{a=\pm}$.
More specifically, for the charged currents we have

\bea
\label{F1F2T2}
2xF_{1t}^{W^{\pm}}&=&\big(\mathcal{N}_{Lt}^{\pm}\big)^2\frac{2\kappa^2}{g_5^2}\times\frac{\pi}{\sqrt{\lambda}}\times\Big(\frac{Q}{\tilde{\kappa}}\Big)^{2-2/\sqrt{\lambda}}\times\Big(\frac{1}{x}\Big)^{1-2/\sqrt{\lambda}}\nonumber\\
&\times&{\rm exp}\bigg[-\frac{\tilde{\xi}^2}{2}\frac{\sqrt{\lambda}}{\log[Q^2/\tilde{\kappa}^2]+\log[1/x]}\bigg]\times\bigg(\frac{\sqrt{\lambda}}{\log[Q^2/\tilde{\kappa}^2]+\log[1/x]}\bigg)^{3/2}\nonumber\\  
&\times &(2\pi)^{1/2}\tilde{\xi}\left(1 + {\cal O}\bigg(\frac{\sqrt{\lambda}}{\log[Q^2/\tilde{\kappa}^2]+\log[1/x]}\bigg) \right)\times\mathcal{F}(j_0,K=0)\times I_{\xi}^T(j_0,Q=Q')\,,\nonumber\\
F_{2t}^{W^{\pm}}&=&\big(\mathcal{N}_{Lt}^{\pm}\big)^2\frac{2\kappa^2}{g_5^2}\times\frac{\pi}{\sqrt{\lambda}}\times\Big(\frac{Q}{\tilde{\kappa}}\Big)^{2-2/\sqrt{\lambda}}\times\Big(\frac{1}{x}\Big)^{1-2/\sqrt{\lambda}}\nonumber\\
&\times&{\rm exp}\bigg[-\frac{\tilde{\xi}^2}{2}\frac{\sqrt{\lambda}}{\log[Q^2/\tilde{\kappa}^2]+\log[1/x]}\bigg]\times\bigg(\frac{\sqrt{\lambda}}{\log[Q^2/\tilde{\kappa}^2]+\log[1/x]}\bigg)^{3/2}\nonumber\\   
&\times &(2 \pi)^{1/2}\tilde{\xi}\left(1 + {\cal O}\bigg(\frac{\sqrt{\lambda}}{\log[Q^2/\tilde{\kappa}^2]+\log[1/x]}\bigg) \right)\times\mathcal{F}(j_0,K=0)\times \Big( I_{\xi}^T(j_0,Q=Q')+I_{\xi}^L(j_0,Q=Q')\Big)\,.\nonumber\\
\eea

\subsection{$F_{3t}^{W^{\pm}}$ structure functions: Reggeon exchange in the t-channel}

Reggeon exchange is also a t-channel contribution stemming from a  spin-1 exchange induced by the Chern-Simons contribution
in (\ref{1}). The latter  allows for the anomalous coupling of $W^-W^+\omega$ in bulk with $\omega_\mu$
a spin-1 flavor singlet U(1) gauge field in bulk (the analogue of the omega-meson). In principle, the Reggeized spin-1 exchange  in bulk
contributes  to the unpolarized and parity odd structure function $F_3$. A similar contribution was observed  for the spin structure function 
in~\cite{KOVENSKY}, following  an earlier  analysis  in~\cite{HATTAX}.
More specifically, the U(1) exchange of $L_{\mu}^{0}$ in bulk stems from  (\ref{fermionAction2}) and  (\ref{CHERNLLL}), with the vertices

\bea
 L\overline\Psi\Psi:\quad &&\int \frac{d^4 p_2  d^4 p_1d^4k}{(2\pi)^{12}}(2\pi)^4 \delta^4(p_2-k-p_1)\,S^k_{L\bar\Psi\Psi}\nonumber\\
LLL:\quad && \int \frac{d^4q'  d^4qd^4k}{(2\pi)^{12}}(2\pi)^4 \delta^4(q-k-q')\,S^k_{LLL}\nonumber\\
\label{vertices33}
\eea
where we have defined

\bea
S^{k}_{L\bar\Psi\Psi}
&=&g_{5}^3\,\kappa_{CS}\int dz\,\bigg[
\epsilon^{\mu\sigma\nu\rho}\epsilon_{\mu}^+(q)\epsilon_{\sigma}^-(q')\partial_z L^{+}(q,z)\,(-ik_{\nu})\epsilon_{\rho}^{0}(k)L^{0}(k,z)L^{-}(q^\prime,z)\nonumber\\
&-&\epsilon^{\sigma\mu\nu\rho}\epsilon_{\sigma}^+(q)\epsilon_{\mu}^-(q')\partial_z L^{-}(q',z)\,(-ik_{\nu})\epsilon_{\rho}^{0}(k)L^{0}(k,z)L^{+}(q,z)\nonumber\\
&+&\epsilon^{\mu\sigma\rho\nu}\epsilon_{\mu}^+(q)\epsilon_{\sigma}^-(q')\partial_z L^{+}(q,z)\,\epsilon_{\rho}^{0}(k)L^{0}(k,z)\,(-iq'_{\nu})L^{-}(q^\prime,z)\nonumber\\
&-&\epsilon^{\sigma\mu\rho\nu}\epsilon_{\sigma}^+(q)\epsilon_{\mu}^-(q')\partial_z L^{-}(q',z)\,\epsilon_{\rho}^{0}(k)L^{0}(k,z)\,(iq_{\nu})L^{+}(q,z)\nonumber\\
&-&\epsilon^{\sigma\rho\mu\nu}\epsilon_{\sigma}^{+}(q)\epsilon_{\rho}^-(q')\epsilon_{\mu}^0(k)\partial_z L^{0}(k,z)\,(-iq'_{\nu})L^{-}(q',z)L^{+}(q,z)\nonumber\\
&+&\epsilon^{\rho\sigma\mu\nu}\epsilon_{\rho}^+(q)\epsilon_{\sigma}^{-}(q^\prime)\epsilon_{\mu}^0(k)\partial_z L^{0}(k,z)(iq_{\nu})L^{+}(q,z)\,L^{-}(q^\prime,z)\bigg]\nonumber\\
&=&g_{5}^3\,\kappa_{CS}(-i)\epsilon^{\mu\sigma\nu\rho}\epsilon_{\mu}^+(q)\epsilon_{\sigma}^-(q')(q'_{\nu}+q_{\nu})\epsilon_{\rho}^{0}(k)\nonumber\\
&\times&\int dz\,\big(\partial_z L^{0}(k,z)L(q',z)L(q,z)-\partial_z L(q,z)L^{0}(k,z)L(q',z)
\label{vertices555}
\eea
with  

\be
L_{\mu}^{+}(x,z)=\epsilon^{+}_{\mu}(q)L^{+}(q,z)e^{iq\cdot x}\,, \qquad L_{\mu}^{-}(x,z)=\epsilon^{-}_{\mu}(q')L^{-}(q',z)e^{-iq'\cdot x}\,,\qquad 
L_{\mu}^{0}(x,z)=\epsilon^{0}_{\mu}(k)L^{+}(k,z)e^{-ik\cdot x}\nonumber\\
\ee
In the last line in (\ref{vertices555}), we have used the fact that $L^{+}(q,z)=L^{-}(q,z)=L(q,z)$. 

Using the relation 

\be
{\rm Im}\,{\cal A}^{tot}_{L  p\rightarrow L p} (s,t=0)=\epsilon^{+\mu}\epsilon^{-\nu}W^{\nu}_{\mu\nu}
\ee
 with $W^{\nu}_{\mu\nu}$  given by (\ref{1}), and ${\cal A}^{tot}_{L  p\rightarrow L p} (s,t)$  given by (\ref{reggeon3222}), we can extract the structure function $F_{3}^{W^\pm}(x,Q^2)$ in the t-channel as  
\be \label{F3T}
F_{3t}^{W^\pm}&=&4\times 4\times g_5\times\frac{1}{g_5^2}\times g_5^3\kappa_{CS}\times\Big(\frac{Q}{\tilde{\kappa}}\Big)^{2-j_0-\Delta(j_0)}\times\Big(\frac{s}{\tilde{\kappa}^2}\Big)^{j_0}\times\frac{e^ {-\sqrt\lambda  \tilde{\xi}^2 / 2\log[s/\tilde{\kappa}^2]}}{(\log[s/\tilde{\kappa}^2])^{3/2}}\nonumber\\
&\times & (\sqrt{\lambda}/ 2 \pi )^{1/2}\; \tilde{\xi}  \;\left(1 + {\cal O}\bigg(\frac{\sqrt{\lambda}}{\log[s/\tilde{\kappa}^2]}\bigg) \right)\times I_{\xi}(j_0,Q,Q'=Q)\times \mathcal{F}_1^{(LN)}(j_0,K=0)\nonumber\\
&=&16\times g_5^2\kappa_{CS}\times\Big(\frac{Q}{\tilde{\kappa}}\Big)^{1-\frac{1}{\sqrt{\lambda}}}\times\Big(\frac{1}{x}\Big)^{1-\frac{1}{\sqrt{\lambda}}}\times\frac{e^ {-\sqrt\lambda  \tilde{\xi}^2 /2(\log[Q^2/\tilde{\kappa}^2]+\log[1/x])}}{(\log[Q^2/\tilde{\kappa}^2]+\log[1/x])^{3/2}}\nonumber\\
&\times & (\sqrt{\lambda}/ 2 \pi )^{1/2}\; \tilde{\xi}  \;\left(1 + {\cal O}\bigg(\frac{\sqrt{\lambda}}{\log[Q^2/\tilde{\kappa}^2]+\log[1/x]}\bigg) \right)\times I_{\xi}(j_0,Q,Q'=Q)\times \mathcal{F}_1^{(LN)}(j_0,K=0)\,,\nonumber\\
\ee
Here  $s\approx \frac{Q^2}{x}$, $j_0=1-1/\sqrt{\lambda}$, $\Delta(j_0)=2$, and $\tilde{\xi}-\pi/2=\gamma=0.55772.....$ is Euler-Mascheroni constant.

\section{Holographic results and comparison to data}

(\ref{5}-\ref{6}) can be inverted to give the unpolarized valence and sea parton distributions in the proton,
in terms of the pertinent holographic structure functions. 
The  ones for the neutron follow  by isospin symmetry. More specifically,  we have for the unpolarized sea
of the proton

\be
\label{7b}
x\overline u(x,Q)=&&\frac 14 \left(F_2^{W^+p}(x,Q)-xF_3^{W^+p}(x,Q)\right)\,,\nonumber\\
x\overline d(x,Q)=&&\frac 14 \left(F_2^{W^-p}(x,Q)-xF_3^{W^-p}(x,Q)\right)\,,\nonumber\\
\ee
and for the unpolarized valence contribution of the proton

\bea
\label{7Xb}
xu_V(x,Q)=x(u(x,Q)-\overline u(x,Q))&=&\frac{1}{4} x\left(F_3^{W^-p}(x,Q)+F_3^{W^+p}(x,Q)\right)\nonumber\\
&+&\frac 14\left(F_2^{W^-p}(x,Q)-F_2^{W^+p}(x,Q)\right)\,,\nonumber\\
xd_V(x,Q)=x(d(x,Q)-\overline d(x,Q))&=&\frac{1}{4} x\left(F_3^{W^-p}(x,Q)+F_3^{W^+p}(x,Q)\right)\nonumber\\
&-&\frac 14\left(F_2^{W^-p}(x,Q)-F_2^{W^+p}(x,Q)\right)\,,\nonumber\\
\eea
with the structure functions receiving contributions from the s- and t-channel

\bea
F_2^{W^{\pm}p}(x,Q)=F_{2s}^{W^{\pm}p}(x,Q)+F_{2t}^{W^{\pm}p}(x,Q)\,,\nonumber\\
F_3^{W^{\pm}p}(x,Q)=F_{3s}^{W^{\pm}p}(x,Q)+F_{3t}^{W^{\pm}p}(x,Q)\,,\nonumber\\
\eea
The charged s-channel even-parity structure functions $F_{2s}^{W^{\pm}p}(x,Q)$ are  given in (\ref{SFWP}) and (\ref{SFWM}), respectively, 
and the t-channel structure functions $F_{2t}^{W^{\pm}p}(x,Q)$  are given in (\ref{F1F2T2}). The charged s-channel odd-parity structure functions 
 $F_{3s}^{W^{\pm}p}(x,Q)$ are   given in  (\ref{SFWP}) and (\ref{SFWM}), respectively. The  charged t-channel odd parity structure functions 
$F_{3t}^{W^{\pm}p}(x,Q)$ are given in  (\ref{F3T}). 

\begin{figure}
  \includegraphics[height=6cm,width=.49\linewidth]{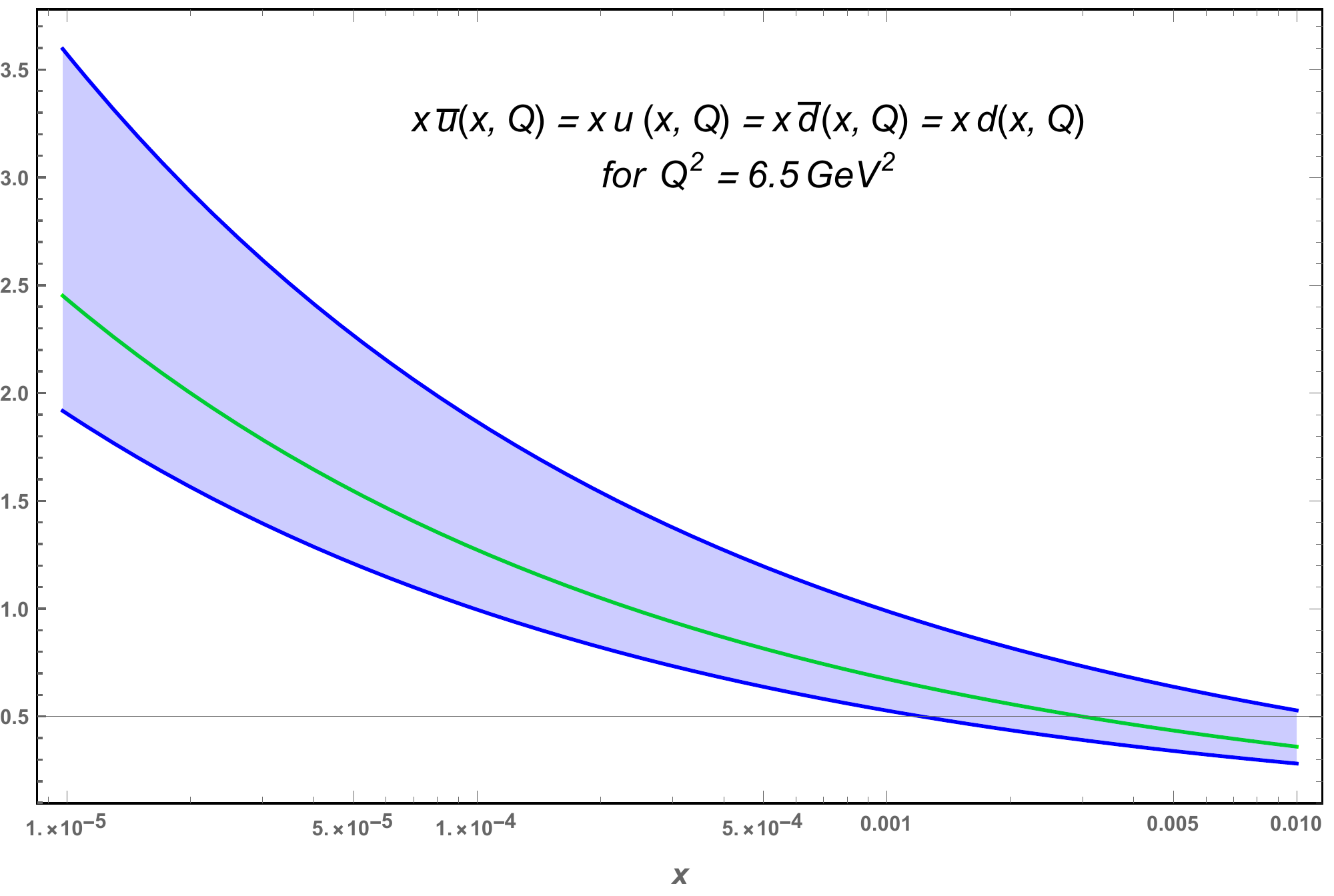}
  \caption{Small-x holographic PDFs as given in (\ref{7bc}) and (\ref{7Xbc}) for: the normalization coefficient $\mathcal{N}_{Lt}^{\pm}=0.309$ (the green curve), the normalization coefficient fixed between $\mathcal{N}_{Lt}^{\pm}=0.274$ and $\mathcal{N}_{Lt}^{\pm}=0.375$ (the blue band), 't Hooft coupling $\lambda=g_{YM}^2N_{c}=9.533$, the twist of proton $\tau=3$ (fixed by the scaling of the electromagnetic form factor of proton in \cite{CARLSON}), and soft-wall IR scale $\tilde{\kappa}=0.350\,\rm{GeV}$ (fixed by the mass of proton and $\rho$ meson in \cite{CARLSON}).}
  \label{fig:xuatqq6pt5}
\end{figure}

\subsection{Results in the small-x regime}

In the small-x regime, the results for the sea $\bar u$ and $\bar d$ distributions are solely due to the t-channel
Pomeron exchange

\be
\label{7bc}
\lim_{x\rightarrow 0} x\overline u(x,Q)&=&\frac 14 F_{2t}^{W^+p}(x,Q)\,,\nonumber\\
\lim_{x\rightarrow 0} x\overline d(x,Q)&=&\frac 14 F_{2t}^{W^-p}(x,Q)\,,
\ee
with no Pomeron contribution to the tail of the valence distributions

\bea
\label{7Xbc}
\lim_{x\rightarrow 0} xu_V(x,Q)&=&0\,,\nonumber\\
\lim_{x\rightarrow 0} xd_V(x,Q)&=&0\,.
\eea
Fig.~\ref{fig:xuatqq6pt5} show the holographic Pomeron contribution at small-x for the $xu$ and $xd$ distributions of the proton at a resolution $Q^2=6.5$ GeV$^2$,
following from  (\ref{7bc}) and (\ref{7Xbc}). The  $^\prime$t Hooft coupling is $\lambda=g_{YM}^2N_{c}=9.533$, the twist of proton is set to $\tau=3$ (fixed by the scaling of the electromagnetic form factor of proton in \cite{CARLSON}), and the soft-wall IR scale is fixed to $\tilde{\kappa}=0.350\,\rm{GeV}$ (to reproduce the mass of the proton and $\rho$ meson
as in~\cite{CARLSON}). The green curve uses the  normalization coefficient $\mathcal{N}_{Lt}^{\pm}=0.309$.  The blue-band corresponds to the 
normalization coefficients fixed between $\mathcal{N}_{Lt}^{\pm}=0.274$ and $\mathcal{N}_{Lt}^{\pm}=0.375$. 


\begin{figure*}
\subfloat[\label{sigmaplusx0pt000016}]{%
  \includegraphics[height=5cm,width=.49\linewidth]{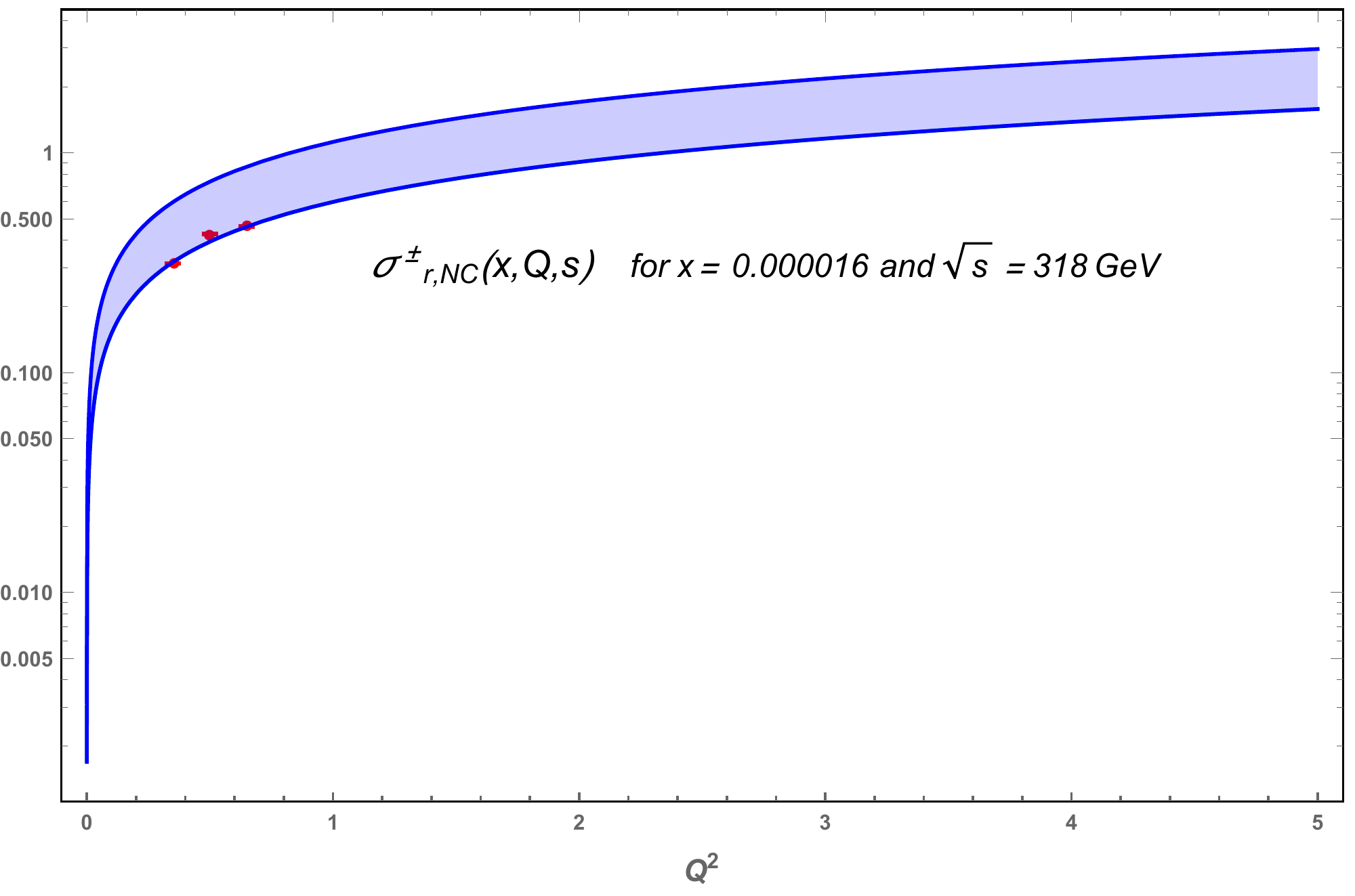}%
}\hfill
\subfloat[\label{sigmaplusx0pt00005}]{%
  \includegraphics[height=5cm,width=.49\linewidth]{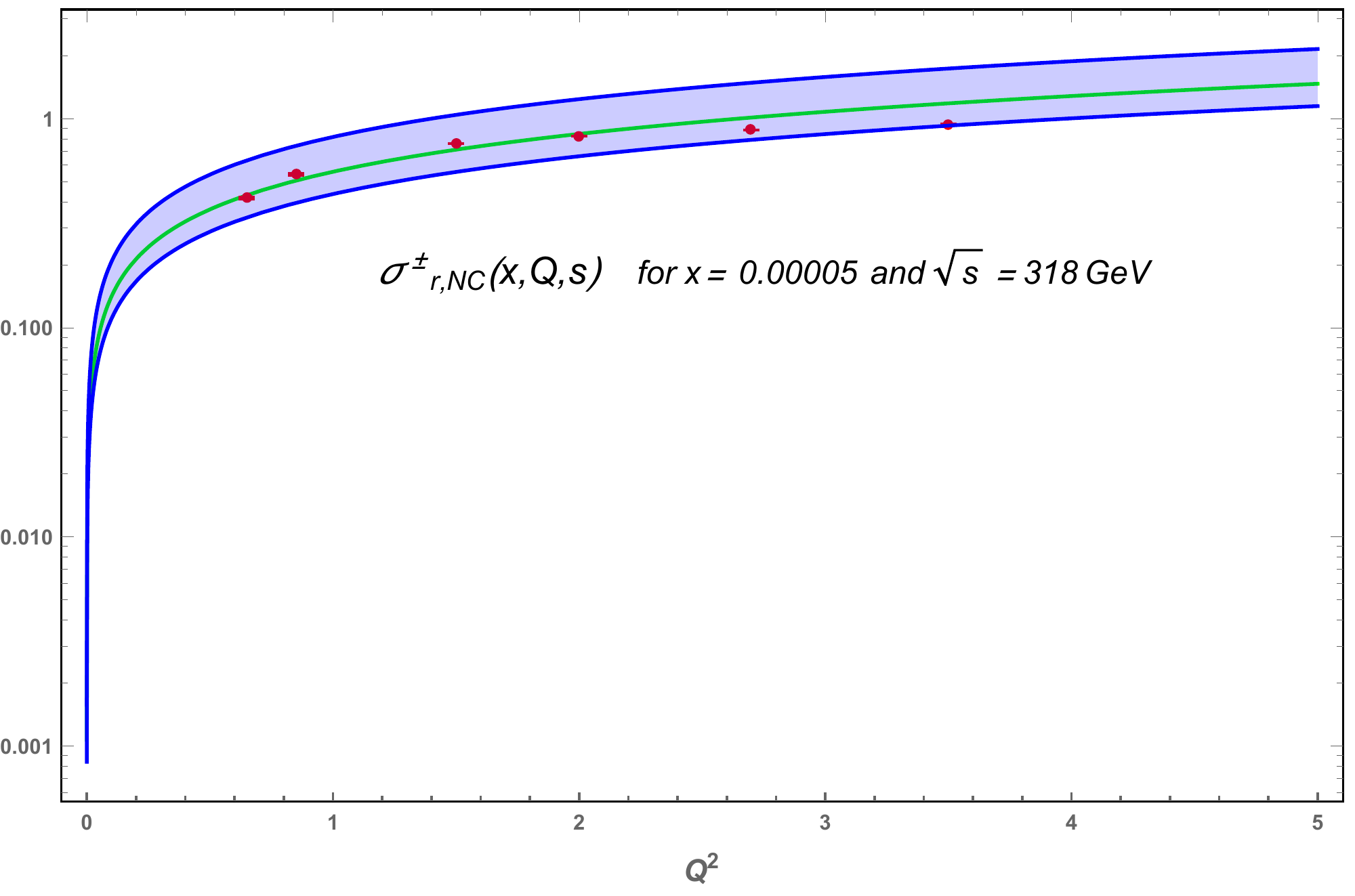}%
}\hfill
\subfloat[\label{sigmaplusx0pt00008}]{%
  \includegraphics[height=5cm,width=.49\linewidth]{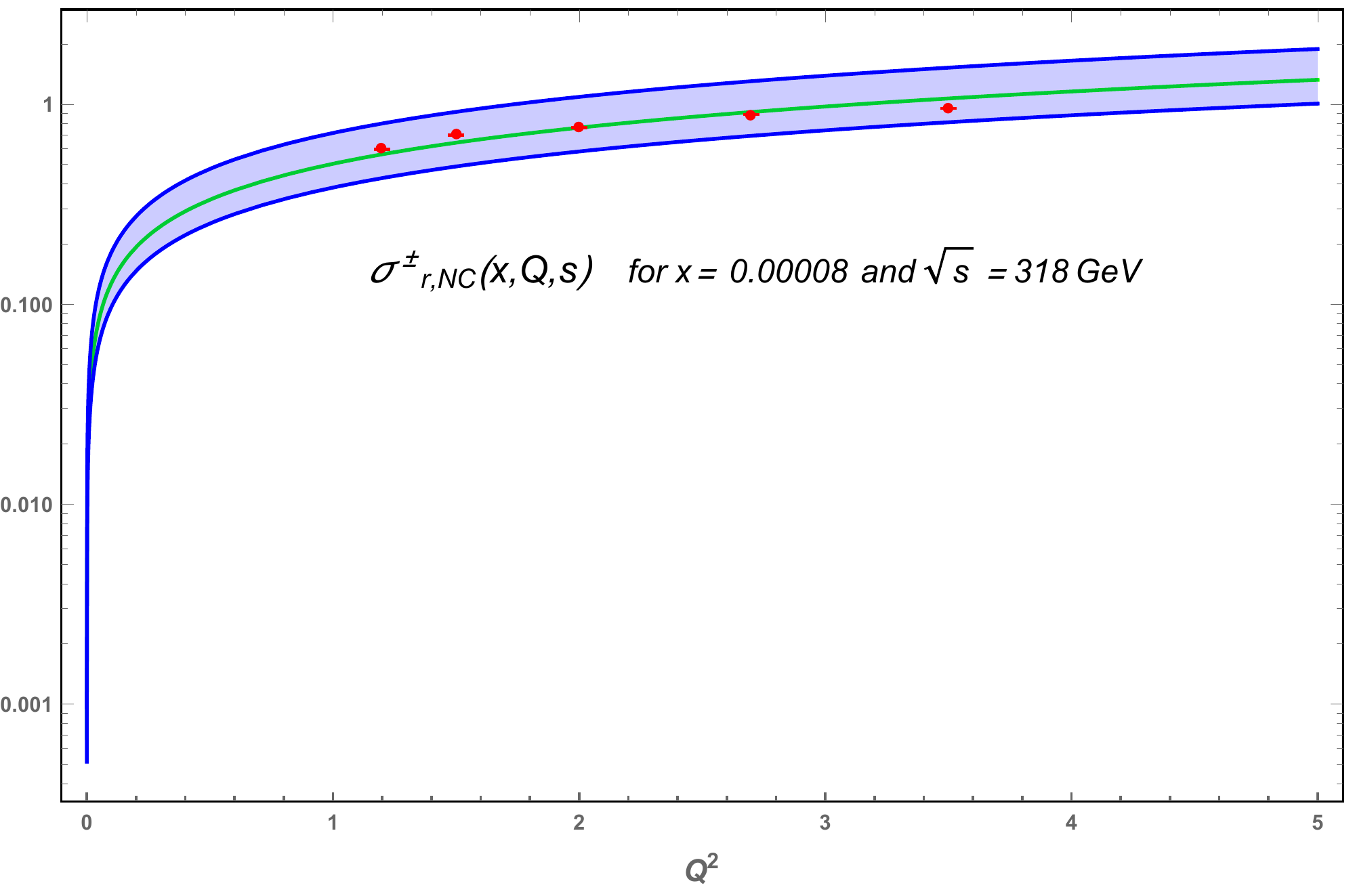}%
}\hfill
\subfloat[\label{sigmaplusx0pt00013}]{%
  \includegraphics[height=5cm,width=.49\linewidth]{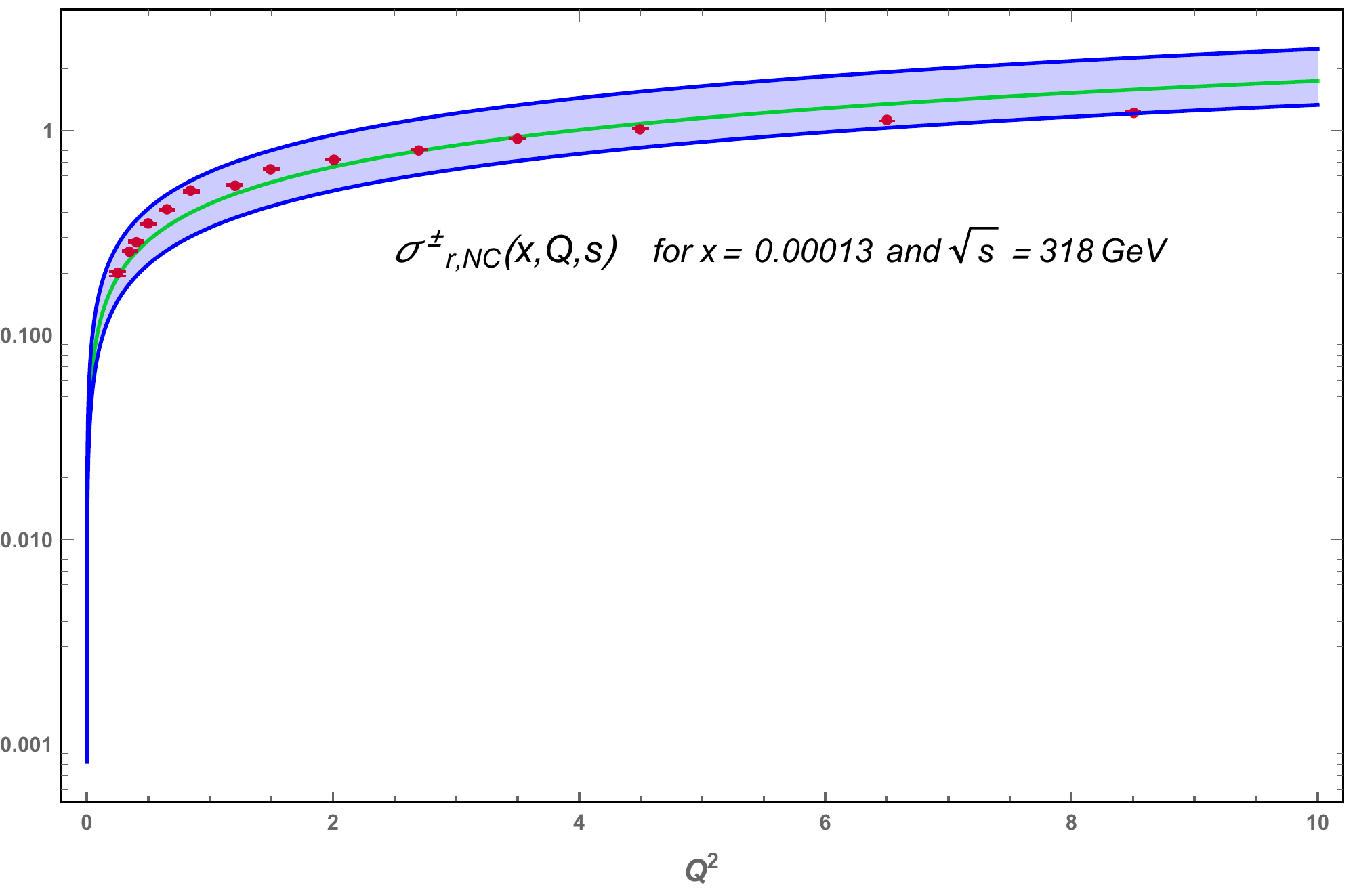}%
}\hfill
\subfloat[\label{sigmaplusx0pt0002}]{%
  \includegraphics[height=5cm,width=.49\linewidth]{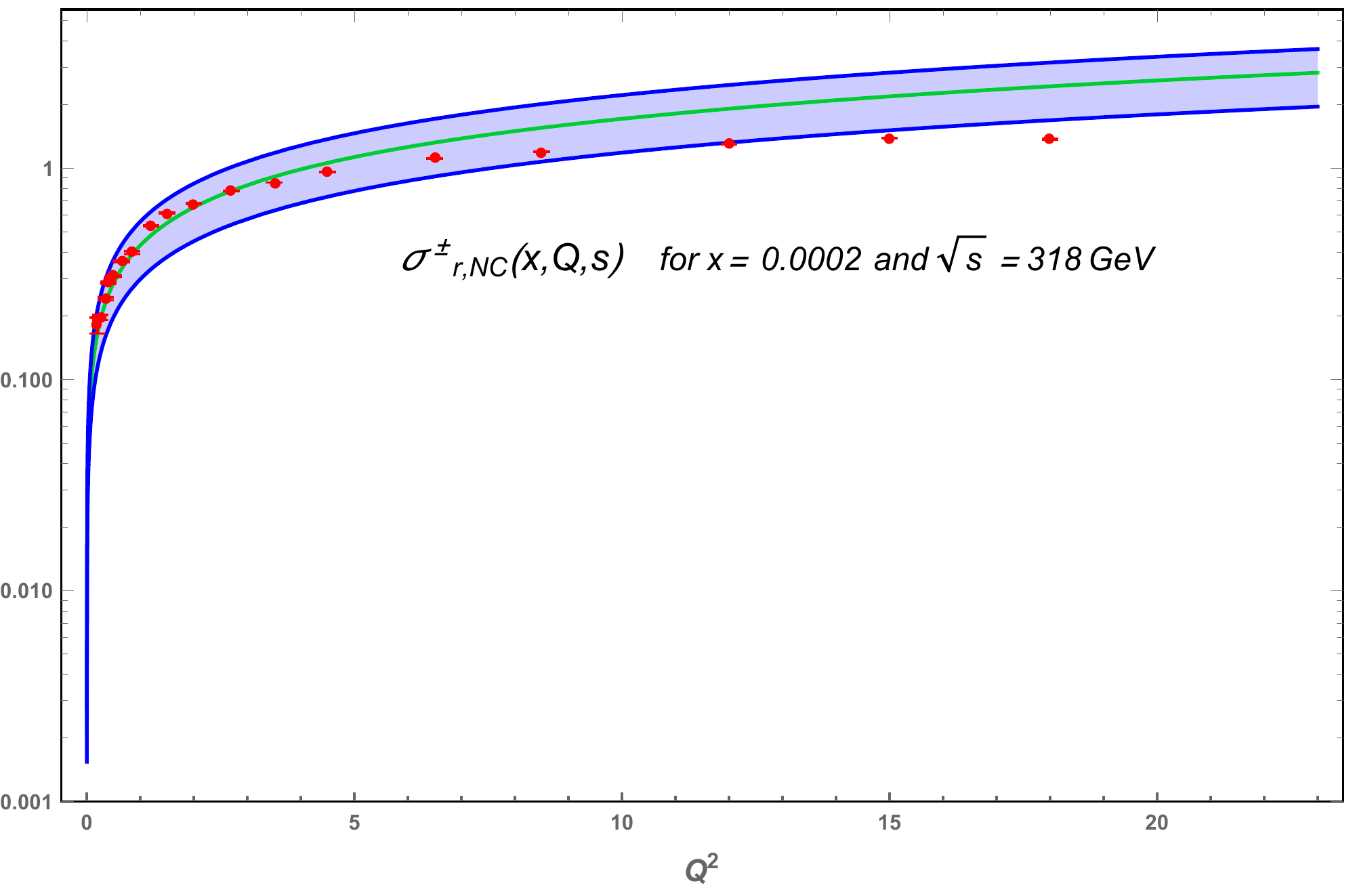}%
}\hfill
\subfloat[\label{sigmaplusx0pt00032}]{%
  \includegraphics[height=5cm,width=.49\linewidth]{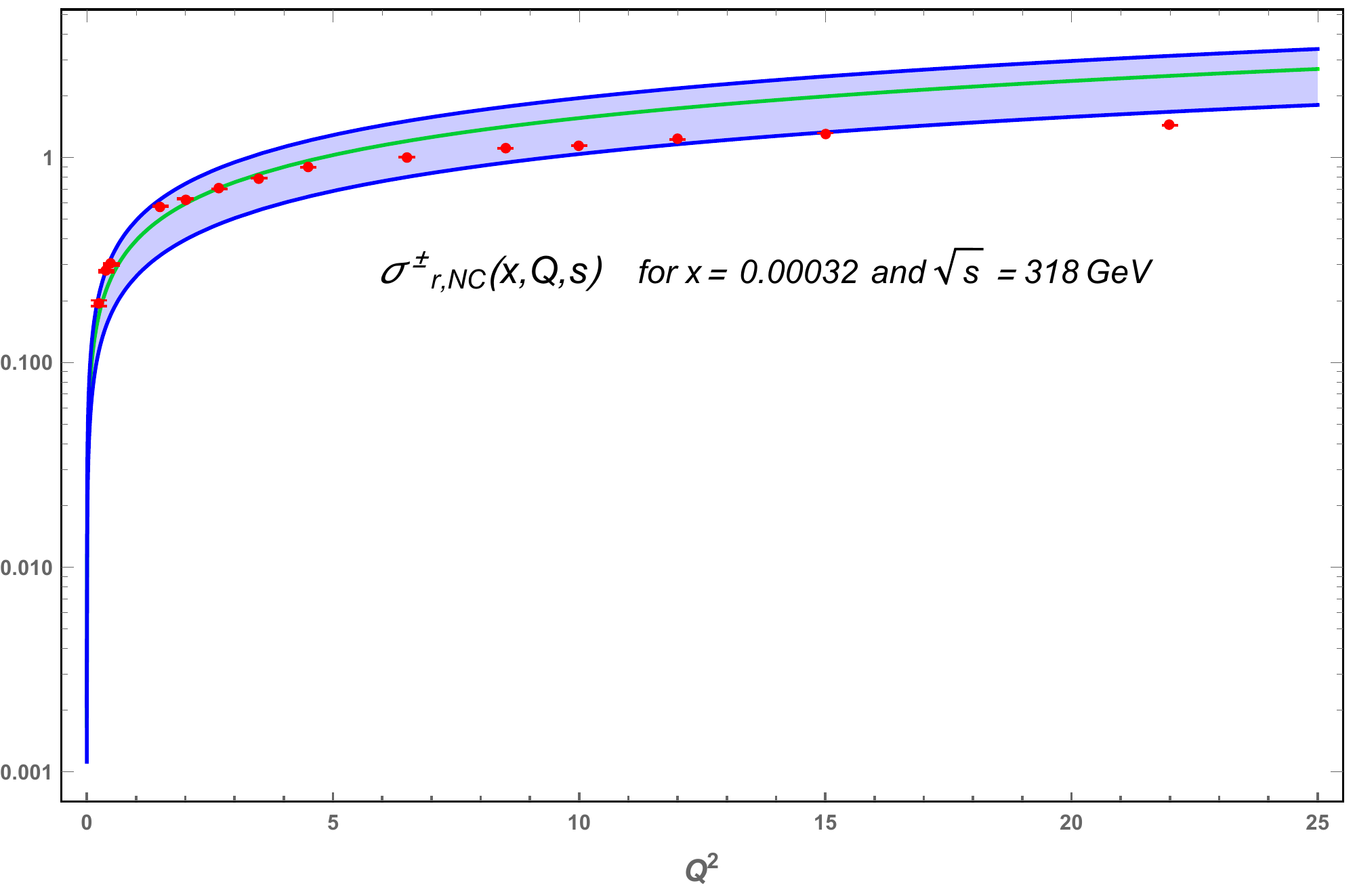}%
}\hfill
\subfloat[\label{sigmaplusx0pt0008}]{%
  \includegraphics[height=5cm,width=.49\linewidth]{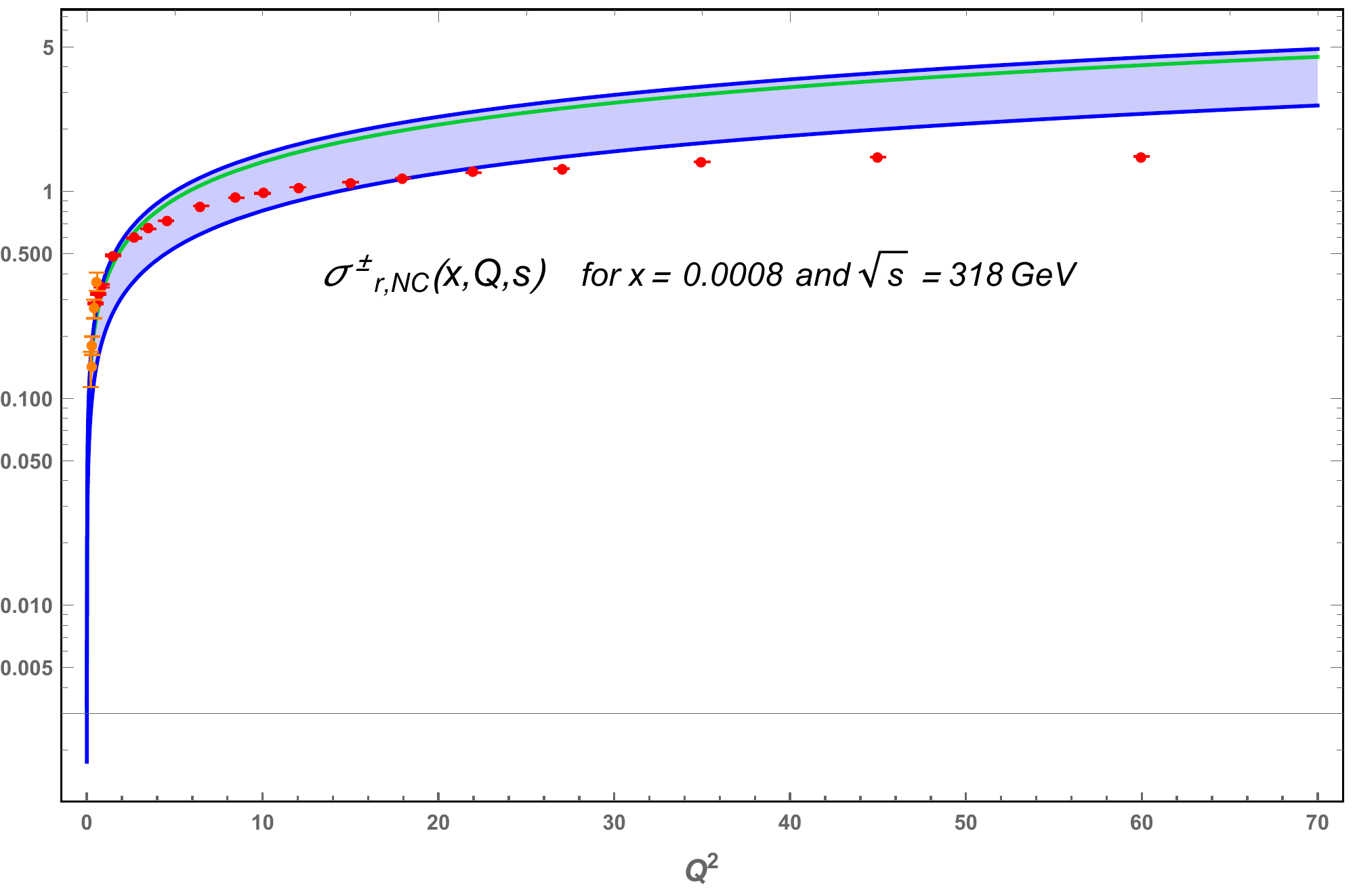}%
}\hfill
\subfloat[  \label{sigmaplusx0pt0013}]{%
  \includegraphics[height=5cm,width=.49\linewidth]{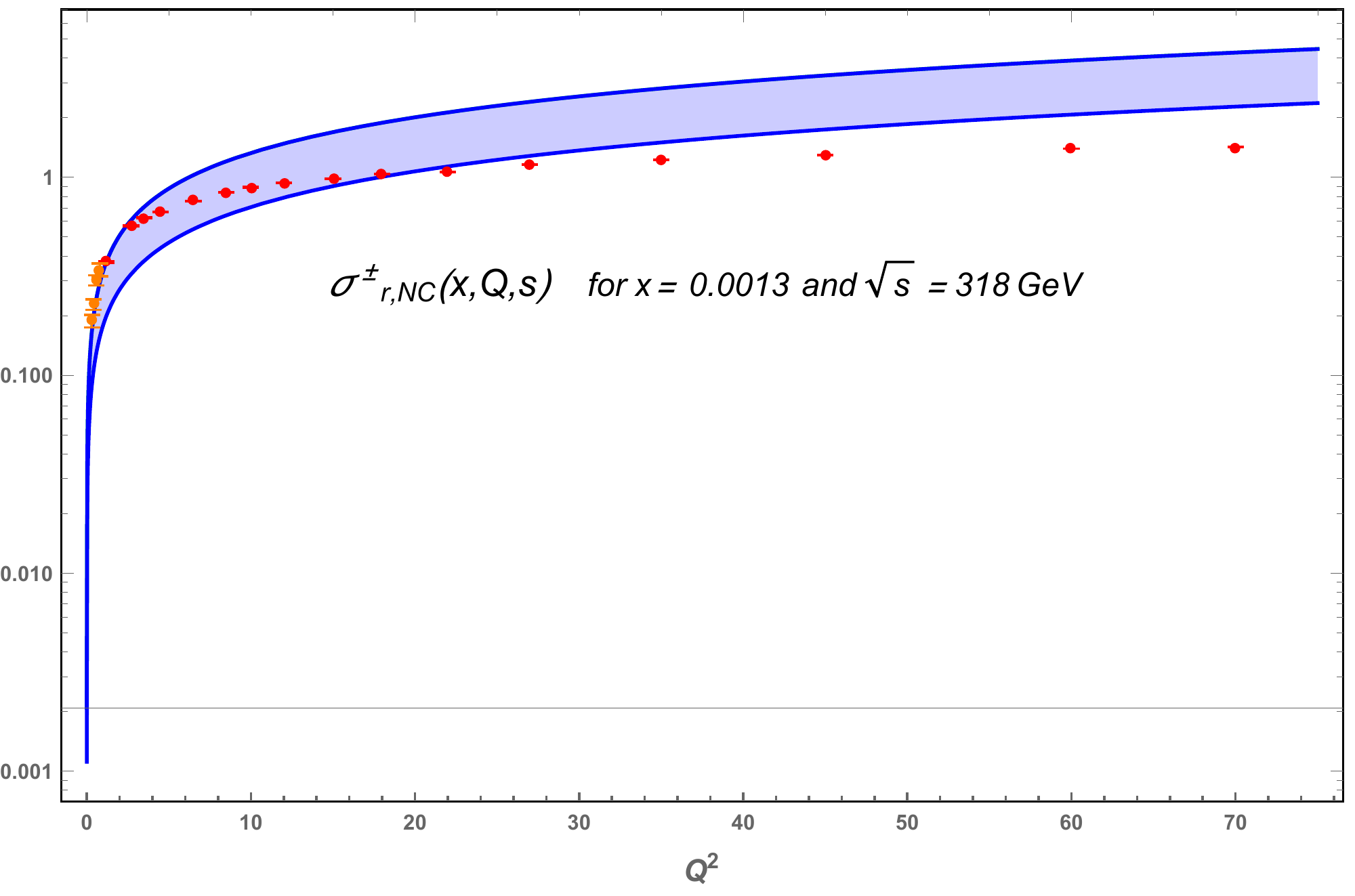}%
}
\caption{The reduced non-charged  and unpolarized deep inelastic  
 $e^{\pm}p$ scattering cross section $\sigma^\pm_{r,NC}(x,Q,s)$ at low-x versus $Q^2$ as given in (\ref{ncsi}) for increasing $x$, at $\sqrt{s}=318$ GeV. The middle-green curve and 
 the blue band are the holographic results for different charge normalizations ${\cal N}_{Lt}^\pm$. The red points are the H1 and ZEUS data~ \cite{Abramowicz:2015mha},
 and the orange points at low $Q^2$ are the data from E665~\cite{Adams:1996gu}. See text.}
\label{sigmarNC}
\end{figure*}

\subsection{Fixing the charged normalization ${\cal N}_{Lt}^\pm$ parameters}

Throughout we will use the same $^\prime$t Hooft coupling $\lambda=g_{YM}^2N_{c}=9.533$ which is within the standard choice in the most holographic constructions. The 
 charged normalizations $\mathcal{N}_{Lt}^{\pm}$  are fixed between $0.274$ and $0.375$ and will be shown as a blue band in all results
 to follow. These normalizations are chosen for a best fit to the reduced non-charged (r,NC) and unpolarized deep inelastic  
 $e^{\pm}p$ scattering cross section $\sigma^\pm_{r,NC}$ at low-x. We recall, that $\sigma^\pm_{r,NC}$
is given by a linear combination of the generalised structure functions 

\begin{eqnarray} \label{ncsi}     
 \ncred =\ncdd \cdot \frac{Q^4 x_{\rm Bj}}{2\pi \alpha^2 Y_+}                                                     
  =            \tilde{F_2} \mp \frac{Y_-}{Y_+} x\tilde{F_3} -\frac{y^2}{Y_+} \tilde{F_{\rm L}}
\end{eqnarray}
with                                                                     
$Y_{\pm}=1 \pm (1-y)^2$.
The overall structure functions, $\boldftwo$, $\boldfl$  and $\boldxft$,
are sums of structure functions, $\boldfX$, $\boldfXgZ$ and $\boldfXZ$,
relating to photon exchange, photon--$Z$ interference and $Z$ exchange, 
respectively, and depend on the electroweak parameters as                           

\begin{eqnarray} \label{strf}                                                   
 \boldftwo &=& F_2 - \kappa_Z v_e  \cdot F_2^{\gamma Z} +                      
  \kappa_Z^2 (v_e^2 + a_e^2 ) \cdot F_2^Z~, \nonumber \\   
 \boldfl &=& F_{\rm L} - \kappa_Z v_e  \cdot F_{\rm L}^{\gamma Z} +                      
  \kappa_Z^2 (v_e^2 + a_e^2 ) \cdot F_{\rm L}^Z~, \nonumber \\                     
 \boldxft &=&  - \kappa_Z a_e  \cdot xF_3^{\gamma Z} +                    
  \kappa_Z^2 \cdot 2 v_e a_e  \cdot xF_3^Z~ \nonumber                                  
\end{eqnarray} 
Here  $v_e$ and $a_e$ are the vector and axial-vector weak couplings of 
the electron to the $Z$ boson, 
and $\kappa_Z(Q^2) =   Q^2 /[(Q^2+M_Z^2)(4\sin^2 \theta_W \cos^2 \theta_W)]$. 
The values of $\sin^2 \theta_W=0.23127$ and $M_Z=91.1876$\,GeV were used
for the electroweak mixing angle and the $Z$-boson mass.
In the quark-parton model (QPM) where the kinematic variable $x_{\rm Bj}$ is equal to
the fractional momentum of the struck quark, $x$, the 
structure functions are given in terms of the PDFs as
\begin{eqnarray} \label{ncfu}                                                   
  (F_2(x,Q), F_2^{\gamma Z}(x,Q), F_2^Z(x,Q)) & \approx &  [(e_u^2, 2e_uv_u, v_u^2+a_u^2)(xu(x,Q)+ x\bar{u}(x,Q))\nonumber \\ 
  &+&  (e_d^2, 2e_dv_d, v_d^2+a_d^2)(xd(x,Q)+ x\bar{d}(x,Q))]~,            
                                 \nonumber \\                                   
  (xF_3^{\gamma Z}(x,Q), xF_3^Z(x,Q)) & \approx & 2  [(e_ua_u, v_ua_u) (xu(x,Q)-x\bar{u}(x,Q))\nonumber \\ 
  &+&  (e_da_d, v_da_d) (xd(x,Q)-x\bar{d}(x,Q))]~,            
                                 \nonumber \\                                   
(F_L(x,Q), F_L^{\gamma Z}(x,Q), F_L^Z(x,Q)) & \approx &  (0,0,0)
\end{eqnarray} 
with  $e_u$ and $e_d$ denoting  the electric charge of up- and
down-type quarks, while $v_{u,d}$ and $a_{u,d}$ are 
the vector and axial-vector weak couplings of the up- and
down-type quarks to the $Z$ boson.



In Fig.~\ref{sigmarNC}  we show the measured $\sigma_{r, NC}^\pm(x,Q,s)$   versus $Q^2$ in GeV$^2$ at $\sqrt{s}$ =318 GeV with increasing x-resolution.
Fig.~\ref{sigmarNC}a follows from (\ref{ncsi}) for $x=0.000016$ using the small-x holographic PDF shown in Fig.~\ref{fig:xuatqq6pt5}.
Fig.~\ref{sigmarNC}b shows the same for $x=0.00005$. The middle green curve corresponds to  $\mathcal{N}_{Lt}^{\pm}=0.309$.
Fig.~\ref{sigmarNC}c follows also from (\ref{ncsi}) for $x=0.00008$, with the middle green curve referring to $\mathcal{N}_{Lt}^{\pm}=0.314$.
Fig.~\ref{sigmarNC}d refers to $x=0.00013$  with the middle-green curve referring to $\mathcal{N}_{Lt}^{\pm}=0.313$.
Fig.~\ref{sigmarNC}e follows again from (\ref{ncsi}) at $x=0.0002$ with the middle-green curve $\mathcal{N}_{Lt}^{\pm}=0.329$.
Fig.~\ref{sigmarNC}f  refers to $x=0.00032$  with the middle-green curve referring to $\mathcal{N}_{Lt}^{\pm}=0.313$.
Fig.~\ref{sigmarNC}f  refers to $x=0.0008$  with the middle-green curve referring to $\mathcal{N}_{Lt}^{\pm}=0.335$.
Fig.~\ref{sigmarNC}f  refers to $x=0.0013$  with the middle-green curve referring to $\mathcal{N}_{Lt}^{\pm}=0.359$.
The data from the H1 and ZEUS collaboration~\cite{Abramowicz:2015mha} are shown in red. 
The   data  shown in orange at very low $Q^2$ are from the E665 collaboration~\cite{Adams:1996gu}.
Note that we have reconstructed the orange data points for the cross section $\sigma^{\pm}_{r,NC}(x,Q,s)$ at $\sqrt{s}=318\,GeV$ from the
E665 data set  for the structure function $F_2(x,Q)$ at $0.0008\leq x\leq 0.001$ using~\cite{Whitlow:1990dr}

\be
\label{RXQ}
R(x,Q)=\frac{F_2}{2xF_1}\bigg(1+\frac{4M_p^2x^2}{Q^2}\bigg)-1=R^{1990}(x.Q)
\ee
As noted above, the blue-band  follows from the 
normalization range fixed by  $\mathcal{N}_{Lt}^{\pm}=0.274$ and $\mathcal{N}_{Lt}^{\pm}=0.375$.


\begin{figure*}
\subfloat[\label{F2qq6pt5}]{%
  \includegraphics[height=6cm,width=.49\linewidth]{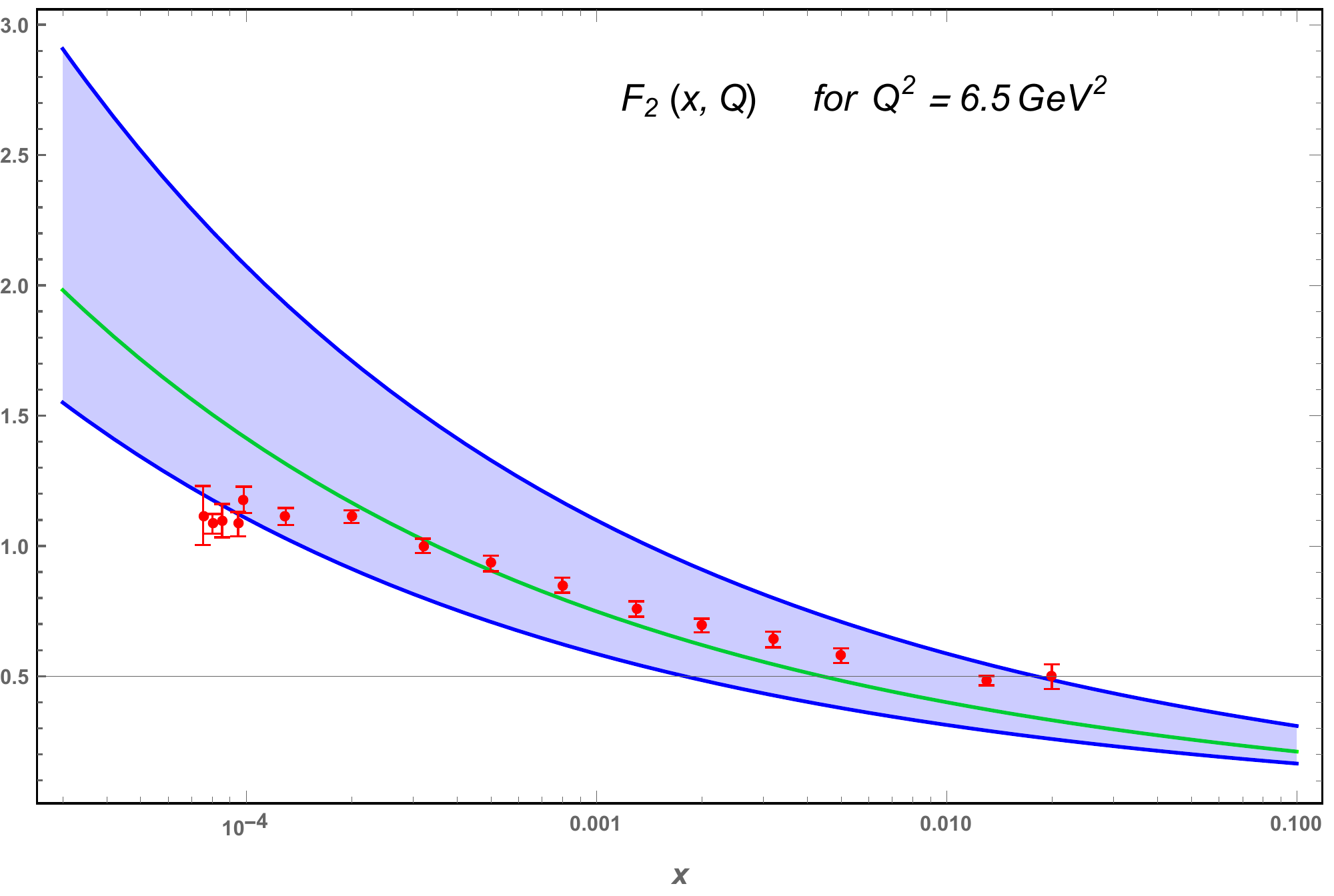}%
}\hfill
\subfloat[\label{F2qq6pt5Zcorrection}]{%
  \includegraphics[height=6cm,width=.49\linewidth]{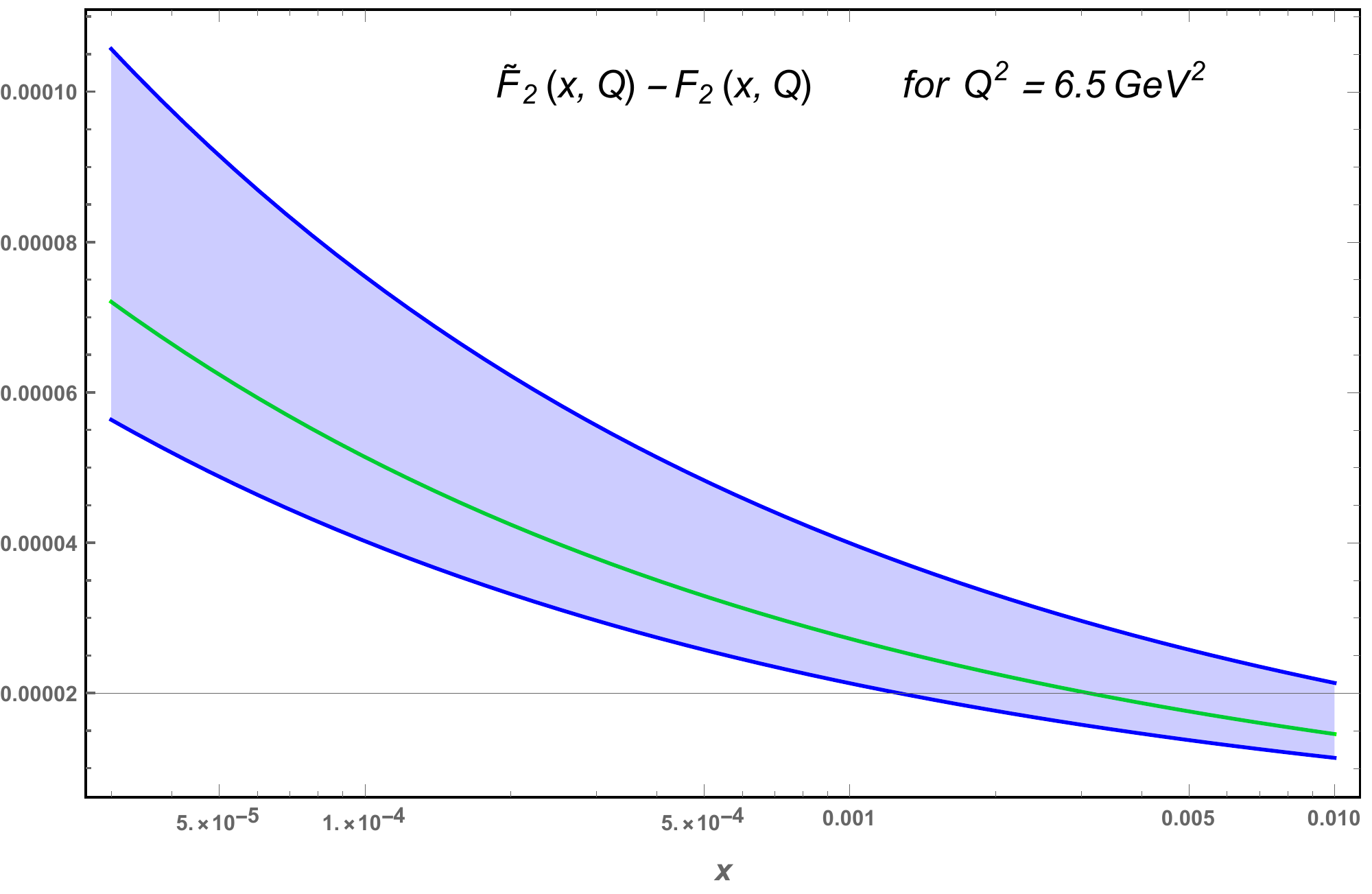}%
}\hfill
\subfloat[\label{F2qq6pt5gZ}]{%
  \includegraphics[height=6cm,width=.49\linewidth]{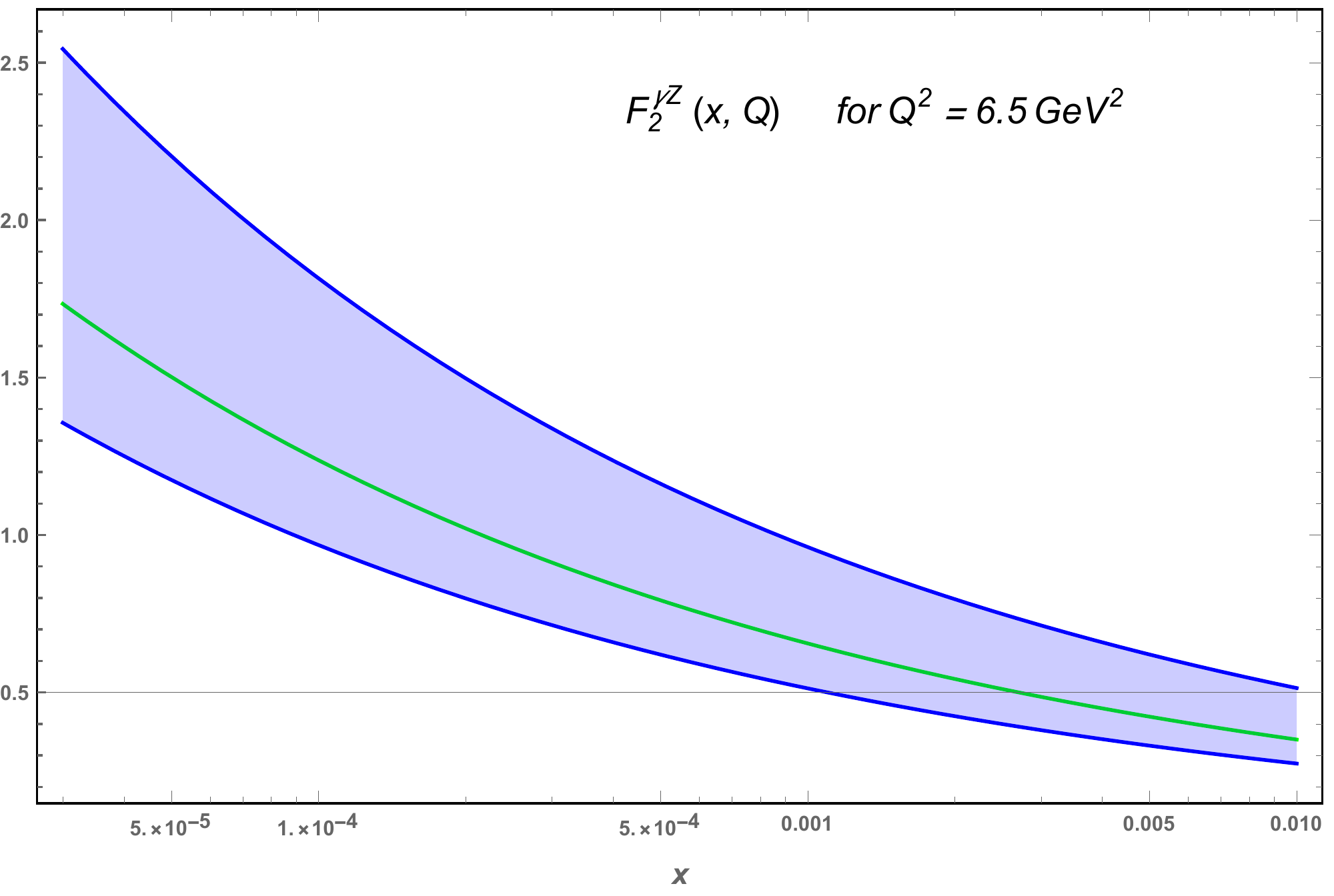}%
}\hfill
\subfloat[\label{F2qq6pt5Z}]{%
  \includegraphics[height=6cm,width=.49\linewidth]{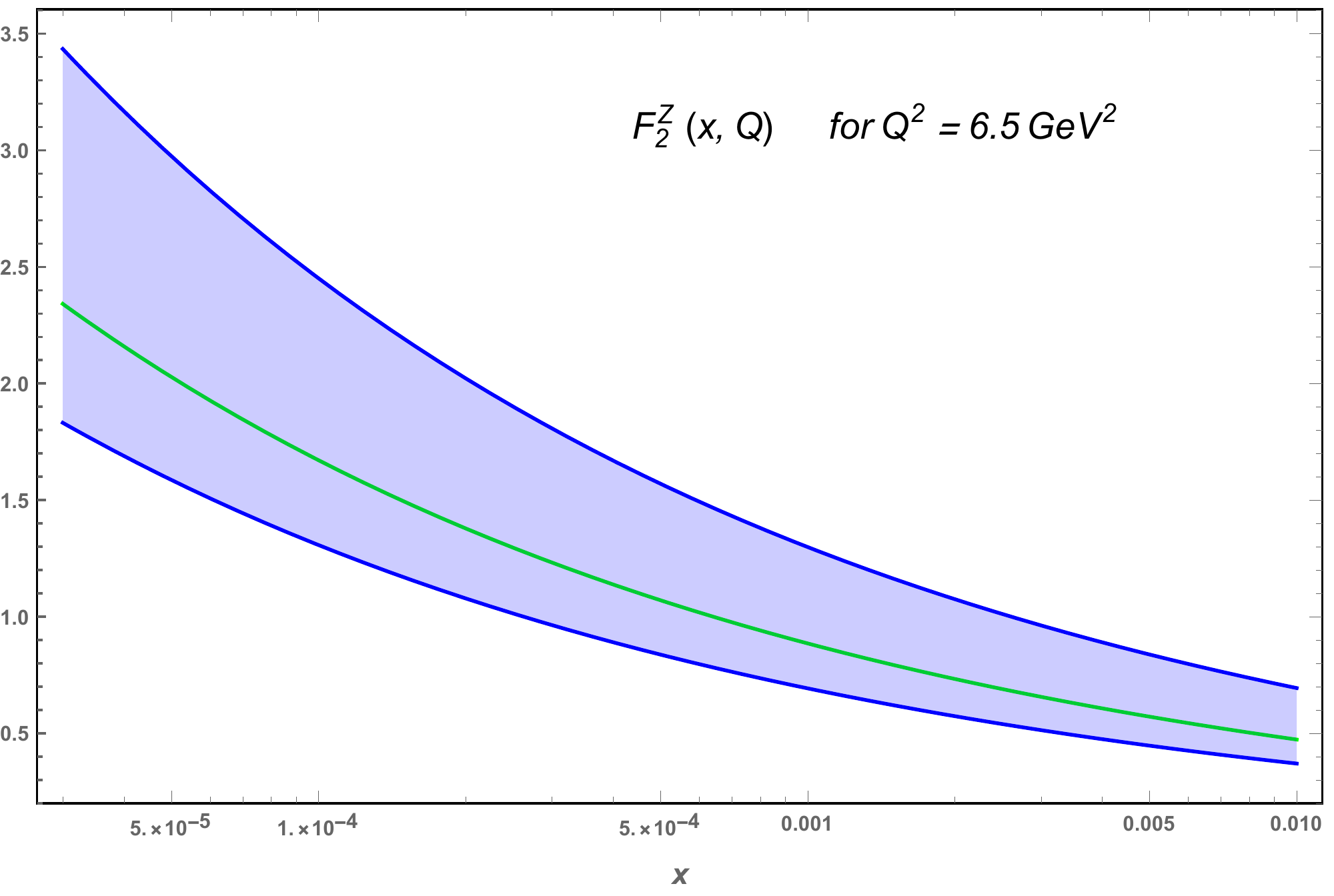}%
}
\caption{{\bf (a)}: $F_2(x,Q)$ as given in (\ref{ncfu}) using the small-x holographic PDFs shown in Fig.~\ref{fig:xuatqq6pt5}. The red data points are from combined H1 and ZEUS collaborations \cite{Abramowicz:2015mha} (reconstructed from their data for the cross section $\sigma^{\pm}_{r,NC}(x,Q,s)$ using our small-x holographic PDFs (\ref{7bc}) and (\ref{7Xbc}) shown in Fig.~\ref{fig:xuatqq6pt5}); {\bf (b)}:
$\tilde{F}_2(x,Q)-F_2(x,Q)$ as given in (\ref{ncfu}) using the small-x holographic PDFs shown in Fig.~\ref{fig:xuatqq6pt5}; 
{\bf (c)}: $F_2^{\gamma Z}(x,Q)$ as given in (\ref{ncfu}) using the small-x holographic PDFs shown in Fig.~\ref{fig:xuatqq6pt5};
{\bf (d)}: $F_2^{Z}(x,Q)$ as given in (\ref{ncfu}) using the small-x holographic PDFs shown in Fig.~\ref{fig:xuatqq6pt5}.}
\label{F2qq6pt5gZF2qq6pt5Z}
\end{figure*}



\begin{figure*}
\subfloat[ \label{xuvalenceatqq6pt5largex}]{%
  \includegraphics[height=6cm,width=.49\linewidth]{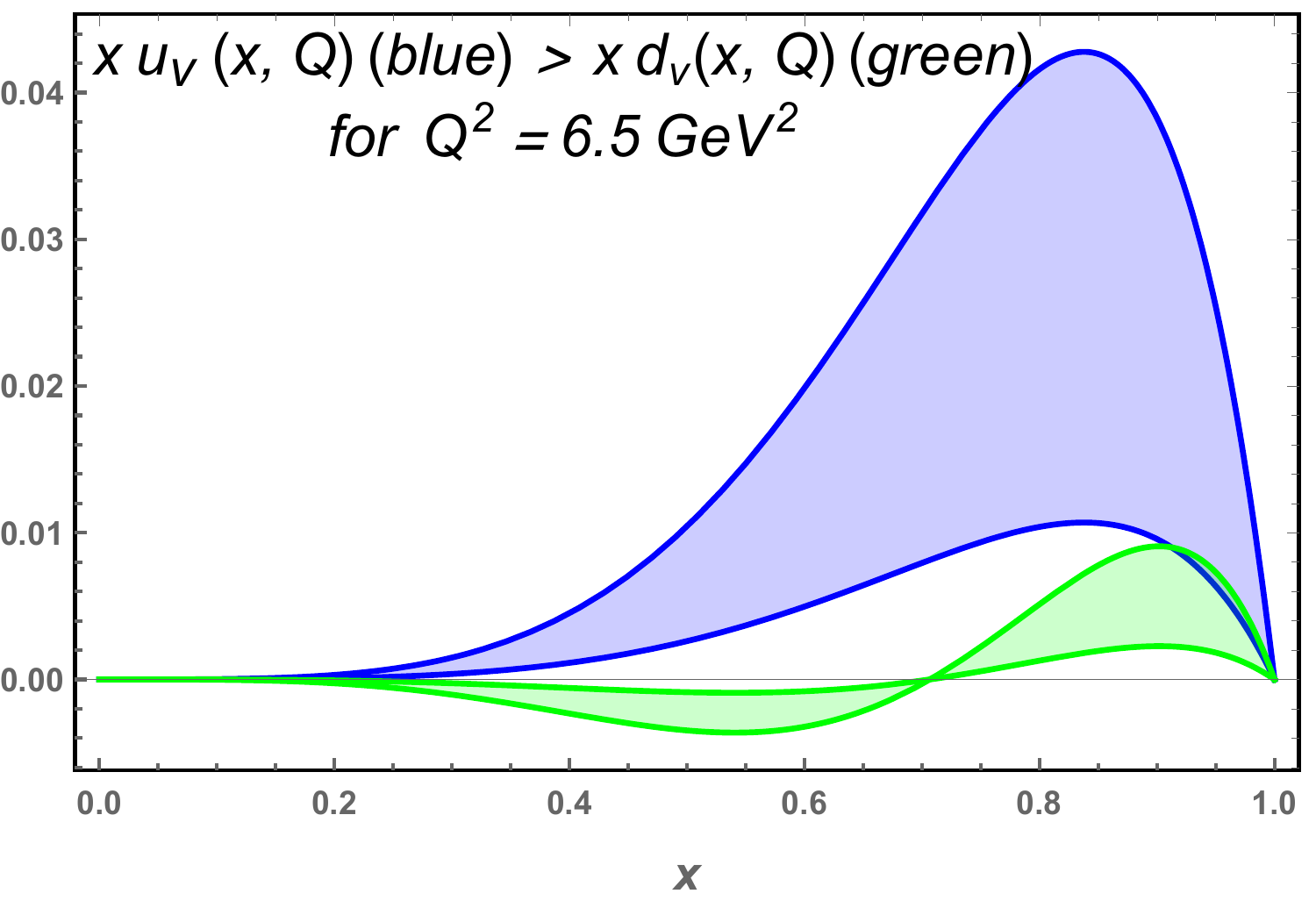}%
}\hfill
\subfloat[\label{xuseaatqq6pt5largex}]{%
  \includegraphics[height=6cm,width=.49\linewidth]{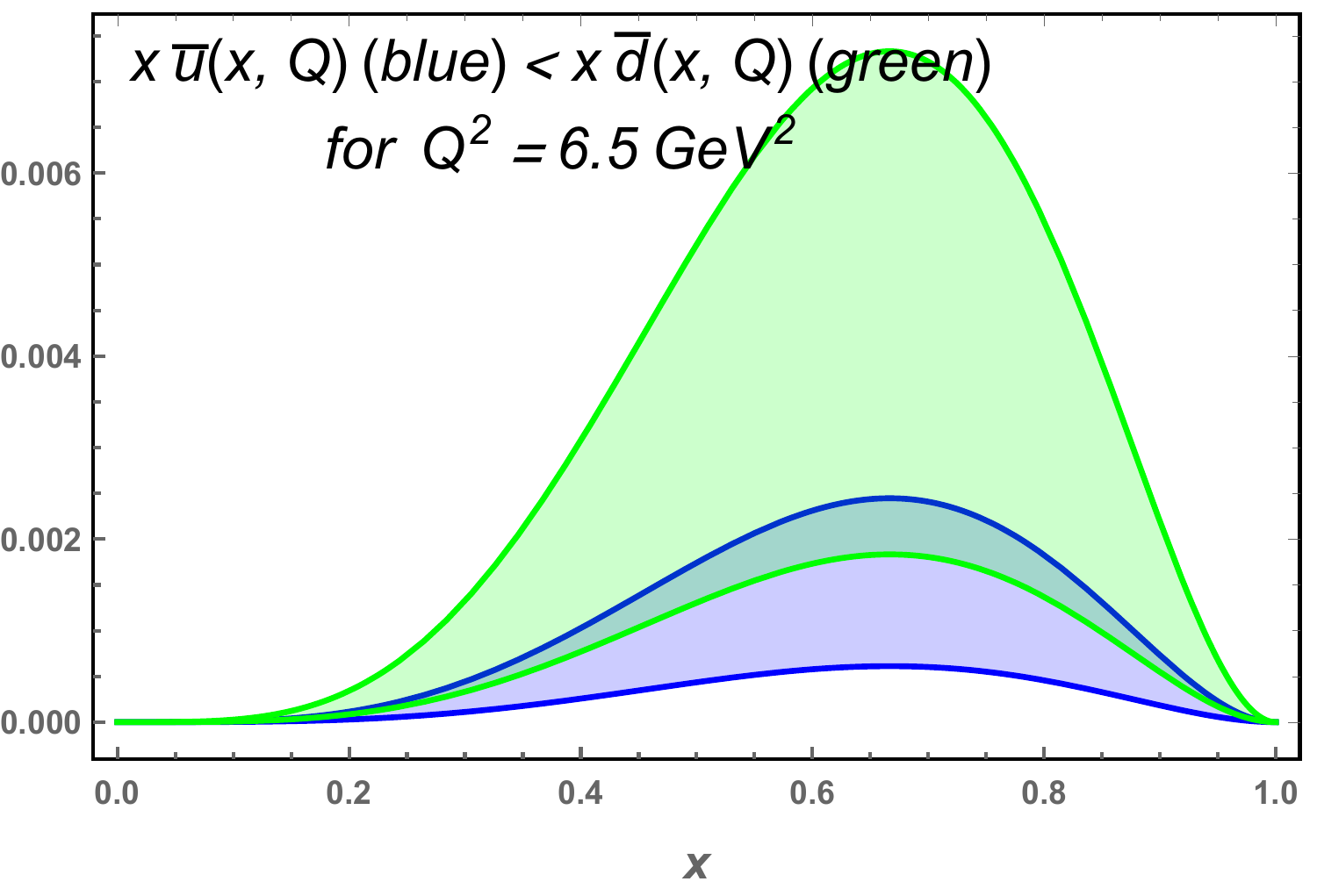}%
}
\caption{
a: large-x holographic PDFs of the valence quarks in the proton as given in (\ref{7Xbclargex});
b: large-x holographic PDFs of the sea quarks in the proton as given in (\ref{7bclargex}). The blue and green bands are determined by varying $\tilde{\tilde{\mathcal{N}}}_{L}^{+}$ between 17.715 and 35.431, and $\tilde{\tilde{\mathcal{N}}}_{L}^{-}$ between 30.667 and 61.335. See text.}
\label{fig:xuvalenceatqq6pt5xuseaatqq6pt5largexnogreen}
\end{figure*}


\subsection{Results in the large-x regime}

As we noted in the introduction, DIS scattering in holographic QCD at moderate values of partonic-x 
involves hadronic and not partonic constituents~\cite{POL}. Indeed, in the large $N_c$ limit, the leading
single trace twist contributions acquire large anomalous dimensions and are suppressed. The dominant
contributions stem from double-trace operators with mesonic quantum numbers. Another way to see this is 
to note that the large gauge coupling at the low renormalization point, causes the color charges to undergo
a rapid depletion into a cascade of even weaker charges, making them visible to hard probes only through 
double trace operators.

At large-x,  DIS scattering is almost off the entire hadron making the 
holographic approach pertinent. In this regime,  the holographic limit enjoys approximate conformal
symmetry,  with the structure functions and form factors exhibiting  various scaling laws including the 
parton-counting rules~\cite{BF,POL}. More specifically,  we have for the sea contribution

\be
\label{7bclargex}
\lim_{x\rightarrow 1} \lim_{\frac{\kappa^2}{Q^2}\rightarrow 0} x\overline u(x,Q)&=&\frac 14 \left(F_{2s}^{W^+p}(x,Q)-xF_{3s}^{W^+p}(x,Q)\right)\,,\nonumber\\
\lim_{x\rightarrow 1} \lim_{\frac{\kappa^2}{Q^2}\rightarrow 0} x\overline d(x,Q)&=&\frac 14 \left(F_{2s}^{W^-p}(x,Q)-xF_{3s}^{W^-p}(x,Q)\right)\,,
\ee
and for the valence contribution

\bea
\label{7Xbclargex}
\lim_{x\rightarrow 1} \lim_{\frac{\kappa^2}{Q^2}\rightarrow 0} xu_V(x,Q)&=&\frac{1}{4} x\left(F_{3s}^{W^-p}(x,Q)+F_{3s}^{W^+p}(x,Q)\right)\nonumber\\
&+&\frac 14\left(F_{2s}^{W^-p}(x,Q)-F_{2s}^{W^+p}(x,Q)\right)\,,\nonumber\\
\lim_{x\rightarrow 1} \lim_{\frac{\kappa^2}{Q^2}\rightarrow 0} xd_V(x,Q)&=&\frac{1}{4} x\left(F_{3s}^{W^-p}(x,Q)+F_{3s}^{W^+p}(x,Q)\right)\nonumber\\
&-&\frac 14\left(F_{2s}^{W^-p}(x,Q)-F_{2s}^{W^+p}(x,Q)\right)\,,
\eea
with the even and odd structure functions given respectively by

\bea \label{F23DiracPauli1}
F_{2s}^{W^{\pm}p}(x,Q)&=&F_{2s, Dirac}^{W^{\pm}p}(x,Q)+F_{2s, Pauli}^{W^{\pm}p}(x,Q)+F_{2s, mixed}^{W^{\pm}p}(x,Q)
\,,\nonumber\\
F_{3s}^{W^{\pm}p}(x,Q)&=&F_{3s, Dirac}^{W^{\pm}p}(x,Q)+F_{3s, Pauli}^{W^{\pm}p}(x,Q)+F_{3s, mixed}^{W^{\pm}p}(x,Q)\,,
\eea

The large-x asymptotic of the Dirac+Pauli+Mixed structure functions following from (\ref{SFWP}),
can be worked out in closed form. For the even-parity structure functions we have

\bea\label{F2DiracPauli}
F_{2s, Dirac}^{W^{\pm}p}(x,Q)&=&\big(\tilde{\mathcal{N}}_{L}^{\pm}\big)^2\times \big(e^{\pm}_{Wnucleon}\big)^2\times\left(\frac{\tilde{\kappa}^2}{Q^2}\right)^{\tau-1} x^{\tau+1}(1-x)^{\tau-2}\,,
\nonumber\\
F_{2s, Pauli}^{W^{\pm}p}(x,Q)\Big)&=&\big(\tilde{\mathcal{N}}_{L}^{\pm}\big)^2\times \eta^2\times 4(\tau-1)^2\times\left(\frac{\tilde{\kappa}^2}{Q^2}\right)^{\tau-1} x^{\tau+1}(1-x)^{\tau-2}\,,
\nonumber\\
F_{2s, mixed}^{W^{\pm}p}(x,Q)\Big)&=&\big(\tilde{\mathcal{N}}_{L}^{\pm}\big)^2\times e^{\pm}_{Wnucleon}\times\eta\times 4(\tau-1)\times\left(\frac{\tilde{\kappa}^2}{Q^2}\right)^{\tau-1} x^{\tau+1}(1-x)^{\tau-2}\,,
\eea
in agreement with a recent analysis in~\cite{Jorrin:2020cil} (see their Eq. 88, Eq 105 and Eq. 132).
For the odd-parity structure functions we also have

\bea\label{F3DiracPauli}
F_{3s, Dirac}^{W^{\pm}p}(x,Q)&=&\big(\tilde{\mathcal{N}}_{L}^{\pm}\big)^2\times \big(e^{\pm}_{Wnucleon}\big)^2\times\left(\frac{\tilde{\kappa}^2}{Q^2}\right)^{\tau-1} x^{\tau+1}(1-x)^{\tau-2}\,,
\nonumber\\
F_{3s, Pauli}^{W^{\pm}p}(x,Q)\Big)&=&\big(\tilde{\mathcal{N}}_{L}^{\pm}\big)^2\times \eta^2\times 4(\tau-1)^2\times\left(\frac{\tilde{\kappa}^2}{Q^2}\right)^{\tau-1} x^{\tau+1}(1-x)^{\tau-2}\,,
\nonumber\\
F_{3s, mixed}^{W^{\pm}p}(x,Q)\Big)&=&\big(\tilde{\mathcal{N}}_{L}^{\pm}\big)^2\times e^{\pm}_{Wnucleon}\times\eta\times 4(\tau-1)\times\left(\frac{\tilde{\kappa}^2}{Q^2}\right)^{\tau-1} x^{\tau+1}(1-x)^{\tau-2}\,,
\eea
also in agreement with the recent results  in~\cite{Jorrin:2020cil} (see their Eq. 88, Eq 104 and Eq. 131).
Note that eventhough the Pauli vertex contribution involves an additional vierbein in comparison to the
Dirac one,  hence an a priori extra suppression with the z-parameter and therefore $Q^2$ by duality, 
it is  balanced  by the extra z-derivative in the magnetic-like coupling $F_{\mu z}$, causing both 
contributions to scale identically with large $Q^2$ at large-x.


\begin{figure*}
\subfloat[ \label{sigmaplusx0pt93}]{%
  \includegraphics[height=6cm,width=.49\linewidth]{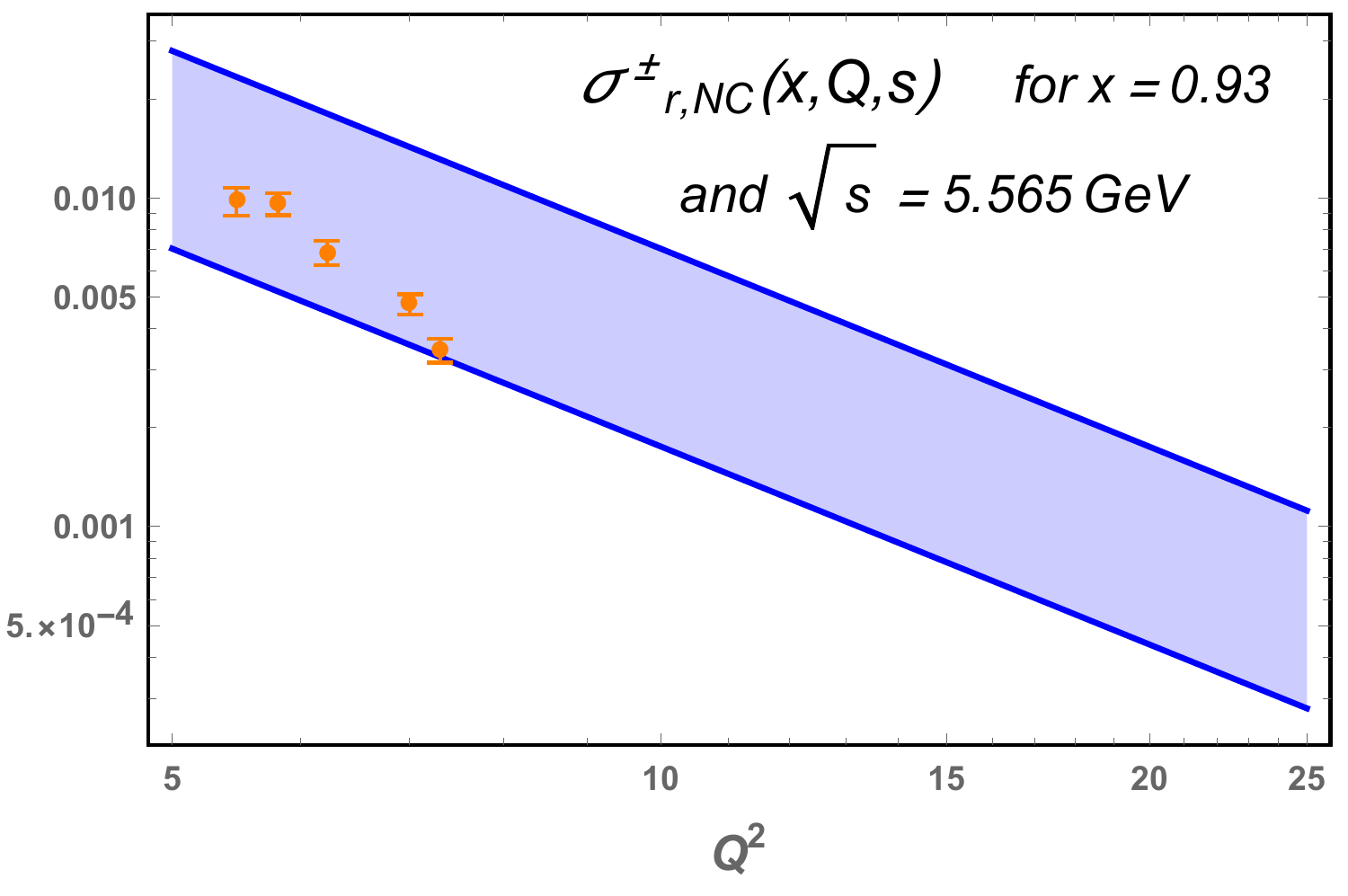}%
}\hfill
\subfloat[\label{sigmaplusx0pt85}]{%
  \includegraphics[height=6cm,width=.49\linewidth]{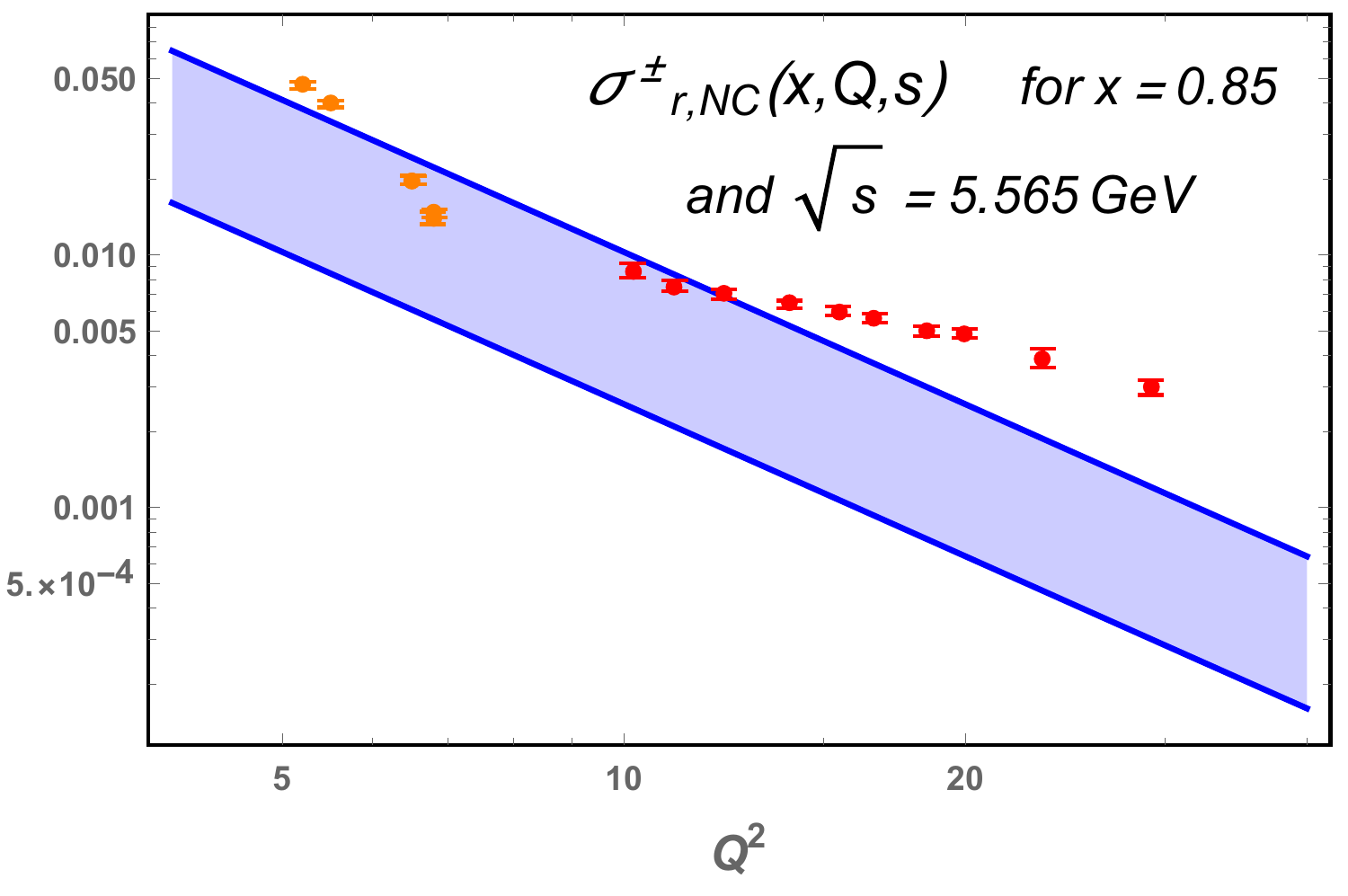}%
}
\caption{
a: $\sigma^{\pm}_{r,NC}(x,Q,s)$ as given in (\ref{ncsi}) for $x=0.93$,
using the large-x holographic PDFs shown in Fig.~\ref{fig:xuvalenceatqq6pt5xuseaatqq6pt5largexnogreen}, with 
the orange data points  from JLAB~\cite{Malace:2009kw};
b: same as in a for $x=0.85$, with the red data points from the combined SLAC and BCDMS collaborations~\cite{Whitlow:1990dr}. See text.}
\label{sigmaplusx0pt93sigmaplusx0pt85}
\end{figure*}



\begin{figure*}
\subfloat[ \label{F2qq6pt5largex}]{%
  \includegraphics[height=6cm,width=.49\linewidth]{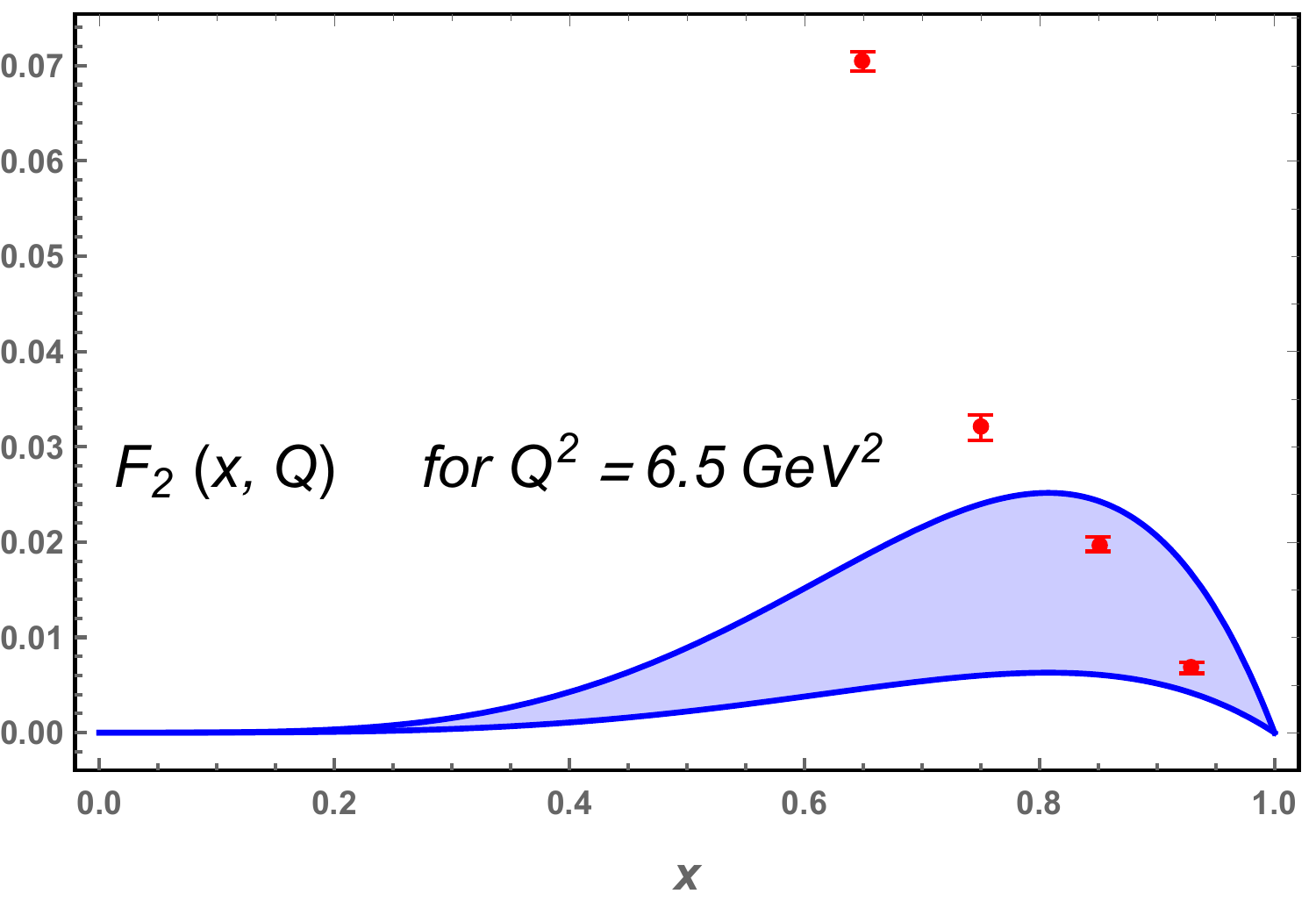}%
}\hfill
\subfloat[\label{F2qq6pt5Zcorrectionlargex}]{%
  \includegraphics[height=6cm,width=.49\linewidth]{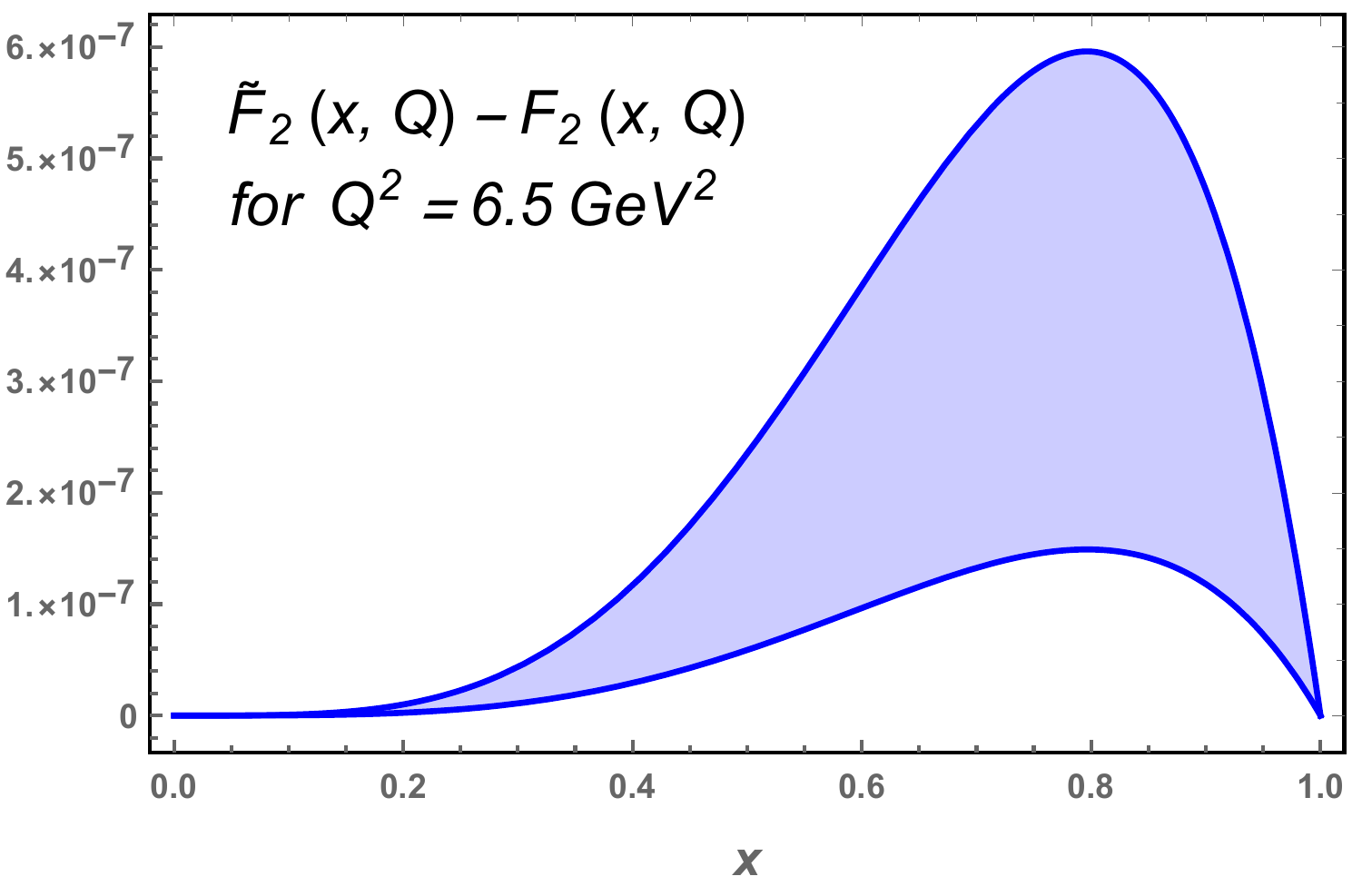}%
}\hfill
\label{F2qq6pt5F2qq6pt5Zcorrectionlargex}
\subfloat[\label{F2qq6pt5gZlargex}]{%
  \includegraphics[height=6cm,width=.49\linewidth]{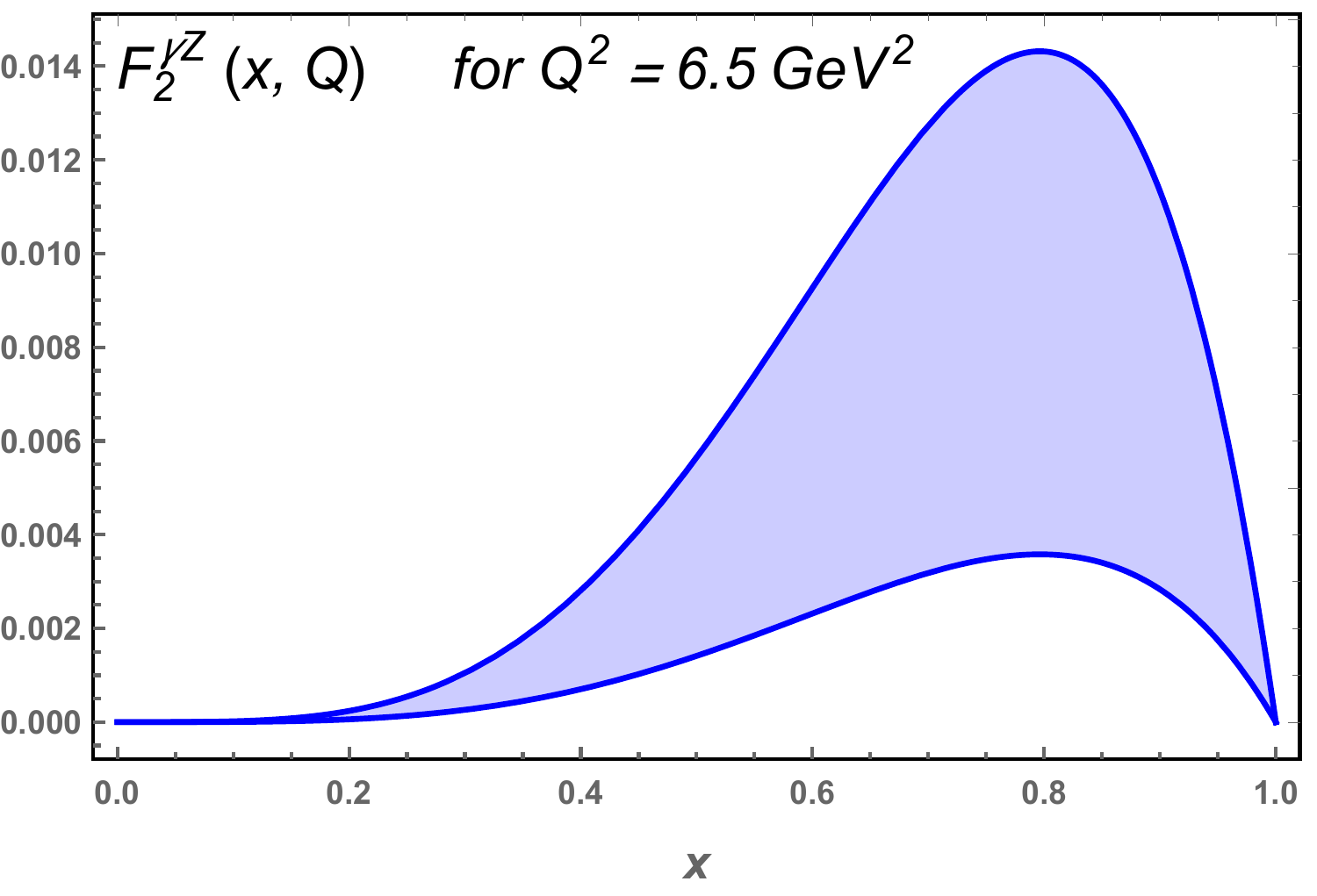}%
}\hfill
\subfloat[\label{F2qq6pt5Zlargex}]{%
  \includegraphics[height=6cm,width=.49\linewidth]{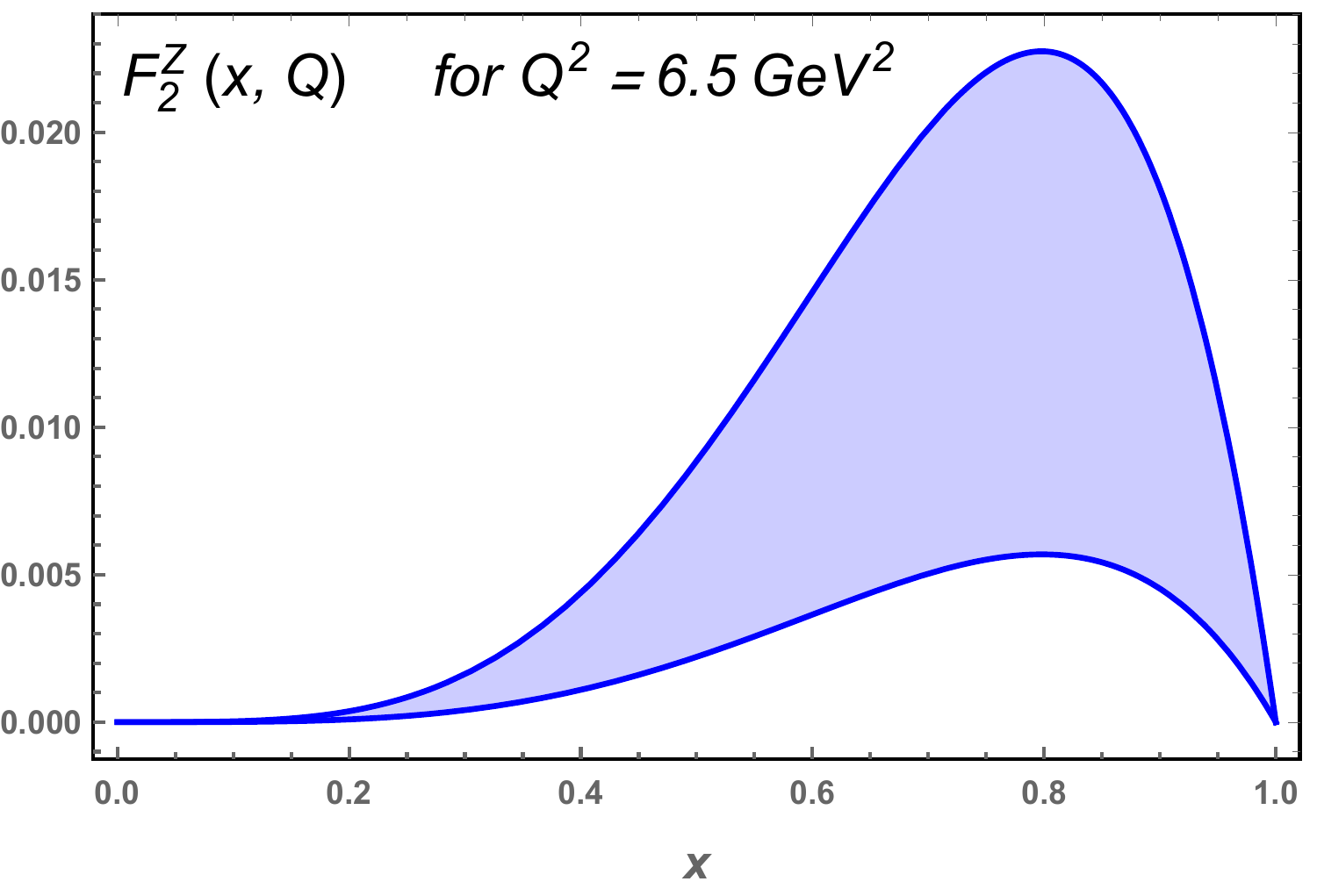}%
}\hfill
\subfloat[\label{F3qq6pt5gZlargex}]{%
  \includegraphics[height=6cm,width=.49\linewidth]{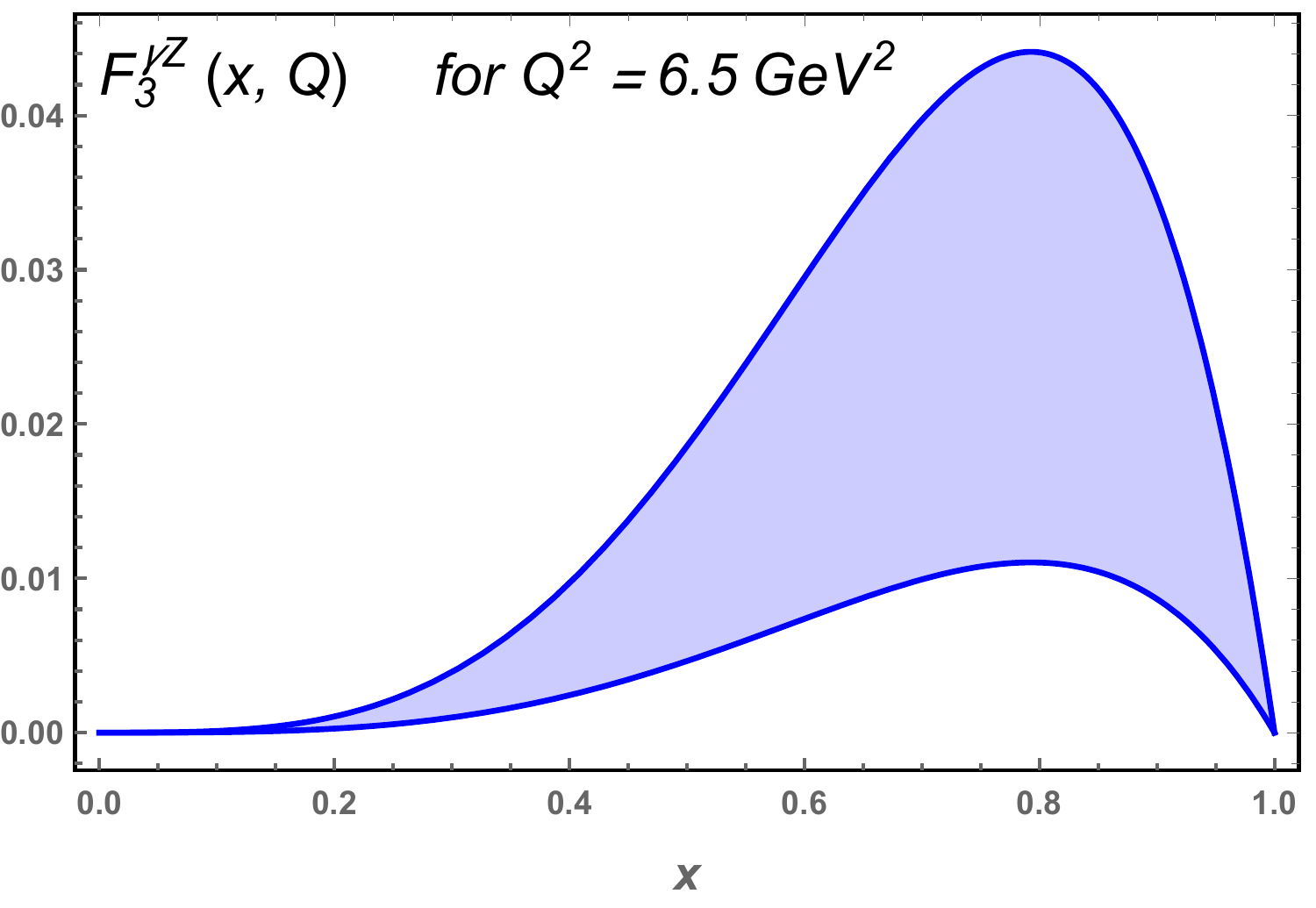}%
}\hfill
\subfloat[\label{F3qq6pt5Zlargex}]{%
  \includegraphics[height=6cm,width=.49\linewidth]{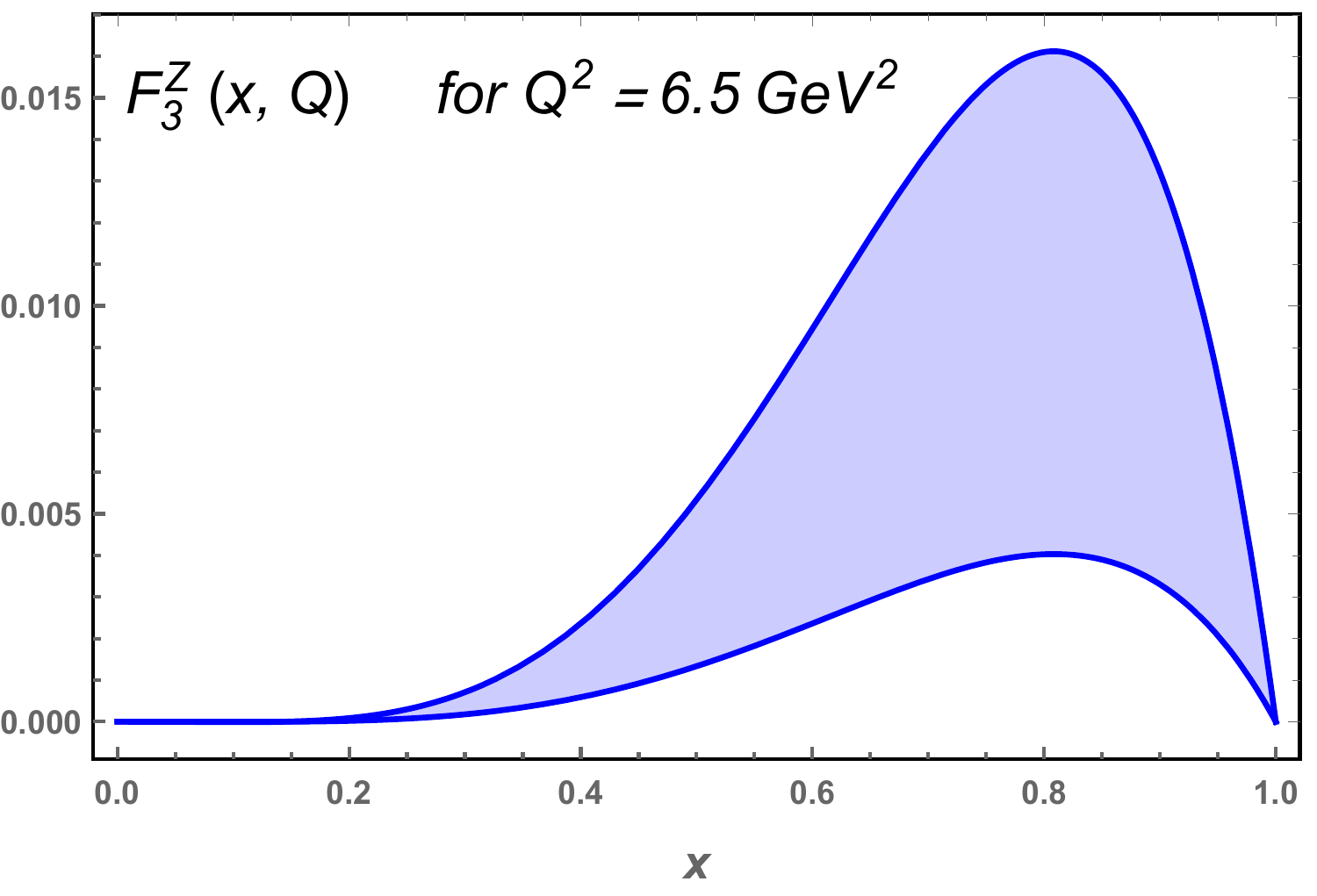}%
}
\caption{
(a)$F_2(x,Q)$ as given in (\ref{ncfu}) using the large-x holographic PDFs shown in Fig.~\ref{fig:xuvalenceatqq6pt5xuseaatqq6pt5largexnogreen}. The red data points are from combined SLAC and BCDMS collaborations \cite{Whitlow:1990dr} for $x=0.65$, $x=0.75$, and $x=0.85$;
(b) $\tilde{F}_2(x,Q)-F_2(x,Q)$ as given in (\ref{ncfu}) using the large-x holographic PDFs shown in Fig.~\ref{fig:xuvalenceatqq6pt5xuseaatqq6pt5largexnogreen};
(c) $F_2^{\gamma Z}(x,Q)$ as given in (\ref{ncfu}) using the large-x holographic PDFs shown in Fig.~\ref{fig:xuvalenceatqq6pt5xuseaatqq6pt5largexnogreen};
(d) $F_2^{Z}(x,Q)$ as given in (\ref{ncfu}) using the large-x holographic PDFs shown in Fig.~\ref{fig:xuvalenceatqq6pt5xuseaatqq6pt5largexnogreen};
(e) $F_3^{\gamma Z}(x,Q)$ as given in (\ref{ncfu}) using the large-x holographic PDFs shown in Fig.~\ref{fig:xuvalenceatqq6pt5xuseaatqq6pt5largexnogreen};
(g) $F_3^{Z}(x,Q)$ as given in (\ref{ncfu}) using the large-x holographic PDFs shown in Fig.~\ref{fig:xuvalenceatqq6pt5xuseaatqq6pt5largexnogreen}.
}
\label{F2XQ}
\end{figure*}


Here $\tilde{\mathcal{N}}_{L}^{\pm}\equiv c(\tau)\times\mathcal{N}_{L}^{\pm}$ with $c(\tau)$ are undetermined 
coefficients,  that can in principle be fixed  rigorously in a specific model.  As we noted earlier, 
strict bulk-to-boundary correspondence implies  ${\cal N}_L^\pm =1$ in the double and dual limit. Here we are
assuming proportionality between the bulk and boundary structure functions with ${\cal N}_L^\pm$  as overall parameters~\cite{BRAGA},
that capture  partially the  finite corrections to the double limit. In the numerical analysis at large-x to follow, 
we fit only the two parameters $\tilde{\tilde{\mathcal{N}}}_{L}^{\pm}$ which are defined as

\be \label{twoparameters}
\tilde{\tilde{\mathcal{N}}}_{L}^{\pm}\, \equiv \, \big(\tilde{\mathcal{N}}_{L}^{\pm}\big)^2\times \big(e^{\pm}_{Wnucleon}\big)^2+\big(\tilde{\mathcal{N}}_{L}^{\pm}\big)^2\times \eta^2\times 4(\tau-1)^2+\big(\tilde{\mathcal{N}}_{L}^{\pm}\big)^2\times e^{\pm}_{Wnucleon}\times\eta\times 4(\tau-1)\,.
\ee 
In terms of (\ref{twoparameters}), the structure functions (\ref{F23DiracPauli1}) simplify 

\bea\label{F23DiracPauli2}
F_{2s}^{W^{\pm}p}(x,Q)=F_{3s}^{W^{\pm}p}(x,Q)&=&\big(\tilde{\tilde{\mathcal{N}}}_{L}^{\pm}\big)^2\times\left(\frac{\tilde{\kappa}^2}{Q^2}\right)^{\tau-1} x^{\tau+1}(1-x)^{\tau-2}\,.
\eea
with  $\tilde{\kappa}=0.350\,\rm{GeV}$ as fixed by the mass of the proton and $\rho$ meson in \cite{CARLSON}.

(\ref{F23DiracPauli2}) scales as $(1/Q^2)^{{\tau-1}}$ asymptotically in agreement with the hard scaling laws expected from strong coupling~\cite{POL},  but vanishes as $(1-x)^{\tau-2}$ 
at large $x$ in contrast to $(1-x)^{2\tau -3}$ suggested in~\cite{DRELL}. The large-x behavior at strong coupling follows from the observation that for the virtual photon with amplitude $1/Q$ to scatter off  the 
nucleon as a Dirac fermion, the latter has to shrink to a size $(1/Q)^\tau$,  with a scattering probability$(1/Q^2)^{\tau-1}$.  As a result, the structure function at large $Q^2$ but fixed 
$s\sim Q^2(1-x)$ scales as 

\bea
F_2(x,Q)\sim Q^2\bigg|\bigg( \frac 1{Q^2}\bigg)^{\tau-1}\bigg|^2\, \bigg(s=Q^2(1-x)\bigg)^\alpha
\eea
To reproduce the hard scaling law asymptotically requires $2\alpha+6-4\tau=2-2\tau$ or $\alpha=\tau-2$, which is the large-x  scaling in (\ref{F23DiracPauli2}).
To recover Bjorken scaling for the structure function requires $2\alpha+6-4\tau=0$ or $\alpha=2\tau-3$ which is the large-x scaling law suggested
in~\cite{DRELL}. We expect the latter to set in at very large $Q^2$.

In Fig.~\ref{fig:xuvalenceatqq6pt5xuseaatqq6pt5largexnogreen}a we show the behavior the valence distributions 
$xu_V(x,Q)$ in the upper blue-dark-band, and $xd_V(x,Q)$ in the  lower green-light-band  at $Q^2=6.5$ GeV$^2$, following from (\ref{7Xbclargex}). In 
 Fig.~\ref{fig:xuvalenceatqq6pt5xuseaatqq6pt5largexnogreen}b we show the behavior of the sea  distributions 
$x\overline{u}(x,Q)$ in the lower blue-dark-band, and $x\overline{d}(x,Q)$ in the upper green-light-band  at the same $Q^2=6.5$ GeV$^2$,  following from (\ref{7bclargex}).
The normalization coefficients delimiting the blue and green bands are determined by varying $\tilde{\tilde{\mathcal{N}}}_{L}^{+}$ between 17.715 and 35.431, and $\tilde{\tilde{\mathcal{N}}}_{L}^{-}$ between 30.667 and 61.335.

To assess the range of validity in parton-x of the holographic results at large-x, we re-asses the reduced
neutral charge  $\sigma^{\pm}_{r,NC}(x,Q,s)$ as given in (\ref{ncsi}) solely in terms of the large-x holographic PDFs. 
The results are  shown in Fig.~\ref{fig:xuvalenceatqq6pt5xuseaatqq6pt5largexnogreen} at low $Q^2$ for $\sqrt{s}=5.565$ GeV.
The orange data points are from JLAB \cite{Malace:2009kw}.
The red data points are from the combined SLAC and BCDMS collaborations~\cite{Whitlow:1990dr} 
(see Figure 5.14 in~\cite{Whitlow:1990dr}). Note that the 
 $F_2(x,Q)$ data of \cite{Whitlow:1990dr} to $\sigma^{\pm}_{r,NC}(x,Q,s)$ where converted using
 (\ref{RXQ}). These results show that  the range of validity of the holographic results  at large-x is  limited to   $0.75\leq x\leq 1$ (see also below).
Our results for the reported DIS $e^\pm p$ in~Fig.~\ref{sigmaplusx0pt93sigmaplusx0pt85} are consistent with the  recent holographic results 
reported in~\cite{FolcoCapossoli:2020pks} in the large-x regime (see Figure 5 in~\cite{FolcoCapossoli:2020pks}).

\subsection{Structure functions at intermediate $Q^2$}

To investigate further the range of validity of the holographic PDFs at intermediate $Q^2$ and 
large-x,  we show in Fig.~\ref{F2XQ}a  the holographic structure function $F_2(x)$ in (\ref{ncfu}) versus $x$
at $Q^2=6.5$ GeV$^2$, evaluated using  the results for the valence distributions shown in Fig.~\ref{fig:xuvalenceatqq6pt5xuseaatqq6pt5largexnogreen}a. 
 The red data points are from the combined SLAC and BCDMS collaborations in~\cite{Whitlow:1990dr} for $x=0.65$, $x=0.75$, and $x=0.85$.
In Fig.~\ref{F2XQ}b  we show  the difference $\tilde{F}_2(x,Q)-F_2(x,Q)$ versus $x$ as given in (\ref{ncfu}) for $Q^2=6.5$ GeV$^2$ 
using  also the large-x holographic PDFs shown in Fig.~\ref{fig:xuvalenceatqq6pt5xuseaatqq6pt5largexnogreen}.
In Figs.~\ref{fig:xuvalenceatqq6pt5xuseaatqq6pt5largexnogreen}c,d,e,f we show respectively, $F_2^{\gamma Z}(x,Q)$, $F_2^{Z}(x,Q)$, $F_3^{\gamma Z}(x,Q)$
and $F_3^{Z}(x,Q)$  versus large-x for $Q^2=6.5$ GeV$^2$ using also the large-x holographic PDFs in Fig.~\ref{fig:xuvalenceatqq6pt5xuseaatqq6pt5largexnogreen}.
Fig.~\ref{F2XQ}a shows that the holographic results are compatible with the SLAC and BCDMS data~\cite{Whitlow:1990dr} in the range
$0.75\leq x\leq 1$ for $Q^2=6.5$ GeV$^2$. 

\subsection{Comparison to the empirical CTEQ, LHAPDF and NNPDF data sets}

We have also compared our holographic PDF sets both for the small-x and large-x regimes for the valence and sea distributions, to 
292 PDF sets from the CTEQ and LHAPDF projects incorporated within the ManeParse Mathematica package~\cite{Clark:2016jgm}.
The global comparison is displayed in  Figs.~\ref{CTEQ-LHAPDF-VALENCE} and Figs.~\ref{CTEQ-LHAPDF-SEA} for $Q^2=6.5$ GeV$^2$. 
Figs.~\ref{CTEQ-LHAPDF-SEA}a,c show the 292 PDF sets   in multiple solid lines  from CTEQ and LHAPDF data set~\cite{Clark:2016jgm}, in comparison
to our holographic PDF sets shown in blue-light-band for the up-distributions, and in green-light-band for the down-distributions. 
The holographic  growth  at intermediate-x overcomes these data sets, but joins with 
a smaller subset of these data at very low-x.   Figs.~\ref{CTEQ-LHAPDF-SEA}b,d show that the holographic sea at large-x is relatively small
but consistent overall with all the data sets in the expected applicability of our approach. A similar comparison for the holographic valence
distributions is shown in Figs.~\ref{CTEQ-LHAPDF-VALENCE}a,c  at low-x and Figs.~\ref{CTEQ-LHAPDF-SEA}b,d at large-x with consistency in both
limits with the CTEQ and LHAPDF data set~\cite{Clark:2016jgm}.

In Figs.~\ref{NNPDF-SEA} we comparre the holographic results for the sea distributions to the NNPDF collaboration PDF sets with error bars (shown in red band) from  the
LHAPDF projects incorporated within the ManeParse Mathematica package \cite{Clark:2016jgm}. Figs.~\ref{NNPDF-SEA}a,c show our results in blue-dark-band for the sea
of up quarks, and in green-light-band for the sea of down quarks. Figs.~\ref{NNPDF-SEA}b,d show also our results for the sea of up and down quarks at large-x. The growth
of the holographic sea at low-x is larger in the intermediate range, but slower at very low-x. At large-x, our holographic results for the sea are relatively small but consistent
with the reported data set. In Figs.~\ref{NNPDF-VALENCE} the valence up and down distributions are compared to the same data set with the same notations as for the sea.
At low-x, the holographic results are consistent with zero for the valence up and down quark distributions. At larger-x, the holographic results are consistently larger for the
up quark valence distribution in the region of validity of the holographic construction.

\begin{figure*}
\subfloat[\label{xuatqq6pt5goined300pdfsetssmallx}]{%
  \includegraphics[height=6cm,width=.49\linewidth]{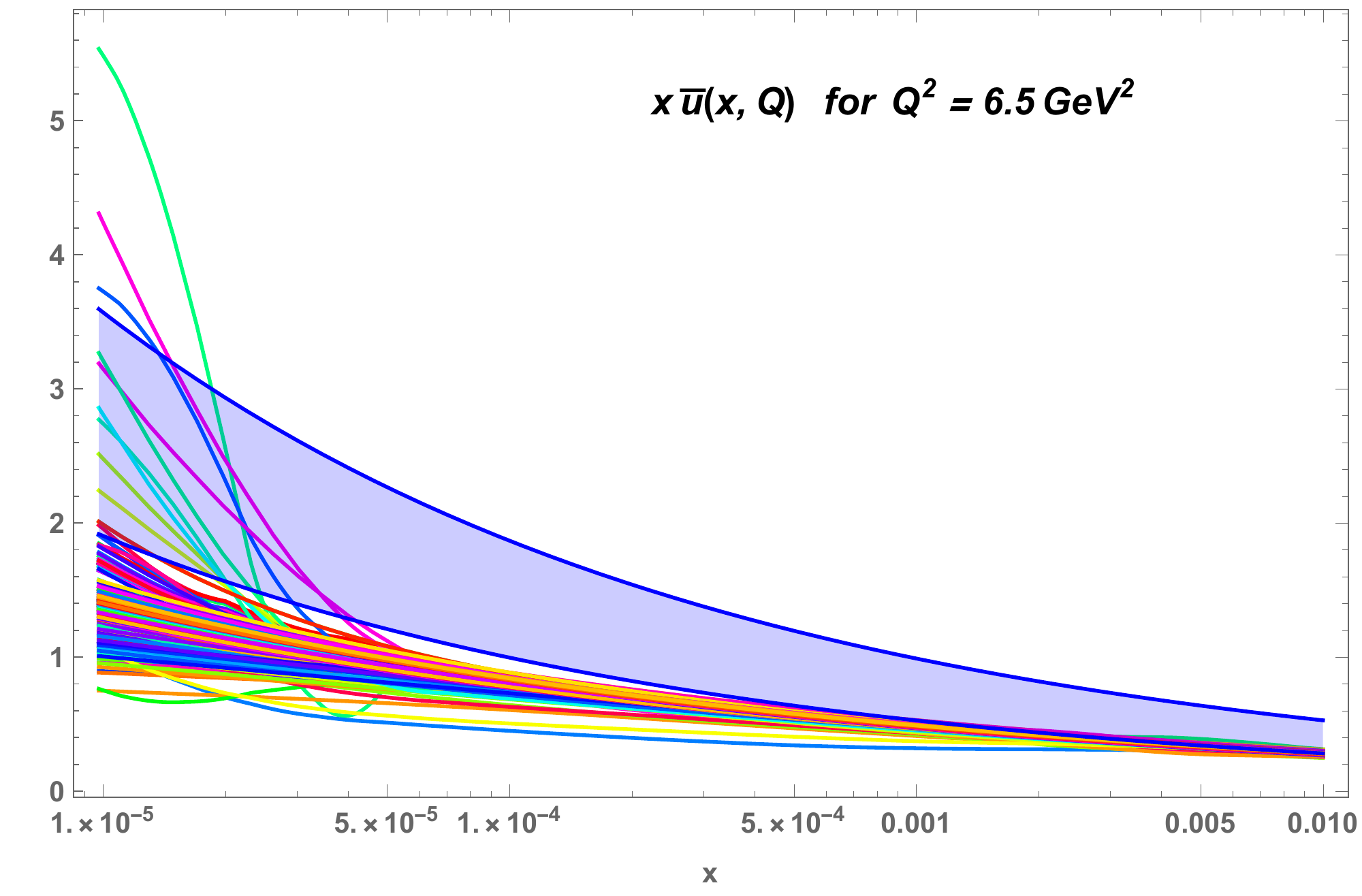}%
}\hfill
\subfloat[\label{xuatqq6pt5goined300pdfsetslargex}]{%
  \includegraphics[height=6cm,width=.49\linewidth]{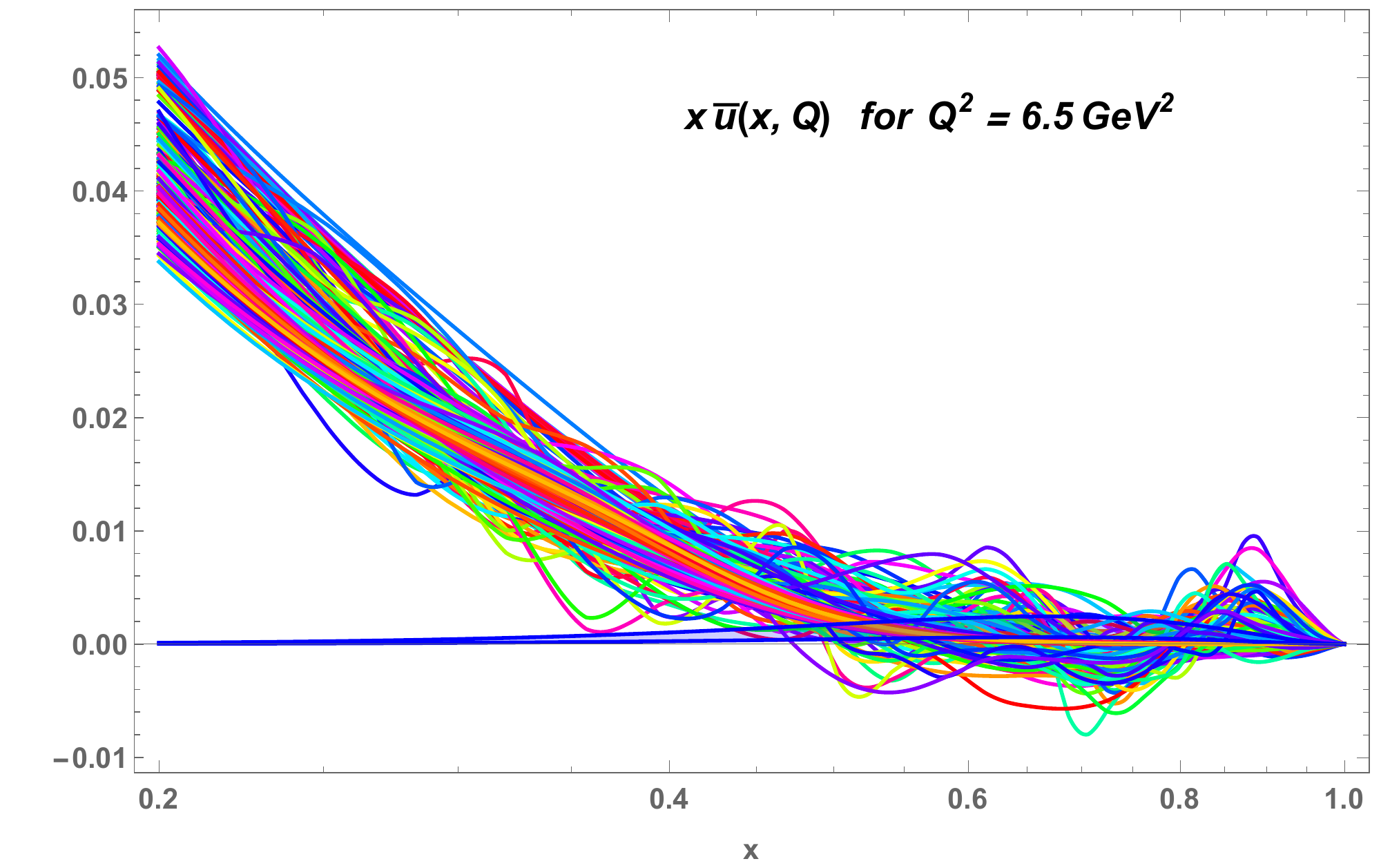}%
}\hfill
\subfloat[\label{xdatqq6pt5goined300pdfsetssmallx}]{%
  \includegraphics[height=6cm,width=.49\linewidth]{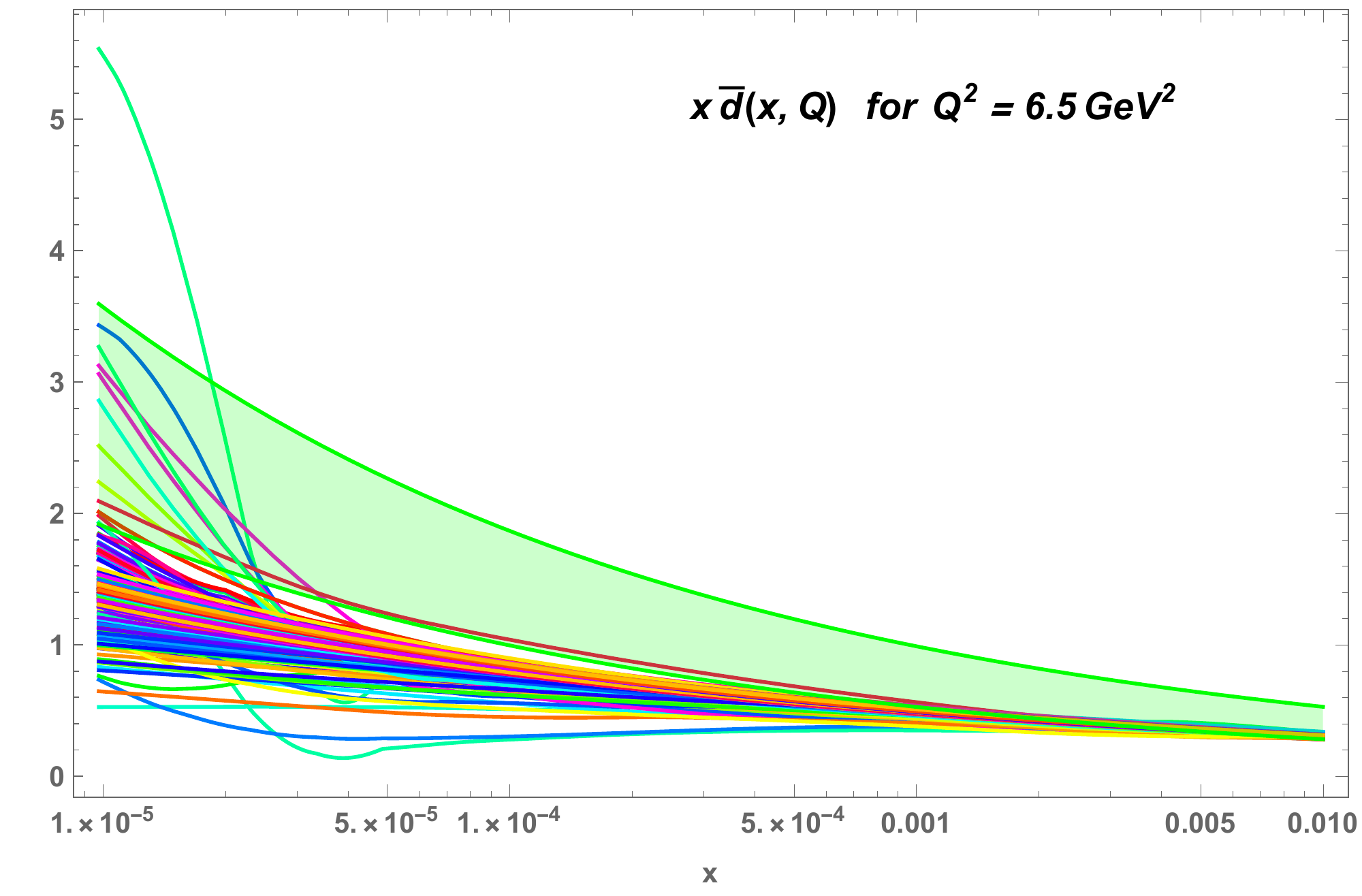}%
}\hfill
\subfloat[\label{xdatqq6pt5goined300pdfsetslargex}]{%
  \includegraphics[height=6cm,width=.49\linewidth]{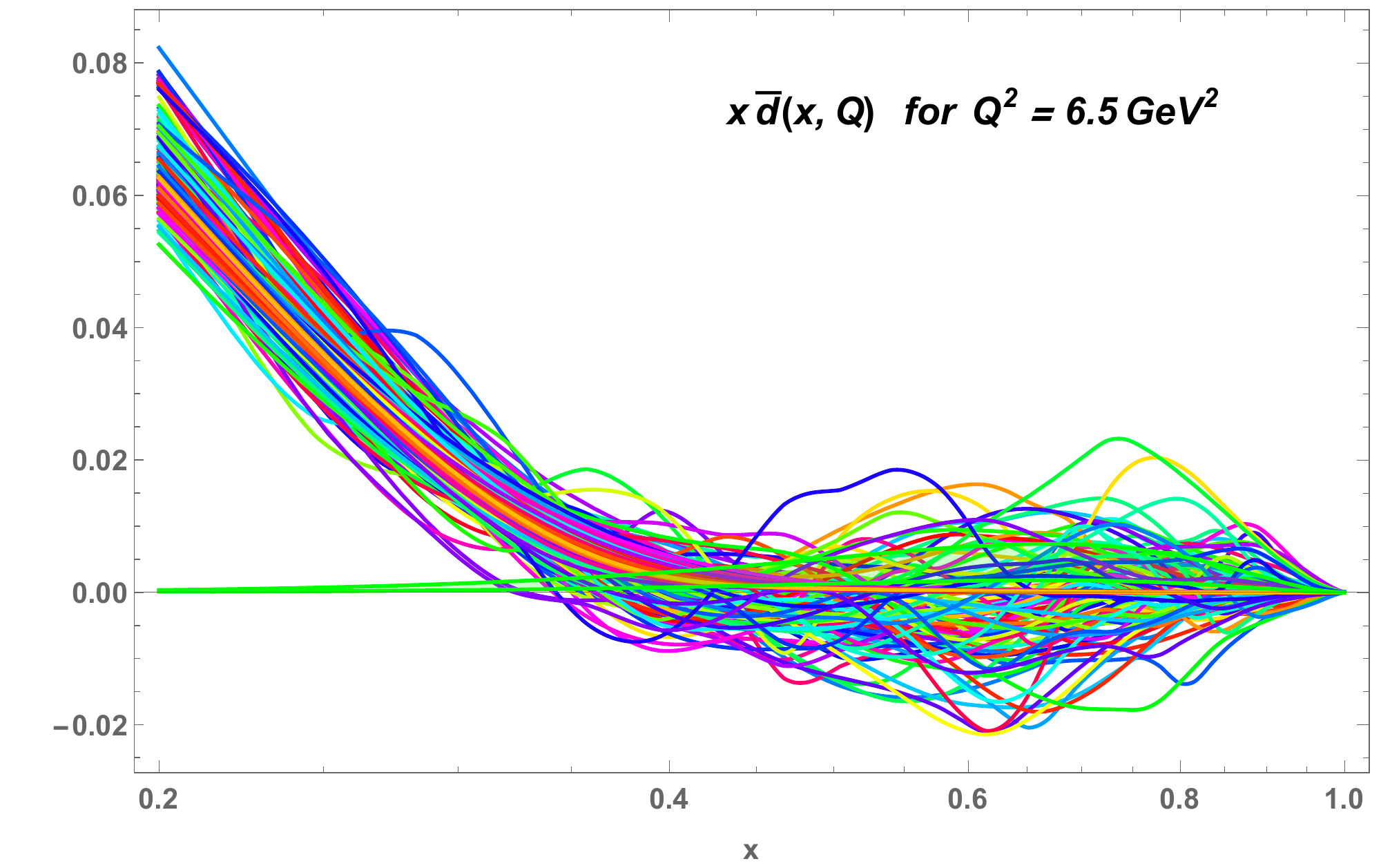}%
}
\caption{
(a,b) Our holographic PDF sets (shown in blue band) compared to 292 PDF sets (shown in multiple solid lines) from CTEQ and LHAPDF projects incorporated within the ManeParse Mathematica package \cite{Clark:2016jgm};
(c,d) Our holographic PDF sets (shown in green band) compared to 292 PDF sets (shown in multiple solid lines) from CTEQ and LHAPDF projects incorporated within the ManeParse Mathematica package \cite{Clark:2016jgm}.}
\label{CTEQ-LHAPDF-VALENCE}
\end{figure*}



\begin{figure*}
\subfloat[\label{xuvalenceatqq6pt5goined300pdfsetssmallx}]{%
  \includegraphics[height=6cm,width=.49\linewidth]{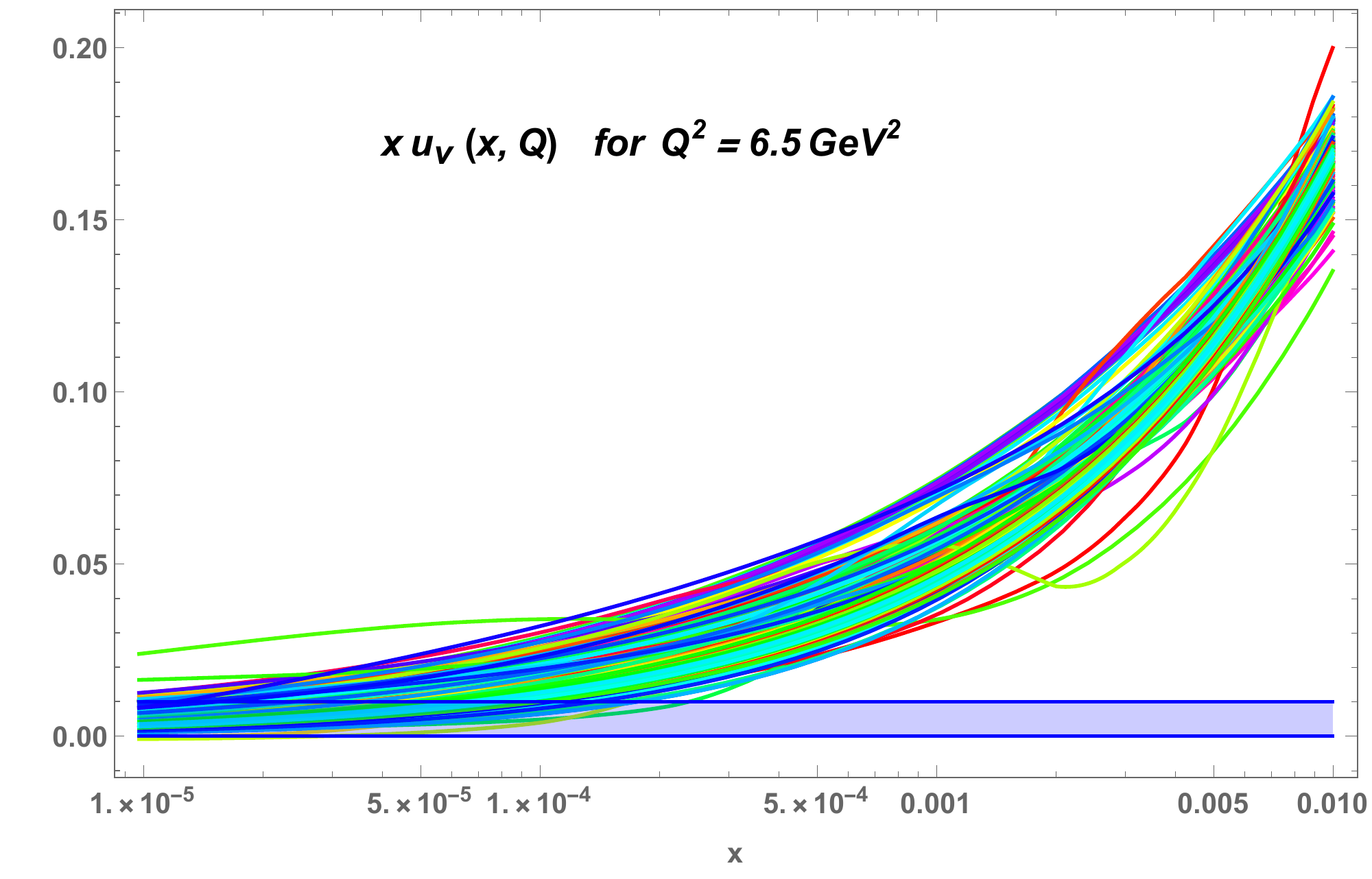}%
}\hfill
\subfloat[\label{xuvalenceatqq6pt5goined300pdfsetslargex}]{%
  \includegraphics[height=6cm,width=.49\linewidth]{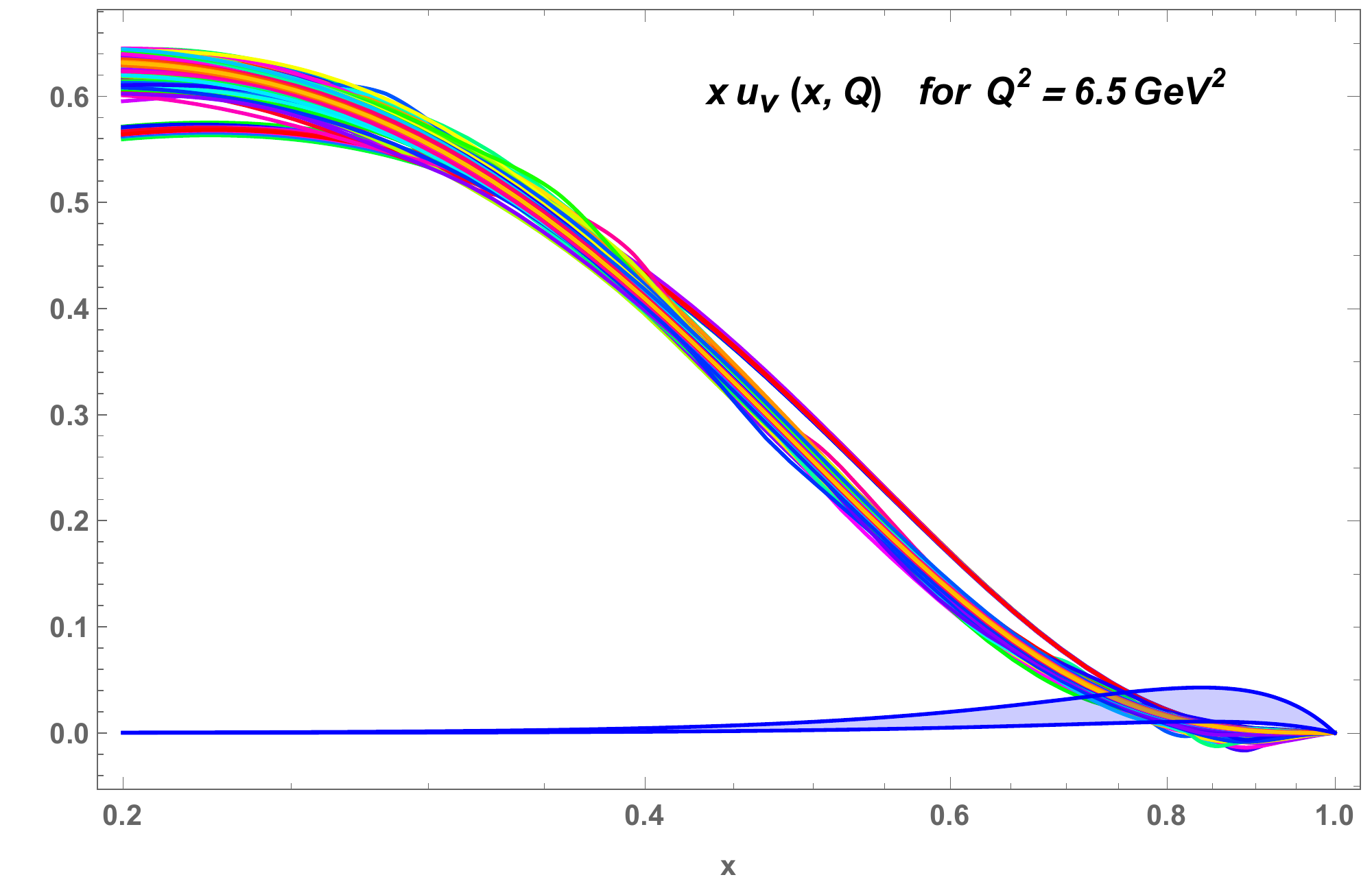}%
}\hfill
\subfloat[\label{xdvalenceatqq6pt5goined300pdfsetssmallx}]{%
  \includegraphics[height=6cm,width=.49\linewidth]{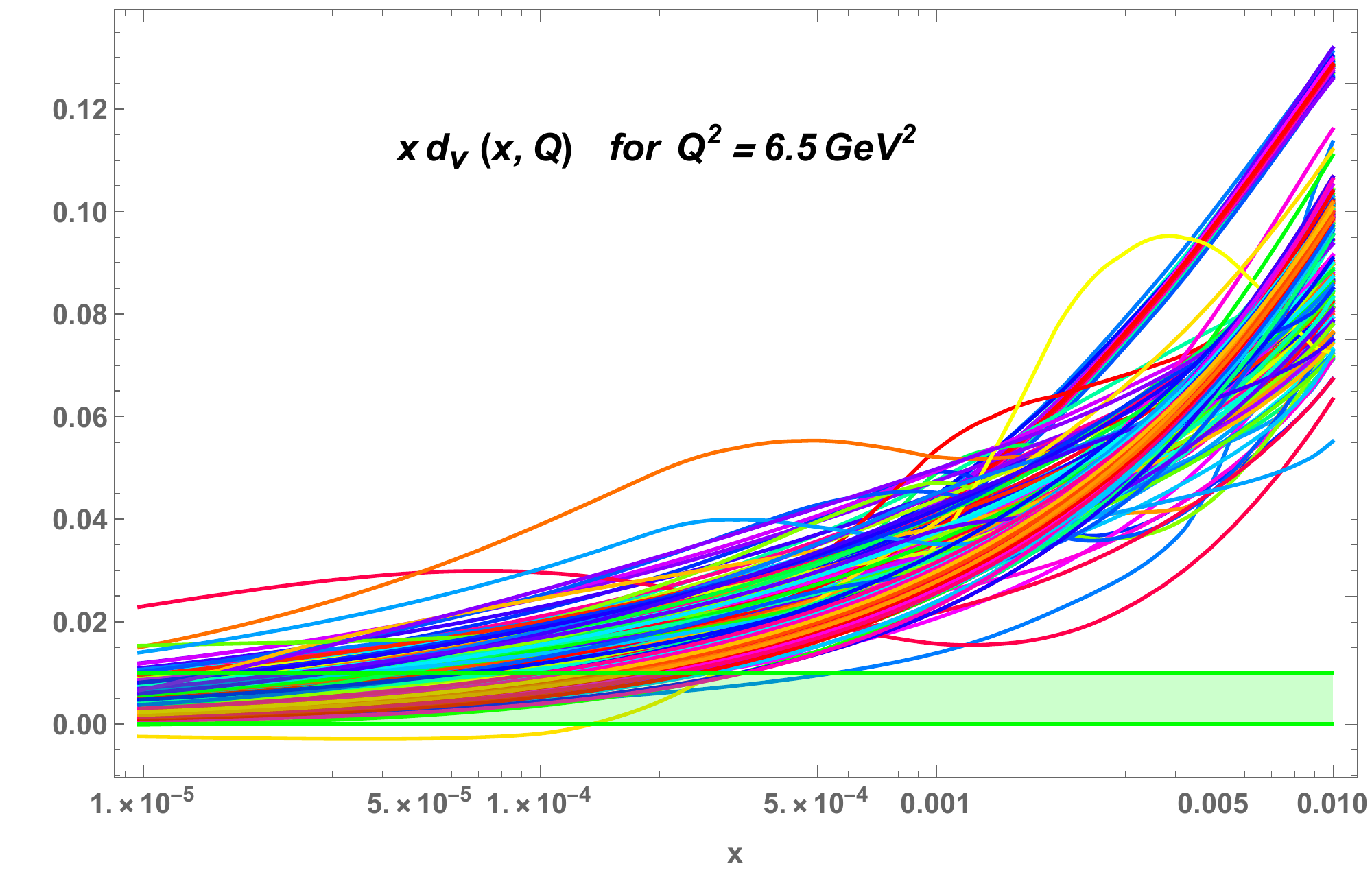}%
}\hfill
\subfloat[\label{xdvalenceatqq6pt5goined300pdfsetslargex}]{%
  \includegraphics[height=6cm,width=.49\linewidth]{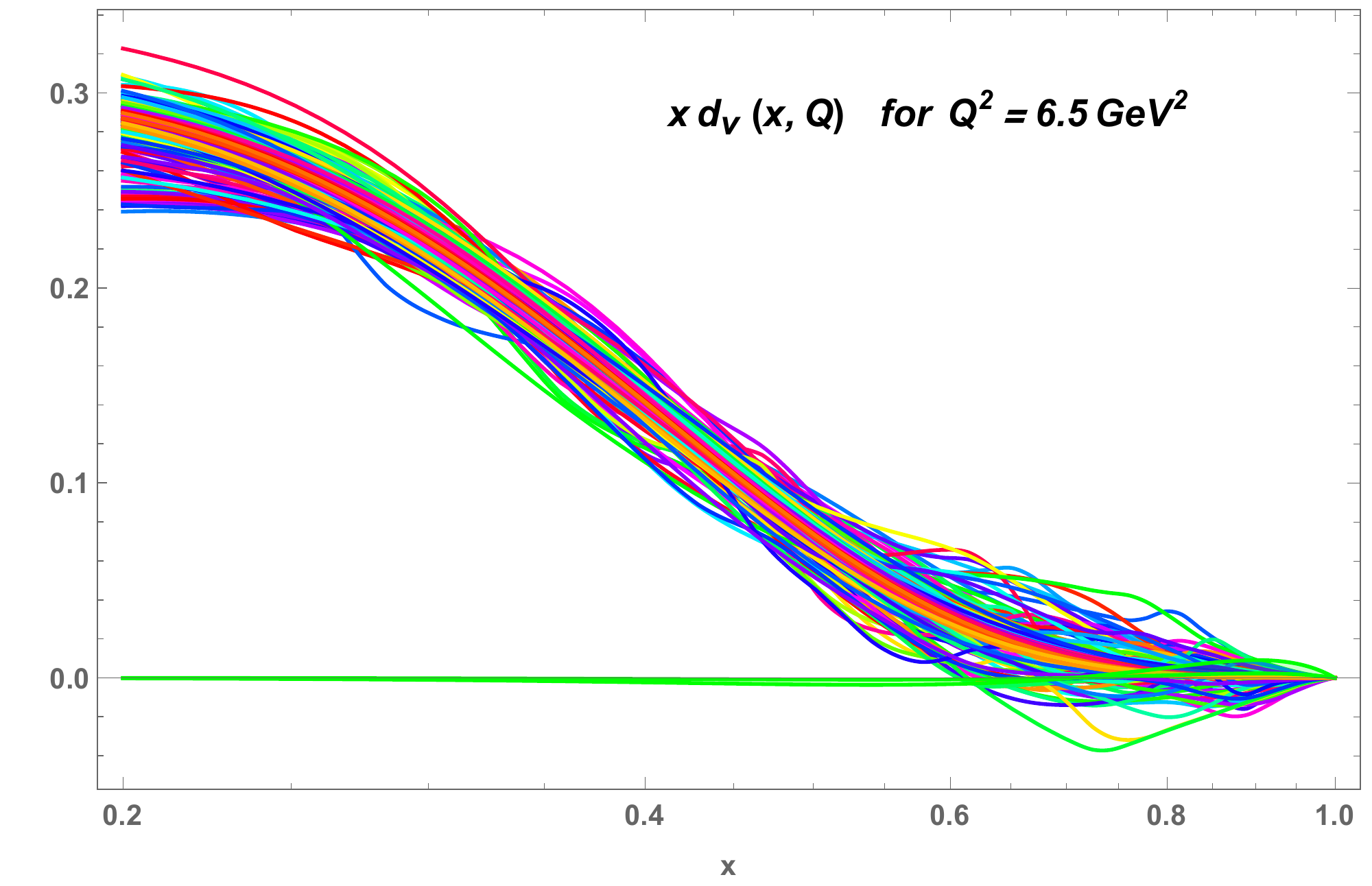}%
}
\caption{
(a,b) Our holographic PDF sets (shown in blue band) compared to 292 PDF sets (shown in multiple solid lines) from CTEQ and LHAPDF projects incorporated within the ManeParse Mathematica package \cite{Clark:2016jgm};
(c,d) Our holographic PDF sets (shown in green band) compared to 292 PDF sets (shown in multiple solid lines) from CTEQ and LHAPDF projects incorporated within the ManeParse Mathematica package \cite{Clark:2016jgm}.}
\label{CTEQ-LHAPDF-SEA}
\end{figure*}




\begin{figure*}
\subfloat[\label{xuatqq6pt5goinedNNpdfsetssmallx}]{%
  \includegraphics[height=6cm,width=.49\linewidth]{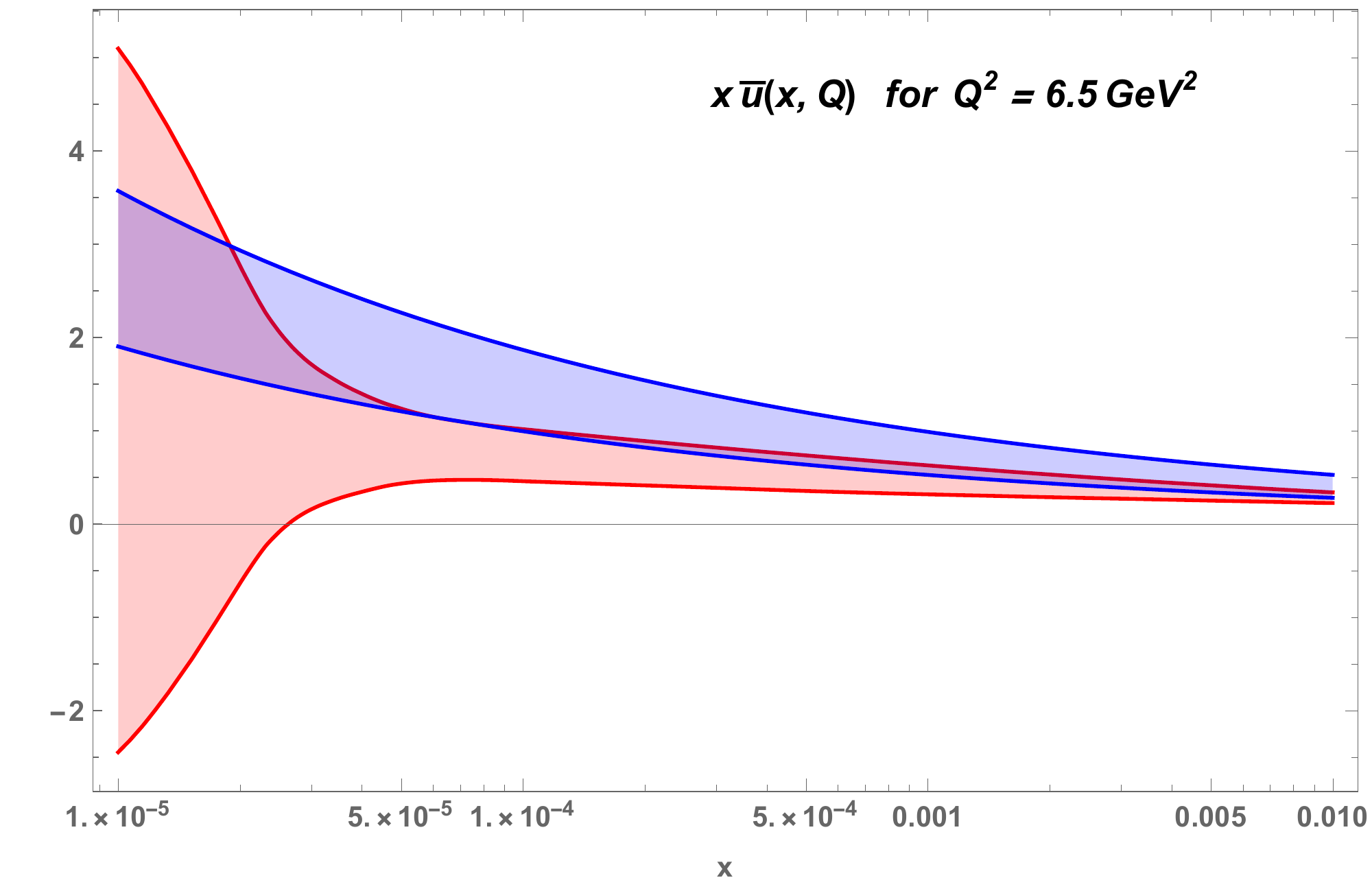}%
}\hfill
\subfloat[\label{xuatqq6pt5goinedNNpdfsetslargex}]{%
  \includegraphics[height=6cm,width=.49\linewidth]{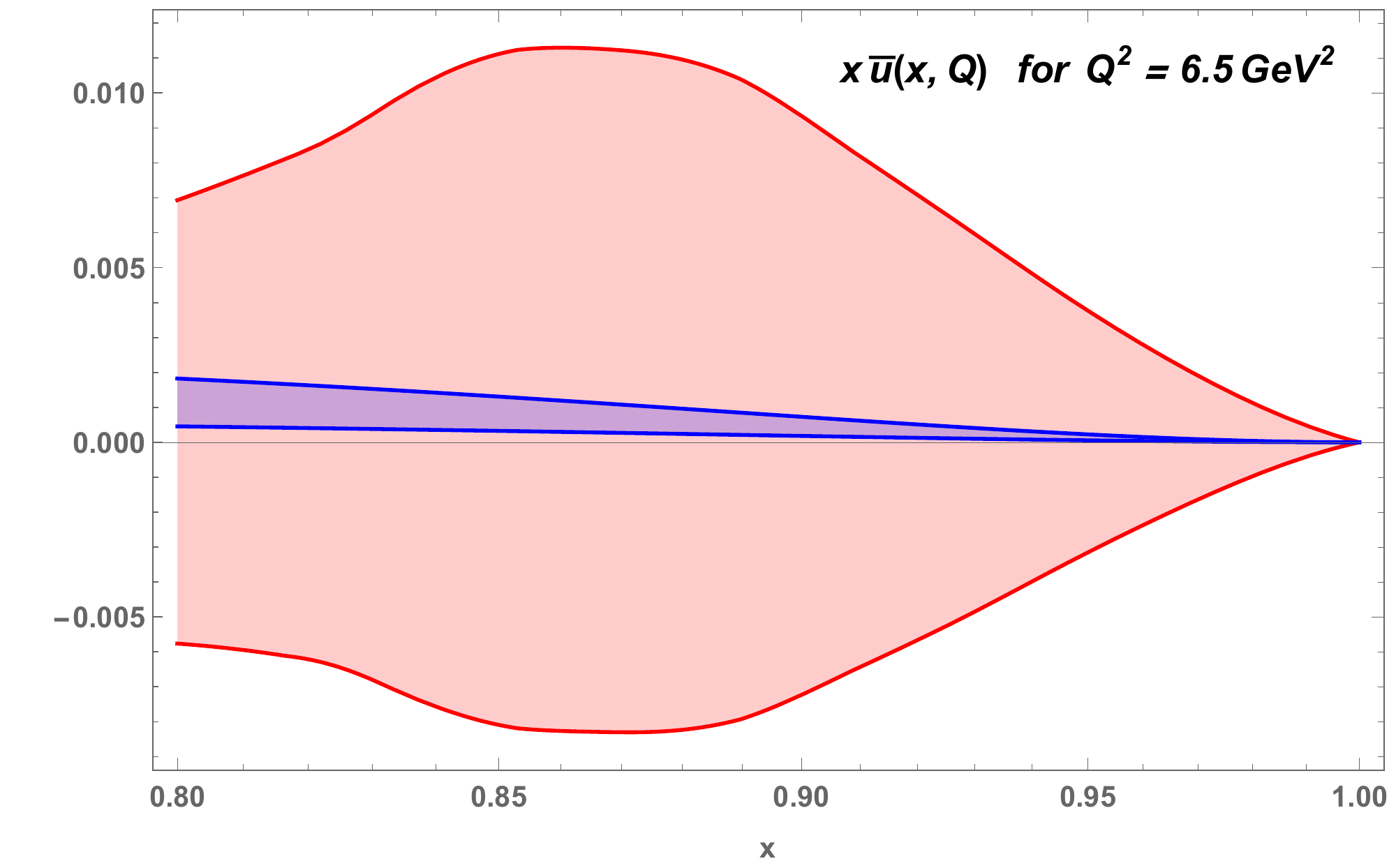}%
}\hfill
\subfloat[\label{xdatqq6pt5goinedNNpdfsetssmallx}]{%
  \includegraphics[height=6cm,width=.49\linewidth]{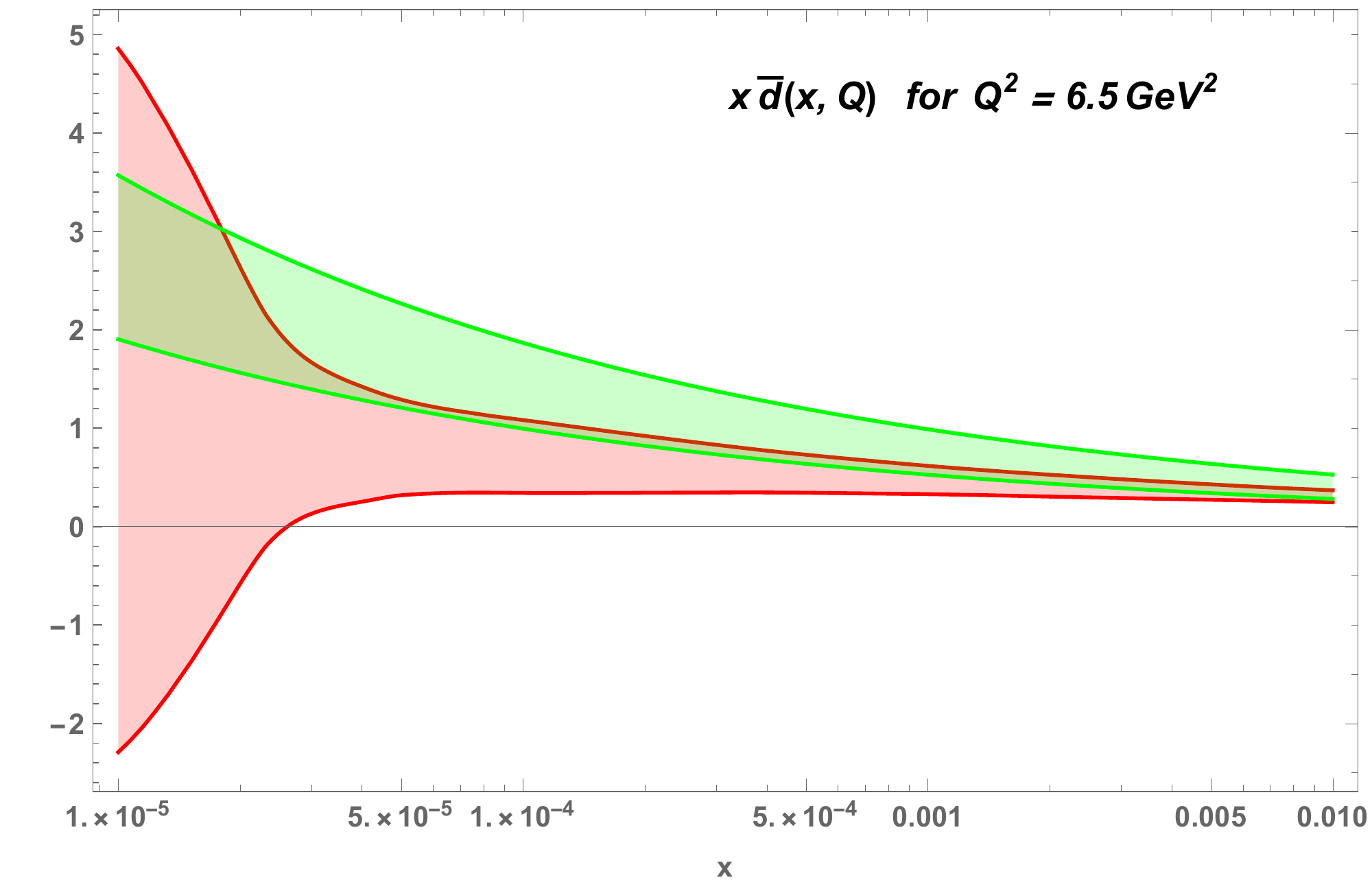}%
}\hfill
\subfloat[\label{xdatqq6pt5goinedNNpdfsetslargex}]{%
  \includegraphics[height=6cm,width=.49\linewidth]{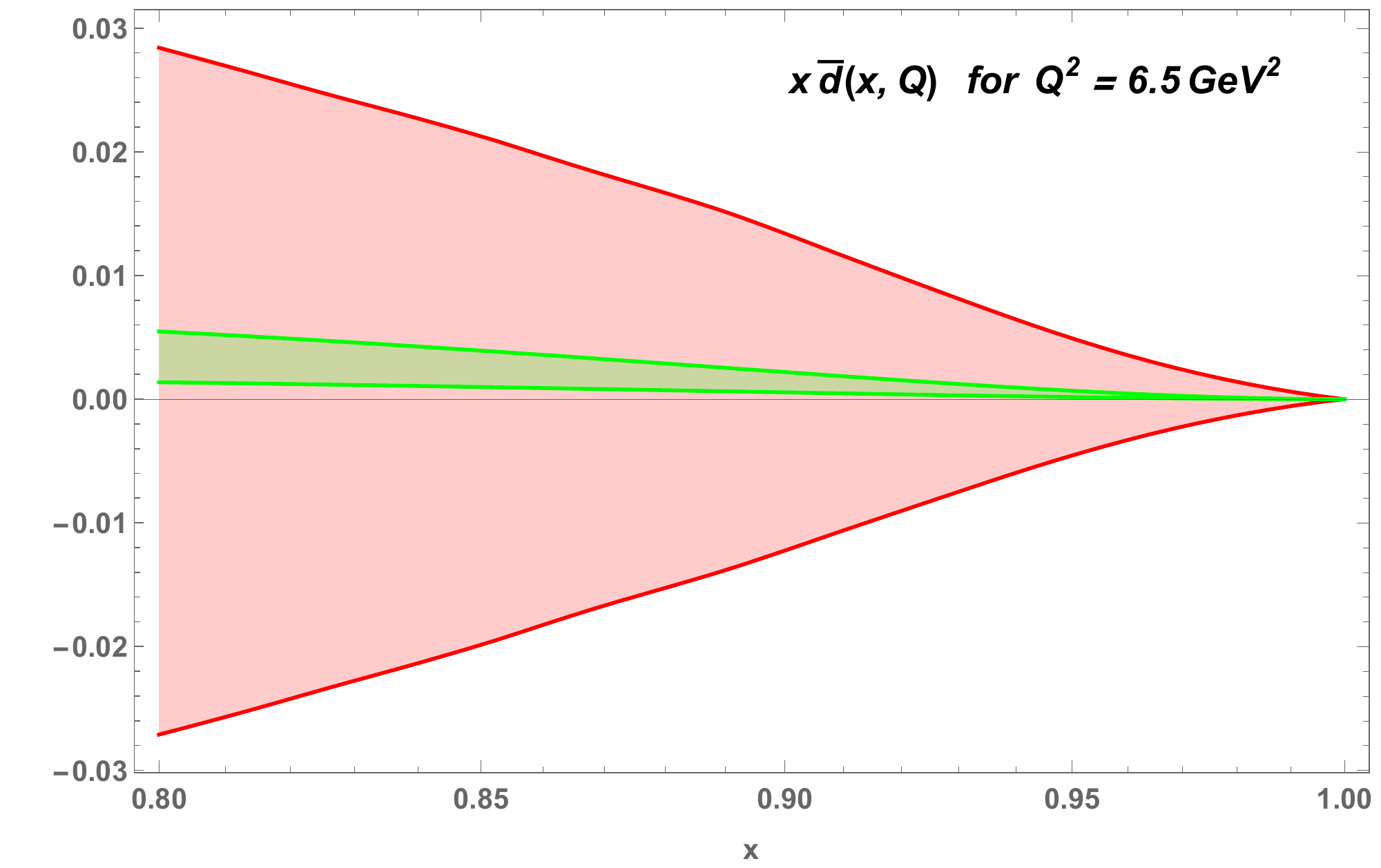}%
}
\caption{
(a,b) Our holographic PDF sets (shown in blue band) compared to NNPDF collaboration PDF sets with error bars (shown in red band) from LHAPDF projects incorporated within the ManeParse Mathematica package \cite{Clark:2016jgm};
(c,d) Our holographic PDF sets (shown in green band) compared to NNPDF collaboration PDF sets with error bars (shown in red band) from LHAPDF projects incorporated within the ManeParse Mathematica package \cite{Clark:2016jgm}.}
\label{NNPDF-SEA}
\end{figure*}


\section{Conclusions}

We presented a comprehensive holographic derivation of DIS neutrino-nucleon scattering process
using AdS/CFT duality formulated in a slice of AdS$_5$ with bulk U(2) valued vector mesons 
to describe  a tower of iso-vector and iso-axial particles, and a fermionic isodoublet  to account 
for the even- and odd-parity proton and neutron excitations. To distinguish the vectors from the axials,
chiral symmetry is broken either through a tachyon in the bi-fundamental representation both for the
hard or soft wall model, or though boundary conditions for the hard wall. The Yukawa coupling
of the bi-fundamental tachyon field between the even-odd bulk fermionic fields was ignored, since
the DIS regime is mostly sensitive to the high lying part of the nuclon spectrum for which the effect
is negligible. 

Using Witten diagrams, we have  derived the pertinent Dirac and Pauli fiorm factors for both the
direct or diagonal currents and the transition or off-diagonal currents. The results for the direct vector 
form factors are consistent with those in~\cite{CARLSON}, and for the transition form factors they are
consistent with those in~\cite{BRAGA}. To our knowledge, the results for the direct and transition 
axial form factors are new. We use them for a holographic estimate of the axial coupling $g_A$
which is directly sensitive to the explicit and implicit breaking of chiral symmetry.

We have used the  transition form factors for the left-currents, to explicitly construct the s-channel contributions to
neutrino and anti-neutrino DIS scattering through the pertinent Witten diagram. This has led to the explicit identification
of the s-channel holographic contributions to the even- and odd-parity structure functions for both $W^\pm$ charged
currents. The t-channel contributions stemming from the Reggeized Pomeron exchange is explicitly constructed
and shown to dominate at low-x the even-parity structure functions. The  Reggeon excchange 
through the bulk Chern-Simons  interaction is shown to dominate the low-x odd-parity structure function.

For intermediate values of Bjorken-x, DIS scattering in holography is very different from QCD.
The leading twist-2 operator in the OPE of the JJ-correlators acquire large anomalous dimension,
and is dwarfed by the double trace operators which carry higher twists but are protected. In this
regime, DIS scattering is off a poin-like hadron which is similar to scattering for $x\sim 1$. When
$x\ll 1$, DIS scattering in QCD is dominated by Reggeon exchange and is dual to spin-2 or spin-1
Reggeized exchange in bulk. So lepton-nucleon scattering in holography allows to probe the partonic
content of the nucleon in the two regimes of $x\sim 1$ and $x\ll 1$. 

We have carried explicit calculations and comparison to the availabel data both for the structure
functions and PDFs from LHAPDF and CTEQ, in both of these regimes. Our results show consistency for large-x  for the
intermediate range of $Q^2<10$ GeV$^2$ where scaling violations are still substantial. At low-x
our results show a somehow larger growth at intermediate but low-x. The results at higher $Q^2$-resolution can be
obtained through standard QCD evolution.

\section{Acknowledgements}
K.M. is supported by the U.S.~Department of Energy, Office of Science, Office of Nuclear Physics, contract no.~DE-AC02-06CH11357, and an LDRD initiative at Argonne National Laboratory under Project~No.~2020-0020. I.Z. is supported by the U.S. Department of Energy under Contract No.~DE-FG-88ER40388.


\begin{figure*}
\subfloat[\label{xuvalenceatqq6pt5goinedNNpdfsetssmallx}]{%
  \includegraphics[height=6cm,width=.49\linewidth]{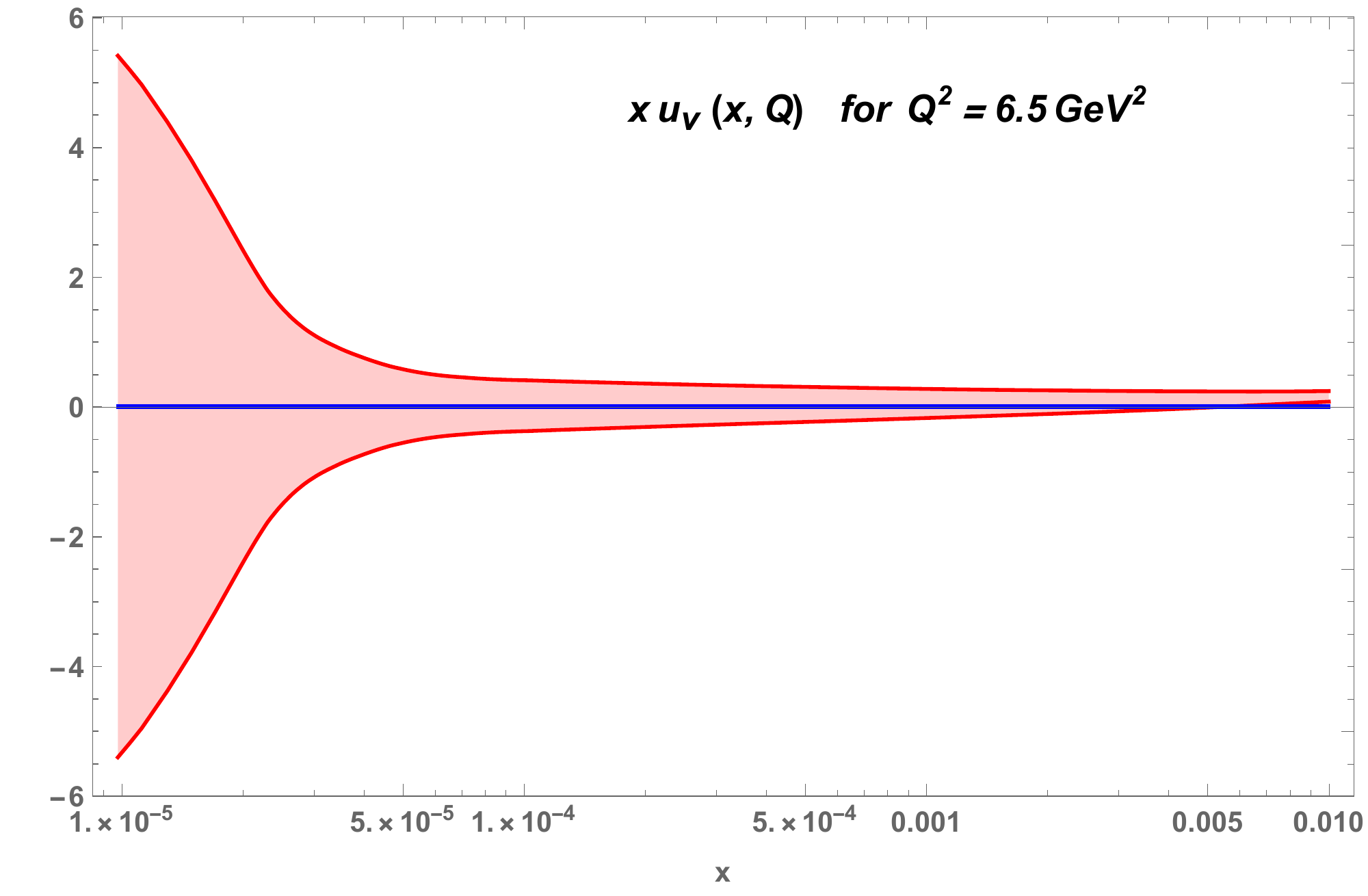}%
}\hfill
\subfloat[\label{xuvalenceatqq6pt5goinedNNpdfsetslargex}]{%
  \includegraphics[height=6cm,width=.49\linewidth]{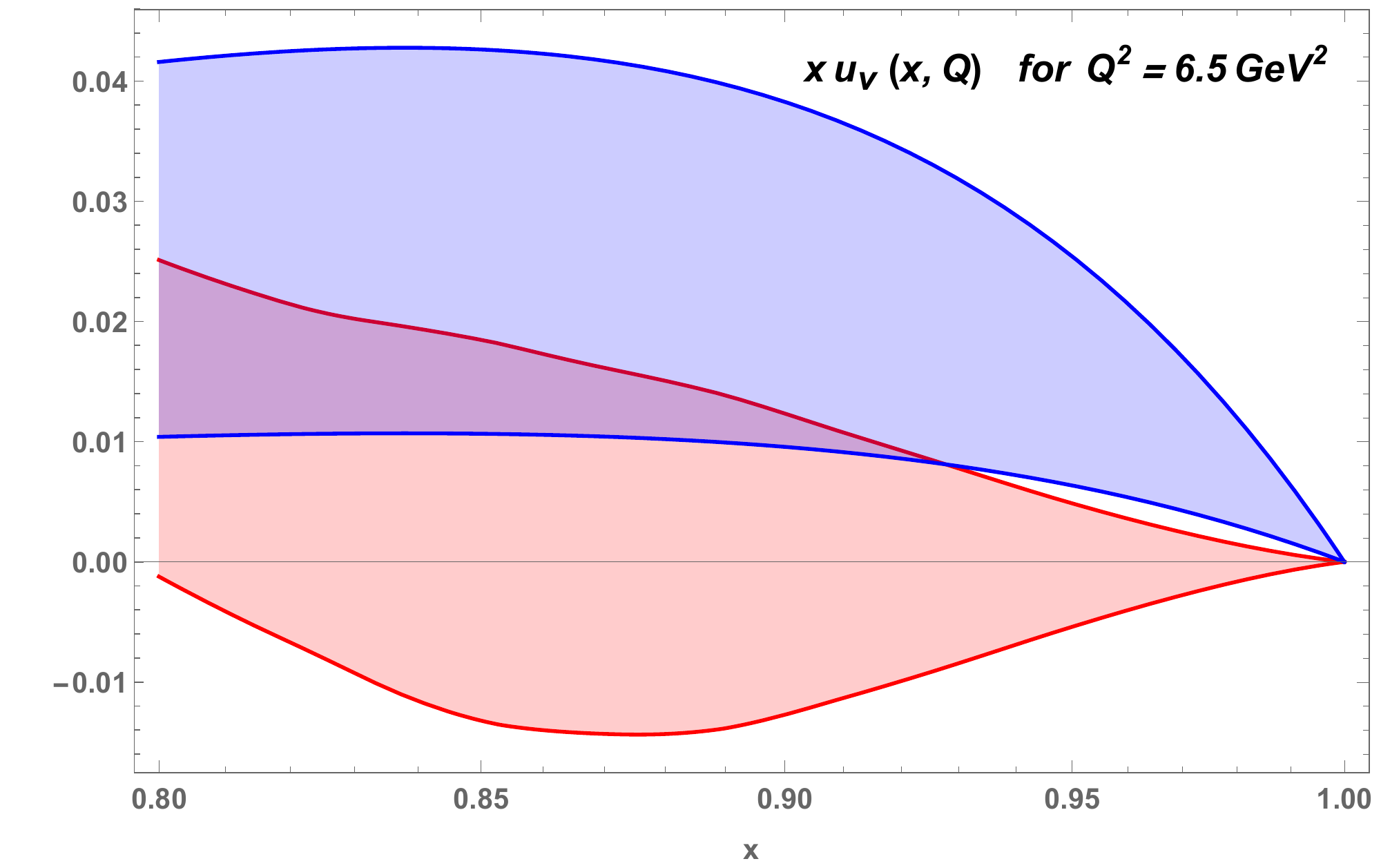}%
}\hfill
\subfloat[\label{xdvalenceatqq6pt5goinedNNpdfsetssmallx}]{%
  \includegraphics[height=6cm,width=.49\linewidth]{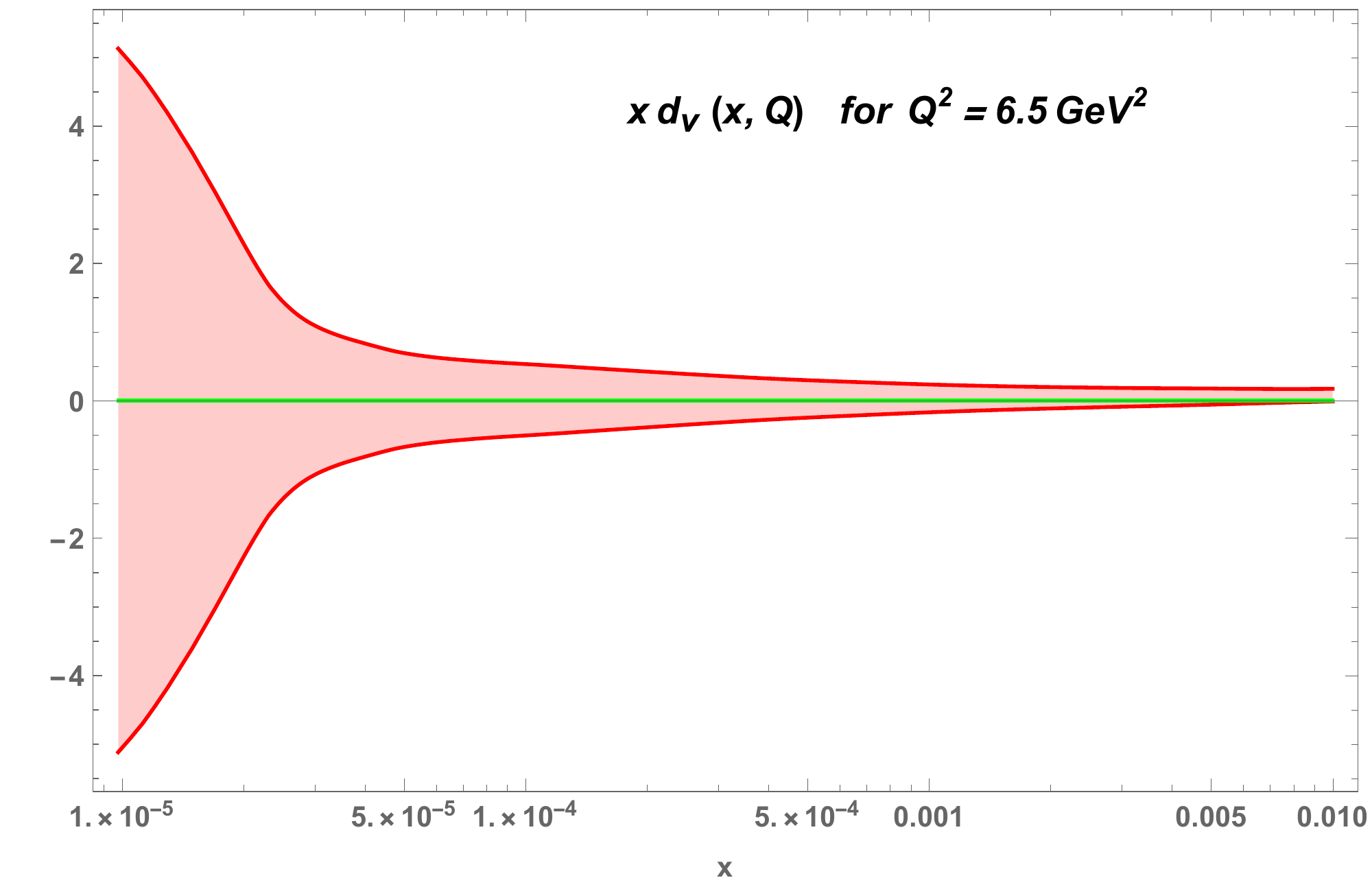}%
}\hfill
\subfloat[\label{xdvalenceatqq6pt5goinedNNpdfsetslargex}]{%
  \includegraphics[height=6cm,width=.49\linewidth]{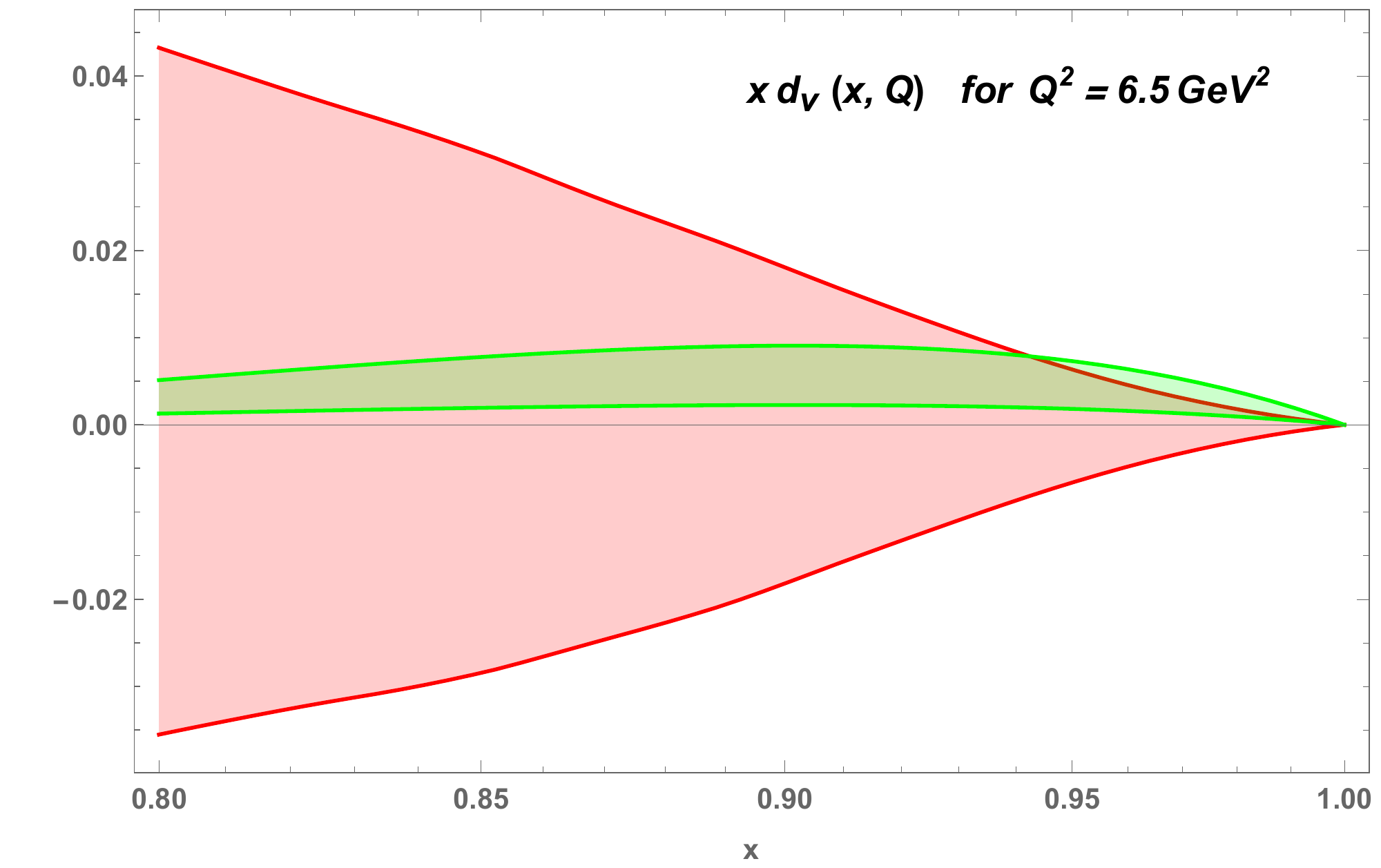}%
}
\caption{
(a,b) Our holographic PDF sets (shown in green band) compared to NNPDF collaboration PDF sets with error bars (shown in red band) from LHAPDF projects incorporated within the ManeParse Mathematica package \cite{Clark:2016jgm};
(c,d) Our holographic PDF sets (shown in blue band) compared to NNPDF collaboration PDF sets with error bars (shown in red band) from LHAPDF projects incorporated within the ManeParse Mathematica package \cite{Clark:2016jgm}.}
\label{NNPDF-VALENCE}
\end{figure*}




\clearpage

\appendix
\renewcommand{\thesection}{\Alph{section}.\arabic{section}}
\setcounter{section}{0}


\section{U$_V$(1) gauge fields  }

The U$_V$(1)  gauge fields solves the equations of motion in bulk

\begin{eqnarray}
\label{Solutionsgauge}
\Box  V^\mu  &+& z e^{\tilde\kappa^2 z^2}  \partial_z \Big( e^{-\tilde\kappa^2 z^2} \frac{1}{z}
\partial_z V^\mu \Big) \,=\, 0\nonumber\\
\Box V_z &-&  \partial_z \Big( \partial_\mu V^\mu \Big) \,=\, 0\,.
\end{eqnarray} 
subject to  the gauge condition

\begin{equation}
\label{gaugechoice}
 \partial_\mu V^\mu \,+\,
z e^{ \kappa^2 z^2} \partial_z \Big( e^{-\kappa^2 z^2} \frac{1}{z}
 V_z \Big) \,=\,0 \,,
\end{equation}
and the boundary condition

\begin{equation}
V_\mu (z, y) \vert_{z\to 0} \,=\, \epsilon_{\mu} (q)\, e^{-iq\cdot y} \,,
\end{equation}
with polarization $\epsilon_\mu(q)$. 
The normalizable solutions in KK modes are

\begin{eqnarray}
V_\mu (z, y) &=& \epsilon_\mu (q)\, e^{-iq\cdot y} \,\Gamma \bigg(1+\frac{Q^2}{4\tilde\kappa^2} \bigg)\,\,\tilde\kappa^2z^2 
\,\,{\cal U} \bigg(1+\frac{Q^2}{4 \tilde\kappa^2} ; 2 ; \tilde\kappa^2 z^2 \bigg)
\nonumber\\
V_z (z, y)  &=& \frac{i}{2} \,  \epsilon(q) \cdot q \,  e^{-iq\cdot y} \,\, \Gamma \bigg(1+\frac{Q^2}{4\tilde\kappa^2 } \bigg)\,\, z \,\, 
{\cal U} \bigg(1+\frac{Q^2}{4 \tilde\kappa^2} ; 1 ; \tilde\kappa^2 z^2 \bigg)\,,\nonumber\\ 
\label{Gauge}
\end{eqnarray} 
for $q^2=Q^2>0$, where $\,{\cal U} (a;b;w) \,$ are the confluent hypergeometric functions of the second kind.  

\subsection{Bulk to boundary propagator}

The non-normalizable wave function for the virtual photon is of the form
$V_{\mu}(z,y)=\mathcal{V}(Q,z)\,\epsilon_{\mu}e^{-iq\cdot y}$, with $q^2=-Q^2<0$ and

\be
\mathcal{V}(Q,z)=g_5\sum_n \frac{F_n\phi_n(z)}{Q^2+m_n^2}\,. \label{vps1}
\ee
For the soft wall

\be
\phi_n(z)=c_{n}\tilde{\kappa}^2z^2 L_n^1( \tilde{\kappa}^2z^2)\equiv J_{A}(m_n,z)  \,,
\ee
with $c_{n}=\sqrt{{2}/{n+1}}$ fixed by  the normalization condition 

\be
\int dz\,\sqrt{g}e^{-\phi}\,(g^{xx})^2\,\phi_n(z)\phi_m(z)=\delta_{nm}\,,\nonumber\\
\ee
Also recall that ${\cal V}(Q,z)=\frac{1}{z'}\partial_{z'}G(z,z')|_{z'=\epsilon}$, ties to  the bulk to boundary propagator with 

\bea
\mathcal{V}(Q,z)
=\tilde{\kappa}^2 z^2 \,\,\Gamma \bigg(1+\frac{Q^2}{4\tilde{\kappa}^2} \bigg)\,\,{\cal U} \bigg(1+\frac{Q^2}{4 \tilde{\kappa}^2} ; 2 ; \tilde{\kappa}^2 z^2 \bigg)
=\tilde{\kappa}^2z^2\int_{0}^{1}\frac{dx}{(1-x)^2}x^a{\rm exp}\Big(-\frac{x}{1-x}\tilde{\kappa}^2z^2\Big)\,,
\label{vps2sw}
\ee
and satisfies  ${\cal V}(0,z)={\cal V}(Q,0)=1$.

\subsection{Bulk to bulk propagator}

The bulk-to-bulk propagator for the massive mesons, for space-like momenta ($q^2=-Q^2$), can be written as

\be \label{bbmeson}
G_{\mu\nu}(z,z^\prime)=i{\cal T}_{\mu\nu}G(z,z^\prime)=i\bigg(-\eta_{\mu\nu} + \frac{q_{\mu} q_{\nu}}{m_n^2}\bigg)G(z, z^{\prime})=\sum_n J_{V}(m_n,z)\,\tilde G_{\mu\nu}(Q,m_n)\,J_{V}(m_n,z')\,,\nonumber\\
\ee
with
\be
G(z,z')=-\sum_n \frac{\phi_n(z)\phi_n(z')}{Q^2+m_n^2}\,, \label{vbbs1}
\ee
and
\be
J_{V}(m_n,z)=\phi_n(z)\,,\,\,\,\,\,\,\,\,\,\,\,\tilde G_{\mu\nu}(Q,m_n)=\bigg(-\eta_{\mu\nu} + \frac{q_{\mu} q_{\nu}}{m_n^2}\bigg)\frac{-i}{Q^2+m_n^2}\,.
\ee

Note that  the bulk to boundary propagator follows by taking $z\rightarrow 0$,  which simplifies (\ref{vbbs1}) as

\bea
\label{G0z}
G(z\rightarrow 0,z')\approx\bigg[ -\frac{\phi_n(z\rightarrow 0)}{-g_5F_n}\bigg]\sum_n \frac{-g_5F_n\phi_n(z')}{Q^2+m_n^2}
=\frac{z^2}{2}\sum_n \frac{g_5F_n\phi_n(z')}{Q^2+m_n^2}=\frac{z^2}{2}\mathcal{V}(Q,z'), \label{vbbs2}
\eea
where we  used the fact that

\be
F_n=\frac 1{g_5}\bigg(-e^{-\phi}\frac{1}{z^\prime}\partial_{z^\prime}\phi_n(z^\prime)\bigg)_{z^\prime=\epsilon}=-\frac{2}{g_5}c_n(n+1)\tilde{\kappa}^2\,,\nonumber\\
\ee
with $\phi_n(z\rightarrow 0)\approx c_n\tilde{\kappa}^2z^2(n+1)$. Note that the ratio in the bracket in (\ref{G0z}) is n-independent!
If we define the meson decay constant as $f_n=-{F_n}/{m_n}$,  we have

\be
\phi_n(z)=\frac{f_n}{m_n}\times 2g_{5}\tilde{\kappa}^2z^2 L_n^1( \tilde{\kappa}^2z^2)\,,
\ee
in line with vector meson dominance (VMD).



\section{Fermionic fields}~\label{FERMIONFIELDS}
\renewcommand{\thesubsection}{\Alph{subsection}}

We start by considering a  general fermionic field in bulk AdS that solves the free Dirac equation in the presence of a soft wall. For that, we remove the dilaton field by rescaling

\begin{equation} 
\Psi (y,z) = e^{ + \tilde{\kappa}^2 z^2/2} \tilde\psi(y,z),
\end{equation} 
with the reduced field solving

\begin{equation}
\label{EOMII}
\biggl[ i\not\!\partial + \gamma^5\partial_z 
- \frac{2}{z} \gamma^5 
-  \frac{1}{z} \Big(M +  \tilde{\kappa}^2 z^2 \Big)  \biggr] \tilde\psi(y,z) = 0 \,, 
\end{equation} 
 where $ \not\!\partial = \gamma^\mu \, \partial_\mu$. 
We now decompose the reduced field into two chiral copies 

\begin{equation}
\tilde\psi(y,z) = \tilde\psi_L(y,z) + \tilde\psi_R(y,z)\,, \quad 
\tilde\psi_{L/R} = P_{L/R}\tilde\psi \,. 
\end{equation} 
 with $(\gamma^5)^2 = 1 $, and $P_{L/R}=P_\mp=\frac{1 \mp \gamma^5}{2} $. The solution to (\ref{EOMII}) follows by further reduction into $L/R$ KK  modes using

\begin{equation}
\tilde\psi_{L/R}(y,z) = e^{-iP\cdot y } \frac12 \Big( 1 \mp \gamma^5 \Big) u_s(P) \, \frac{z^2}{R^2}
\tilde f_{L/R}(z) \,, 
\end{equation} 
with $u_s(P)$ a free Dirac spinor in 4-dimensions, and the $L/R$ KK modes $\tilde f_{L/R}(z)$ now satisfying

\begin{equation}
\label{EqzLR}
\biggl[ -\partial_z^2 
+ \tilde{\kappa}^4 z^2 + 2 \tilde{\kappa}^2 \Big(M \mp \frac{1}{2} \Big) 
+ \frac{M (M\pm 1)}{z^2} \biggr] \tilde f_{L/R}(z) = P^2  \, \tilde f_{L/R}(z) \,.
\end{equation}
\end{widetext}

\subsection{Spectrum and modes}

The equation of motion (\ref{EqzLR}) has normalizable solutions only when $P^2$ has

\begin{equation} 
\label{Masses}
P^2_n = M_n^2 = 4 \tilde{\kappa}^2 \Big( n + M+ \frac{1}{2} \Big) \,, 
\end{equation}  

\noindent with $ n= 0,1,2,..., $, which are

\begin{widetext}
\begin{eqnarray} 
\tilde f^n_{L}(z) &=& \sqrt{\frac{2\Gamma(n+1)}{\Gamma(n+M+3/2)}} \ \tilde{\kappa}^{M+3/2}
\ z^{M+1} \ e^{-\tilde{\kappa}^2 z^2/2} \ L_n^{M+1/2}(\tilde{\kappa}^2z^2) \,, \nonumber\\
\tilde f^n_{R}(z) &=& \sqrt{\frac{2\Gamma(n+1)}{\Gamma(n+M+1/2)}} \ \tilde{\kappa}^{M+1/2}
\ z^{M} \ e^{-\tilde{\kappa}^2 z^2/2} \ L_n^{M-1/2}(\tilde{\kappa}^2z^2)\nonumber\\
\end{eqnarray}
\end{widetext} 
 with the normalization condition 
 
 \be
 \int\limits_0^\infty dz \, \tilde f^{n^\prime}_{L/R}(z) \tilde f^n_{L/R}(z)=\delta_{n^{\prime}n}
 \ee  
We can rewrite the normalized wavefunctions for the bulk Dirac fermion (or proton and neutron which correspond to the $n=0$ states) as

\bea
\Psi(p,z)&=&\psi_R(z)\Psi^0_{R}(p)+ \psi_L(z)\Psi^0_{L}(p)\,,\nonumber\\
\bar\Psi(p,z)&=&\psi_R(z)\bar\Psi^0_{R}(p)+ \psi_L(z)\bar\Psi^0_{L}(p)\,,\nonumber\\
\eea
where for the {soft-wall}

\be
&&\psi_R(z)=\frac{\tilde{n}_R}{\tilde{\kappa}^{\tau-2}} z^{\frac{5}{2}}\xi^{\frac{\tau-2}{2}}L_0^{(\tau-2)}(\xi)\,,\nonumber\\
&&\psi_L(z)=\frac{\tilde{n}_L}{\tilde{\kappa}^{\tau-1}} z^{\frac{5}{2}}\xi^{\frac{\tau-1}{2}}L_0^{(\tau-1)}(\xi)\,,\nonumber\\
\ee
with $\xi=\tilde\kappa^2z^2$ and  the generalized Laguerre polynomials $L_n^{(\alpha)}(\xi)$, $\tilde{n}_R=\tilde{n}_L \tilde{\kappa}^{-1}\sqrt{\tau-1}$, and $\tilde{n}_L=\tilde{\kappa}^{\tau}\sqrt{{2}/{\Gamma(\tau)}}$. The normalization of the bulk wave functions can also be rewritten as

\be
\int_{0}^{\infty} dz\,\sqrt{g}\,e^{-\phi}\,e^{\mu}_{a}\,\psi_{R/L}^2(z)=\delta^{\mu}_a\,,\nonumber\\
\ee
with $\phi=\tilde{\kappa}^2z^2$, and the inverse vielbein $e^{\mu}_{a}=\sqrt{\abs{g^{\mu\mu}}}\delta^{\mu}_a$ (no summation intended in $\mu$).
The leading twist parameter is $\tau=3$, and the free Weyl spinors $\Psi^0_{R/L}(p)= P_{\pm}u(p)$ and $\bar\Psi^0_{R/L}(p)=\bar u(p)P_{\mp}$ are tied to the free spinor normalization at the boundary

\begin{widetext}

\be
\bar u(p)u(p)=2m_N\qquad\qquad 
2m_N\times\bar u(p')\gamma^{\mu}u(p)=\bar u(p')(p'+p)^{\mu}u(p)\,.
\ee

\subsection{Bulk to boundary propagator}
\renewcommand{\thesubsection}{\Alph{subsection}}

The bulk to boundary propagator (or the nonrenormalizable mode) of the bulk Dirac fermion is given by

\be \label{bbf}
\Psi(p,z)=\sum_n\frac{\sqrt{2g_5^2}F_n^R(p)\psi^R_n(z)}{p^2-M_n^2}\times\Psi^0_{R}(p)+\sum_n\frac{\sqrt{2g_5^2}F_n^L(p)\psi^L_n(z)}{p^2-M_n^2}\times\Psi^0_{L}(p)\,,
\ee
where the fermionic decay functions are defined as 

\bea \label{fermiondc}
F_n^R(p)&=&\frac{1}{\sqrt{2g_5^2}}\times p\,\tilde\kappa\sqrt{\frac{2\Gamma(\tau-1+n)}{\Gamma(\tau-1)\Gamma(n+1)}}\,,\nonumber\\ 
F_n^L(p)&=&\frac{1}{\sqrt{2g_5^2}}\times M_n\tilde\kappa\sqrt{\frac{2\Gamma(\tau-1+n)}{\Gamma(\tau-1)\Gamma(n+1)}}\,.\nonumber\\
\eea 
Note that the on-shell bulk Dirac fermions are just the residues at the poles $p^2-m_n^2$  in (\ref{bbf}). The in-out baryonic states  used in our DIS analysis
throughout are

\begin{eqnarray}
\label{SolutionFermions}
\Psi_i &=& e^{-ip\cdot y } e^{ + \tilde{\kappa}^2 z^2/2} \, \frac{z^2}{R^2} \Big[ \Big(\frac{1 - \gamma^5}{2} \Big) 
u_{s_{i}}(p) \, 
\tilde f^0_{L}(z) \,+\, \Big( \frac{1 + \gamma^5}{2} \Big) u_{s_{i}} (p) \,\tilde f^0_{R}(z) \Big]\times\sqrt{2g_5^2} \times F_{N}(p) \cr \cr
\Psi_X &=& e^{-iP_X \cdot y } e^{ + \tilde{\kappa}^2 z^2/2} \, \frac{z^2}{R^2} \Big[ \Big(\frac{1 - \gamma^5}{2} \Big) 
u_{s_{X}}(P_X) \, \tilde f^{n_X}_{L}(z) \,+\, \Big( \frac{1 + \gamma^5}{2} \Big) u_{s_{X}} (P_X) \,\tilde f^{n_X}_{R}(z) \Big]\times\sqrt{2g_5^2} \times F_{X}(P_X) \,\,,
\cr
& &
\end{eqnarray}
where $s_i $ and $s_X$ label they in-out spin, and identified $F_{0}^R=F_{0}^L=F_N(p)$ (ground state) and $F_{n}^R=F_{n}^L=F_X(P_X)$ (excited state).


\subsection{Yukawa coupling through the tachyon}

The inclusion of a bi-fundamental tachyon field $X(x,z)$ in bulk to lift the degeneracy between the vector and axial-vector mesons, would also imply
a Yukawa coupling between the even-odd  bulk fermionic fields $\Psi_{1,2}$ which we have not considered here. More specifically~\cite{YEE}

\be
\label{YUKAWA}
{\cal S}_{12,X}=\frac{1}{2g_5^2}\frac{g_X}{2}\int d^5x\sqrt{g}\,\left(\overline\Psi_1(x,z)X(x,z)\Psi_2(x,z)+\overline\Psi_2(x,z)X^\dagger(x,z) \Psi_1(x,z)\right)
\ee
which would mix $1,2$ and lifts the degeneracy of the low-lying  even and odd parity states in the nucleon sector, $(\Psi_1\pm \Psi_2)/\sqrt{2}$. A  first order estimate
in perturbation theory gives  for the nucleon ground state with $n=0$

\be
\Delta M_\pm=\mp\frac {g_X}{4}\int \frac {dz}{z}X_0(z)\left(\left|\tilde{f}_L^0(z)\right|^2-\left|\tilde{f}_R^0(z)\right|^2\right)
=\mp \frac{g_X\sigma}{8\tilde\kappa^2}
\ee
with $X_0(z)$ given in (\ref{X0Z}).
(\ref{YUKAWA}) through
the expansion around the vev, $X(x,z)\approx X_0(z)e^{i\Pi(x,z)}$, would also
generate a contribution to the pion-nucleon coupling  and also the axial-charge of the direct and transition axial form factors~\cite{YEE}. 
Since our central interest is neutrino DIS scattering we can neglect this coupling and its effects on our results, as most of our analysis
involves the behavior near the UV boundary where the effects of $\sigma$ is negligible both in the nucleonic wavefunctions and the spectrum.

\section{Details of the Reggeon exchange }\label{DRE}

The bulk gauge field $L^0_{\mu}(k,z)$ exchange contribution to the diffractive Compton scattering amplitude in the t-channel is given by 

\be
\label{AmpL}
&&i{\cal A}^{L}_{Lp\rightarrow  Lp} (s,t)=\sum_n i\tilde{{\cal A}}^{L}_{Lp\rightarrow L p} (m_n,s,t)\nonumber\\
&&i\tilde{{\cal A}}^{L}_{Lp\rightarrow L p} (m_n,s,t)=(-i)V_{LLL}^{\mu}(q,q',k,m_n)\times \tilde{G}_{\mu\nu}(m_n,t)\times
 (-i)V_{L\bar\Psi\Psi}^{\nu}(p_1,p_2,k,m_n)\,,\nonumber\\
\ee
with the bulk vertices ($k=p_2-p_1=q-q'$)

\bear
V_{LLL}^{\mu}(q,q',k,m_n)\equiv&&\left(\frac{\delta S_{LLL}^k}{\delta (\epsilon^{0}_{\mu}\partial_z L^0(k,z))}\right)\,J_{L}(m_n,z)+\left(\frac{\delta S_{LLL}^k}{\delta (\epsilon^{0}_{\mu}L^0(k,z))}\right)\,J_{L}(m_n,z)\nonumber\\
=&&g_5^3\kappa_{CS}B^{\mu}(q,q',\epsilon^{\pm})\int dz\,\Big(\mathcal{V}(Q,z)\mathcal{V}(Q',z)\partial_z J_{L}(m_n,z)-\partial_z \mathcal{V}(Q,z)\mathcal{V}(Q',z)J_{L}(m_n,z)\Big)\,,\nonumber\\
V_{L\bar\Psi\Psi}^{\nu}(p_1,p_2,k,m_n)\equiv &&\left(\frac{\delta S_{L\bar\Psi\Psi}^k}{\delta (\epsilon^{0}_{\nu}L^0(k,z))}\right)\,J_{L}(m_n,z)=g_5\int dz\sqrt{g}\,e^{-\phi}z\bar\Psi(p_2,z)\gamma^\nu\Psi(p_1,z)J_{L}(m_n,z)\,,\nonumber\\ \label{vL}
\eear
We have defined  $p=({p_1+p_2})/{2}$, $t=-K^2$, $\mathcal{V}(Q,z)\equiv L(q=\sqrt{-Q^2},z)$ as given  in  (\ref{vps2sw}), and used the bulk-to-bulk gauge field propagator (\ref{bbmeson}) with the  substitutions  $q\rightarrow k $, $Q\rightarrow K$, and $V\rightarrow L$. We have also used the vertices in (\ref{vertices33}), and defined

\be
B^{\mu}(q,q',\epsilon^{\pm})\equiv (-i)\epsilon^{\rho\sigma\nu\mu}\epsilon_{\rho}^+(q)\epsilon_{\sigma}^-(q')(q'_{\nu}+q_{\nu})\,.\nonumber\\
\ee
For $z'\rightarrow 0$, we can use (\ref{vbbs2}) and simplify  (\ref{AmpL}) as

\be
\label{nAmph}
&&i{\cal A}^{L}_{Lp\rightarrow L p} (s,t)\approx(-i)\mathcal{V}^{\mu}_{LLL}(q_1,q_2,k_z)\times \big(-i\eta_{\mu\nu}\big)\times(-i)\mathcal{V}^{\nu}_{L\bar\Psi\Psi}(p_1,p_2,k_z)\,,	\nonumber\\
\ee
with
\be
&&\mathcal{V}^{\mu}_{LLL}(q_1,q_2,k_z)=g_5^3\kappa_{CS}B^{\mu}(q,q',\epsilon^{\pm})\int dz\,\Big(\mathcal{V}(Q,z)\mathcal{V}(Q',z)z-\partial_z \mathcal{V}(Q,z)\mathcal{V}(Q',z)\frac{z^2}{2}\Big)\,,\nonumber\\
&&\mathcal{V}^{\nu}_{L\bar\Psi\Psi}(p_1,p_2,k_z)=g_5\times\frac{3}{2}\times\int dz\,\sqrt{g}\,e^{-\phi}z\,\bar\Psi(p_2,z)\gamma^\nu\Psi(p_1,z)\mathcal{V}(K,z)=g_5F_1^{(LN)}(K)\,,
\nonumber\\
\ee
where $\mathcal{V}(K,z)\equiv L^0(k=\sqrt{-K^2},z)$, and $F_1^{(LN)}(K)$ is the form factor of the the nucleon due to $L_{\mu}^0$.

The Reggeization of the bulk spin-1 gauge field $L_{\mu}^0(k,z)$ exchange can be obtained, in a similar way
to the Reggezation of the spin-2 graviton exchange, through the substitution

\be \label{spinjwf}
J_L(m_n(j),z) \rightarrow z^{-(j-1)}\phi_n(j,z)=z^{-(j-1)}\frac{\tilde\phi_n(j,z)}{z}
\ee
followed by the summation over all spin-j meson exchanges using the Sommerfeld-Watson formula

\be
\label{SPIN}
\frac 12\sum_{j\geq 1}(s^{j-1}+(-s)^{j-1})\rightarrow  -\frac {\pi} 2\int_{\mathbb C}\frac{dj}{2\pi i}\left(\frac{s^{j-1}+(-s)^{j-1}}{{\rm sin}\,\pi j}\right)\nonumber\\
\ee
The contour ${\mathbb C}$ is to the left of all odd poles $j=1,3, ...$ (in contrast to the Reggeized graviton where the contour is chosen to the left of 
the even poles),  and requires the analytical continuation of the
exchanged amplitudes to the complex j-plane. 

The spin-j  normalized meson wavefunctions $J_L(m_n(j),z)$ (\ref{spinjwf}) are expressed in terms of the wavefunctions of massive scalar 
fields $\tilde\phi_n(j,z)$ which are given, for the soft wall model, in terms of the generalized Laguerre polynomials as

\be
 \label{wfSW}
&&\tilde\phi_{n}(j,z)=c_n(j)\,z^{\Delta(j)}L_{n}^{\Delta(j)-2}(w)\,,
\ee
with $w=\tilde{\kappa}^2z^2$.  The normalization coefficients are

\be
c_n(j)=\Big(\frac{2\tilde{\kappa}^{2(\Delta(j)-1)}\Gamma(n+1)}{\Gamma(n+\Delta(j)-1)}\Big)^{\frac 12}\,,
\ee
and the dimension of the massive scalar fields (with an additional mass coming from the massive open string states attached to the  D9 or D7-branes) $\Delta(j)$ is given by
\bea
\Delta(j)&=&2+\sqrt{4+m^2R^2+\frac{R^2}{\alpha'}(j-1)}\nonumber\\
&=&2+\sqrt{\sqrt{\lambda}(j-j_0)}\,,
\eea
where, in the last line, we have used the fact that $m^2R^2=-3$.
The spin-1 transverse bulk gauge field defined as $zL_\mu^0(m_n,z)$ obeys  the same bulk equation of motion as a bulk massive scalar field $\tilde\phi_n(j=1,z)$ with $m^2R^2=-3$ which is manifest in (\ref{XP5}). We have also used the open string quantized mass spectrum $m_j^2R^2=(j-1)({R^2}/{\alpha^\prime})=\sqrt{\lambda}(j-1)$ for open strings attached to the D9 or D7-branes in bulk, and we have defined $j_0=1-{1}/{\sqrt{\lambda}}$.

We now recall that the non-normalized bulk-to-boundary propagators of massive scalar fields are given in terms of Kummer's (confluent hypergeometric) function of the second kind,  and their integral representations  are (for space-like momenta $k^2=-K^2$)

\bea
\mathcal{\tilde V}(j,K,z)
=&&z^{\Delta(j)}U\Big(a_K+\frac{\Delta(j)}{2},\Delta(j)-1;w\Big)
=z^{\Delta(j)}w^{2-\Delta(j)}U\Big(\tilde{a}(j),\tilde{b}(j);w\Big)\nonumber\\
=&&z^{\Delta(j)}w^{2-\Delta(j)}\frac{1}{\Gamma(\tilde{a}(j))}
\int_{0}^{1}dx\,x^{\tilde{a}(j)-1}(1-x)^{-\tilde{b}(j)}{\rm exp}\Big(-\frac{x}{1-x}w\Big)\,,
\label{BBSWj}
\eea
with $w=\tilde{\kappa}^2z^2$
\be
a_K=a=\frac{K^2}{4\tilde{\kappa}^2}\qquad
\tilde{a}(j)=a_K+2-\frac{\Delta(j)}{2}\qquad
\tilde{b}(j)=3-\Delta(j)
\ee
after using the identity  $U(m,n;y)=y^{1-n}U(1+m-n,2-n,y)$.
Therefore, the bulk-to-bulk propagator of spin-j mesons 

\be
J_L(m_n(j),z)\,G(j,z,z')\,J_{L}(m_n(j),z')=z^{-(j-1)}\,G(j,z,z')\,z'^{-(j-1)}\nonumber\\ 
\ee
can be approximated at the boundary as (for space-like momenta $k^2=-K^2$)

\be
 G(j,z\rightarrow 0,z')\approx &&-\bigg[\frac{\phi_{n}(j,z\rightarrow 0)}{-g_5\mathcal{F}_n(j)}\bigg]\times\sum_n \frac{-g_5\mathcal{F}_n(j)\phi_n(j,z')}{K^2+m_n^2(j)}\nonumber\\
 =&&(\tilde\kappa^2)^{\Delta(j)-2}\frac{z^{\Delta(j)-1}}{\Delta(j)-1}\frac{\Gamma(\Delta(j)-2+a)}{\Gamma(\Delta(j)-2)}\,\mathcal{V}(j,K,z')\nonumber\\
\ee
where $\phi_n(j,z\rightarrow 0)=\frac{1}{z}\tilde\phi_n(j,z\rightarrow 0)$. We have defined the non-normalized bulk-to-boundary propagator of spin-j mesons
\be
\mathcal{V}(j,K,z')=\sum_n \frac{-g_5\mathcal{F}_n(j)\phi_n(j,z')}{K^2+m_n^2(j)}=\frac{1}{z}\mathcal{\tilde V}(j,K,z')\vert_{a+\frac{\Delta(j)}{2}\rightarrow a+1+(\Delta(j)-3)}\,,\nonumber\\
\ee
with the shift $a+\frac{\Delta(j)}{2}\rightarrow a+1+(\Delta(j)-3)$ defined in such a way that the mass spectrum of massive scalar fields $m_n^2=4\tilde\kappa^2(n+\frac{\Delta(j=1)}{2})$ and the mass spectrum of spin-1 gauge fields $m_n^2=4\tilde\kappa^2(n+1)$ match, i.e., we shift $n+\frac{\Delta(j=1)}{2}\rightarrow n+1+(\Delta(j=1)-3)$, giving the mass spectrum of spin-j mesons
\be
m_n^2(j)=4\tilde\kappa^2(n+1+(\Delta(j)-3))\,.
\ee
We have also used

\bea 
\mathcal{F}_n(j)=&&\frac{\mathcal{C}(j,K,\epsilon)}{g_5}\bigg(-\sqrt{g}\,e^{-\phi}\,\big(g^{xx}\big)^2\,\partial_{z'}\phi_n(j,z^\prime)\bigg)_{z^\prime=\epsilon}\,,\nonumber\\
\mathcal{C}(j,K,\epsilon)=&&\mathcal{V}(j,K,\epsilon)
\eea
and the substitution $\phi_n(j,z\rightarrow 0)=\frac{1}{z}\tilde\phi_n(j,z\rightarrow 0)\approx c_n(j)\,z^{\Delta(j)-1}L_{n}^{\Delta(j)-2}(0)$ for the soft wall model.

After the Reggeization, the scattering amplitude for the spin-j meson exchange becomes
\be
\label{nAmph}
&&i{\cal A}^{L}_{Lp\rightarrow L p} (j,s,t)\approx(-i)\mathcal{V}^{\mu}_{LLL}(j,q_1,q_2,k_z)\times \big(-i\eta_{\mu\nu}\big)\times(-i)\mathcal{V}^{\nu}_{L\bar\Psi\Psi}(j,p_1,p_2,k_z)\,,	\nonumber\\
\ee
with

\bea \label{pvertices}
\mathcal{V}^{\mu}_{LLL}(j,q_1,q_2,k_z)=&&\frac{1}{g_5^2}\times g_5^3\kappa_{CS}B^{\mu}(q,q',\epsilon^{\pm})\int dz\,z^{2(j-1)}\nonumber\\
&&\times \bigg(\mathcal{V}(Q,z)\mathcal{V}(Q',z)\times \frac{(\Delta(j)-1-(j-1))z^{\Delta(j)-1-(j-1)-1}}{\Delta(j)-1}\nonumber\\
&&-\partial_z \mathcal{V}(Q,z)\mathcal{V}(Q',z)\times\frac{z^{\Delta(j)-1-(j-1)}}{\Delta(j)-1}\bigg)
\times (\tilde\kappa^2)^{\Delta(j)-2}\frac{\Gamma(\Delta(j)-2+a)}{\Gamma(\Delta(j)-2)}\nonumber\\
=&&\mathcal{V}_{LLL}(j,Q,Q')\times B^{\mu}(q,q',\epsilon^{\pm})\,,\nonumber\\
\mathcal{V}^{\nu}_{L\bar\Psi\Psi}(p_1,p_2,k_z)=&&g_5\times\frac{3}{2}\times\int dz\,\sqrt{g}\,e^{-\phi}z^{1+2(j-1)}\,\bar\Psi(p_2,z)\gamma^\nu\Psi(p_1,z)z^{-(j-1)}\mathcal{V}(j,K,z)\nonumber\\
=&&g_5F_1^{(LN)}(j,K)\times\bar u(p_2)\gamma^\nu u(p_1)\,,
\nonumber\\
\ee
We have defined 

\bea
\mathcal{V}_{LLL}(j,Q,Q')=&&\frac{1}{g_5^2}\times  g_5^3\kappa_{CS}\int dz\,z^{2(j-1)}\nonumber\\
&&\times \bigg(\mathcal{V}(Q,z)\mathcal{V}(Q',z)\times \frac{(\Delta(j)-1-(j-1))z^{\Delta(j)-1-(j-1)-1}}{\Delta(j)-1}\nonumber\\
&&-\partial_z \mathcal{V}(Q,z)\mathcal{V}(Q',z)\times\frac{z^{\Delta(j)-1-(j-1)}}{\Delta(j)-1}\bigg)(\tilde\kappa^2)^{\Delta(j)-2}\frac{\Gamma(\Delta(j)-2+a)}{\Gamma(\Delta(j)-2)}\nonumber\\
=&&\frac{1}{g_5^2}\times g_5^3\kappa_{CS}\times Q^{2-j-\Delta(j)}\times I_{\xi}(j,Q,Q')\times\tilde\kappa^{2\Delta(j)-4}\times\frac{\Gamma(\Delta(j)-2+a)}{\Gamma(\Delta(j)-2)}\times\frac{1}{\Delta(j)-1}\nonumber\\
\eea
with
\bea
I_{\xi}(j,Q,Q')&=&\int d\xi\,\xi^{j-2+\Delta(j)}\bigg(\mathcal{V}(\xi)\mathcal{V}(\xi Q'/Q)\times (\Delta(j)-1-(j-1))\xi^{-1}-\partial_\xi \mathcal{V}(\xi)\mathcal{V}(\xi Q'/Q)\bigg)\nonumber\\
&\approx & (\Delta(j)-1-(j-1)) \times 2^{\Delta(j)+j-3}\,\frac{Q'}{Q} \,G_{2,2}^{2,2}\left(\frac{Q^2}{Q'^2}\bigg |
\begin{array}{c}
 \frac{1}{2},\frac{3}{2} \\
 \frac{1}{2} (j+\Delta(j)-1),\frac{1}{2} (j+\Delta(j) +1) \\
\end{array}
\right)\nonumber\\
&+& 2^{\Delta(j)+j-2}\,\frac{Q'}{Q} \,G_{2,2}^{2,2}\left(\frac{Q^2}{Q'^2}\bigg |
\begin{array}{c}
 \frac{1}{2},\frac{3}{2} \\
 \frac{1}{2} (j+\Delta(j)+1),\frac{1}{2} (j+\Delta(j) +1) \\
\end{array}
\right)\,,\nonumber\\
\eea
where $G_{p,q}^{m,n}\left(z\bigg |
\begin{array}{c}
 a_1,...,a_p \\
 b_1,...,b_q \\
\end{array}
\right)$ is the Meijer G-function. We  have used the identities

\bea
\label{LIM}
\lim_{\frac{Q}{\tilde{\kappa}}\rightarrow\infty}\mathcal{V}(\xi)=\xi K_1(\xi)\qquad {\rm and}\qquad \partial_\xi(\xi^\nu K_\nu(\xi))=-\xi^\nu K_{\nu-1}(\xi)
\eea
 to evaluate the integrals with $\xi=Qz$. The function $F_1^{(LN)}(j,K)$  in (\ref{pvertices}) admits the integral representation
 
\bea
\label{Fj1}
F_1^{(LN)}(j,K)&=&\frac{3}{2}\times\frac{1}{2}\frac{\tilde{\kappa}^{-(j-1)-\Delta(j)-1}}{\Gamma(a)}\int_{0}^{1}dx\,x^{a-1}(1-x)^{-\tilde{b}(j)}\nonumber\\&&\times\bigg(\bigg(\frac{\tilde{n}_R}{\tilde{\kappa}^{\tau-1}}\bigg)^2\times\Gamma(c(j))\bigg(\frac{1}{1-x}\bigg)^{-c(j)}
+\bigg(\frac{\tilde{n}_L}{\tilde{\kappa}^{\tau}}\bigg)^2\times\Gamma(c(j)+1)\bigg(\frac{1}{1-x}\bigg)^{-(c(j)+1)}\bigg)\,,\nonumber\\
&=&\frac{3}{2}\times\frac{1}{2}\tilde{\kappa}^{-(j-1)-\Delta(j)-1}\nonumber\\&&\times\bigg(\bigg(\frac{\tilde{n}_R}{\tilde{\kappa}^{\tau-1}}\bigg)^2\times\frac{\Gamma(c(j))\Gamma(1-\tilde{b}(j)+c(j))}{\Gamma(1-\tilde{b}(j)+c(j)+a)}
+\bigg(\frac{\tilde{n}_L}{\tilde{\kappa}^{\tau}}\bigg)^2\times\frac{\Gamma(c(j)+1)\Gamma(2-\tilde{b}(j)+c(j))}{\Gamma(2-\tilde{b}(j)+c(j)+a)}\bigg)\nonumber\\
\eea
where

\be
\tilde{b}(j)=3-\Delta(j)\,,\qquad\qquad
c(j)=(\tau+1)+\frac{j-1}{2}-\frac{\Delta(j)}{2}-\frac{1}{2}\,.
\ee

After summing over all contributions from the spin-j mesons, the total amplitude ${\cal A}^{tot}_{L p\rightarrow L p} (s,t)$ is given by

\bea\label{reggeon}
{\cal A}^{tot}_{L p\rightarrow L p} (s,t)=&&-\int_{\mathbb C}\frac{dj}{2\pi i}
\left(\frac{s^{j-1}+(-s)^{j-1}}{{\rm sin}\,\pi j}\right){\cal A}^{L}_{L p\rightarrow L p} (j,s,t)\nonumber\\
{\cal A}^{L}_{L p\rightarrow L p} (j,s,t)=&&\mathcal{V}_{LLL}(j,Q,Q')\times B^{\mu}(q,q',\epsilon^{\pm})\times g_5 F_1^{(LN)}(j,K)\times\bar u(p_2)\gamma_\mu u(p_1)\,,
\eea
The contour $\mathbb C$ is at the rightmost of the branch-point of $F_1^{LN}(j,K)$ and the leftmost of $j=1,3, ...$.
From (\ref{reggeon}), we determine the single Reggeon amplitude (total amplitude) in momentum space,
after wrapping the j-plane contour ${\mathbb C}$ to the left,

\bea\label{pomeron}
{\cal A}^{tot}_{L  p\rightarrow L p} (s,t)=-s^{j_{0}-1}\int_{-\infty}^{j_0}\frac{dj}{\pi}
\left(\frac{1 +e^{-i\pi}}{{\rm sin}\,\pi j}\right)s^{j-j_{0}}\,\text{Im}[{\cal A}^{L}_{L p\rightarrow L p} (j,s,t)]
\eea
The imaginary part follows from the discontinuity of the $\Gamma$-function

\bea
\label{disc}
\text{Im}[{\cal A}^L_{L p\rightarrow L p} (j,s,t)]&\approx & 
\bigg(\Gamma(\Delta(j)-2)\mathcal{V}_{LLL}(j,Q,Q')\times B^{\mu}(q,q',\epsilon^{\pm})
\times g_5F_{1}^{(LN)}(j,K)
\times\bar u(p_2)\gamma_\mu u(p_1)\bigg)\bigg\vert_{j\rightarrow j_0, \Delta(j)\rightarrow 2}\nonumber\\
&\times &\text{Im}\bigg[\frac{1}{\Gamma(\tilde{\Delta}(j))}\bigg]\nonumber\\
\eea
with the complex argument

\be
\tilde{\Delta}(j)=\Delta(j)-2=i\sqrt{\sqrt{\lambda}(j_0-j)}\equiv iy\nonumber\\
\ee
and $j_0=1-{1}/{\sqrt{\lambda}}$. For $y\rightarrow 0$, we may approximate $1/\Gamma(iy)\approx  iy\,e^{i\gamma y}$,
 with the Euler-Mascheroni constant $\gamma=0.55772...$.
 The single Reggeon amplitude (total amplitude) in momentum space (\ref{reggeon}) can now be cast in block form

\be
\label{reggeon2}
{\cal A}^{tot}_{L  p\rightarrow L p} (s,t)= I(j_0,s)\times G_{5}(j_0,s,t)
\ee
with

\bea
\label{IJ}
&&I(j_0,s)=-\tilde{s}^{j_{0}}\int_{-\infty}^{j_0}\frac{dj}{\pi}
\left(\frac{1 +e^{-i\pi}}{{\rm sin}\,\pi j}\right)\tilde{s}^{j-j_{0}}\,\sin\left[\tilde{\xi}\sqrt{\sqrt{\lambda}(j_0-j)}\right]\nonumber\\
&&G_{5}(j_0,s,t)=\frac{1}{\tilde s}\bigg(\tilde\kappa^{2(j-1)}\Gamma(\Delta(j)-2)\mathcal{V}_{LLL}(j,Q,Q')\times B^{\mu}(q,q',\epsilon^{\pm})\times g_5F_1^{(LN)}(j,K)
\times\bar u(p_2)\gamma_{\mu} u(p_1)\bigg)\bigg\vert_{j\rightarrow j_0,\,\Delta(j)\rightarrow 2}\nonumber\\
\eea

We have set $\tilde{s}\equiv {s}/{\tilde{\kappa}^2}$, and $\tilde{\xi}-\pi/2=\gamma=0.55772.....$ is Euler-Mascheroni constant. We note that the apparent pole
in the Gamma-function at the Reggeon intercept, cancels out in the combination  $\Gamma(\Delta(j_0)-2){\cal V}_{LLL}(j_0,Q,Q')$.

 In the block form
(\ref{reggeon2}), the spin-j integral $I(j_0,s)$ is similar to the spin-j integral in~\cite{POLX}  (see Eq. 4.19),
with the identifications $\mathcal{K}(s,b^{\perp},z,z')\leftrightarrow {\cal A}^{tot}_{L p\rightarrow L p} (s,t)$, $(zz'/R^4)G_3(j_0,v)  \leftrightarrow G_5(j_0,s,t)$, $\xi(v)\leftrightarrow\tilde{\xi}$, and $\widehat s\leftrightarrow \tilde{s}$. We then follow~\cite{POLX}
to evaluate the spin-j integral by closing the j-contour appropriately. In the high energy limit $\sqrt{\lambda}/\tilde{\tau}\rightarrow 0$
($\tilde{\tau}\equiv\log\tilde{s}$),  the  single Reggeon contribution to the amplitude is

\bea\label{reggeon322}
{\cal A}^{tot}_{L  p\rightarrow L p} (s,t)\simeq e^{j_0\tilde{\tau}} \left[(\sqrt{\lambda}/\pi)+ i\right] ( \sqrt{\lambda}/ 2 \pi )^{1/2}\; \tilde{\xi}  \; \frac{e^ {-\sqrt\lambda  \tilde{\xi}^2 / 2\tilde{\tau}}}{\tilde{\tau}^{3/2}}\left(1 + {\cal O}\bigg(\frac{\sqrt{\lambda}}{\tilde{\tau}}\bigg) \right)
\times  G_{5}(j_0,s,t)\,.
\eea

We can rewrite the amplitude (\ref{reggeon322}) as

\bea\label{reggeon3222}
{\cal A}^{tot}_{L  p\rightarrow L p} (s,t)&\simeq & 4\times 4\times g_5\times\frac{1}{g_5^2}\times g_5^3\kappa_{CS}\times\Big(\frac{Q}{\tilde{\kappa}}\Big)^{2-j_0-\Delta(j_0)}\times\Big(\frac{s}{\tilde{\kappa}^2}\Big)^{j_0}\times\frac{e^ {-\sqrt\lambda  \tilde{\xi}^2 / 2\log[s/\tilde{\kappa}^2]}}{(\log[s/\tilde{\kappa}^2])^{3/2}}\nonumber\\
&\times &\left[(\sqrt{\lambda}/\pi)+ i\right]\times (\sqrt{\lambda}/ 2 \pi )^{1/2}\; \tilde{\xi}  \;\left(1 + {\cal O}\bigg(\frac{\sqrt{\lambda}}{\log[s/\tilde{\kappa}^2]}\bigg) \right)\times \tilde{G}_{5}(j_0,t,Q,Q')\nonumber\\
\eea
where 
\bea
\tilde G_{5}(j_0,s,t)&\equiv &\frac{1}{4}\times \frac{1}{4}\times\frac{1}{g_5}\times\frac{1}{\frac{1}{g_5^2}\times g_5^3\kappa_{CS}}\times\frac{1}{\tilde{\kappa}^2}\times\frac{1}{Q^{2-j-\Delta(j)}}\times\frac{1}{\kappa^{2\Delta(j)-4}}\times\frac{1}{\tilde{\kappa}^{-(j-1)-\Delta(j)-1}}\times\frac{1}{\tilde\kappa^{2(j-1)}}\times G_{5}(j_0,s,t)\nonumber\\
&=&I_{\xi}(j_0,Q,Q')\times \mathcal{F}_1^{(LN)}(j_0,K)
\times s^{-1}\times \frac{1}{4}\times B^{\mu}(q,q',\epsilon^{\pm})\times\frac{1}{4}\times\bar u(p_2)\gamma_{\mu} u(p_1)
\eea
with
\be
\mathcal{F}_1^{(LN)}(j_0,K)\equiv \frac{\Gamma(\Delta(j_0)-2+a)}{\Delta(j_0)-1}\times \frac{1}{\tilde{\kappa}^{-(j_0-1)-\Delta(j_0)-1}}\times F_1^{(LN)}(j_0,K)\,.
\ee

\end{widetext}

\section{Details of the Pomeron exchange }

The transverse and traceless part of the graviton  ($\eta_{\mu\nu}\rightarrow\eta_{\mu\nu}+h_{\mu\nu}$) follows from the quadratic part of the Einstein-Hilbert action in 
 de-Donder gauge, 

\bea
\label{Action2EH}
S=&&\int d^{5} x \sqrt{g}\,e^{-2\phi}\,\mathcal{L}_{h}\,,\nonumber\\
\mathcal{L}_{h}=&& -\frac{1}{4\tilde{g}_5^2}\,g^{\mu\nu}\,\eta^{\lambda\rho}\eta^{\sigma\tau}\partial_{\mu}h_{\lambda\sigma}\partial_{\nu}h_{\rho\tau}\,,
\eea
with Newton constant $16\pi G_N={8\pi^2}/{N_c^2}=\tilde{g}_5^2=2\kappa^2$. The massive glueball spectrum 
 is determined by solving the equation of motion for $h_{\mu\nu}$ following from (\ref{Action2EH}), with for
 spin-2 glueballs


\bea
\label{SOFTMFG2}
m_n^2=8\tilde\kappa^2_N(n+1)\qquad
\tilde{g}_5f_n=2\tilde\kappa_N
\eea

\subsection{Graviton coupling in bulk}

For the graviton in the axial gauge $h_{\mu z}=h_{zz}=0$.
Using  $\eta_{\mu\nu}\rightarrow\eta_{\mu\nu}+h_{\mu\nu}$ in the linearized bulk action gives

  \be
 h\overline\Psi\Psi:\quad &&-\frac{\sqrt{2\kappa^2}}{2}\int d^5x\,\sqrt{g}\,h_{\mu\nu}T_F^{\mu\nu}\nonumber\\
  h LL:\quad && -\frac{\sqrt{2\kappa^2}}{2}\int d^5x\,\sqrt{g}\,h_{\mu\nu}T_L^{\mu\nu}\nonumber\\
  \label{vertices1}
 \ee
with the energy-momentum tensors for the fermions and left gauge fields

\begin{widetext}
 \bea
T_F^{\mu\nu}&=&e^{-\phi}\frac{i}{2}\,z\,\overline\Psi\gamma^\mu\overset{\leftrightarrow}{\partial^\nu}\Psi-\eta^{\mu\nu}\mathcal{L}_F\,,\nonumber\\
T_L^{\mu\nu} &=&-e^{-\phi}\Big(z^4\eta^{\rho\sigma}\eta^{\mu\beta}\eta^{\nu\gamma}\,F^L_{\beta\rho}F^L_{\gamma\sigma}
-z^4\,\eta^{\mu\beta}\eta^{\nu\gamma}\,F^L_{\beta z}F^L_{\gamma z}\Big)-\eta^{\mu\nu}\mathcal{L}_L\,.
  \label{EMT}
 \eea
and the rescaling

\bea
\label{SUBX}
\Psi\rightarrow \sqrt{2g_5^2}\Psi\qquad L_{N}\rightarrow g_5L_{N}\qquad h_{\mu\nu}\rightarrow\sqrt{2\kappa^2}\,h_{\mu\nu}\,.
\eea
\end{widetext}

Evaluating the couplings or the vertices (\ref{vertices1}) on the solutions, Fourier transforming the fields to momentum space, and integrating by part the trace-full part for the fermions, we find for the couplings to the fermions ($h\overline \Psi \Psi$)  to the left gauge fields  ($hLL$)

\bea
 h\overline\Psi\Psi:\quad &&\int \frac{d^4 p_2  d^4 p_1d^4k}{(2\pi)^{12}}(2\pi)^4 \delta^4(p_2-k-p_1)
\big(S^k_{h\bar\Psi\Psi}\big)\nonumber\\
h LL:\quad && \int \frac{d^4q'  d^4qd^4k}{(2\pi)^{12}}(2\pi)^4 \delta^4(q-k-q')
\big(S^k_{hLL}\big)\nonumber\\
\label{vertices3}
\eea

with

\begin{widetext}
\bea
S^{k}_{h\bar\Psi\Psi}&=&-\frac{\sqrt{2\kappa^2}}2\int dz\sqrt{g}\,e^{-\phi}z\,\epsilon^{TT}_{\mu\nu}h(k,z)\bar\Psi(p_2,z)\gamma^\mu p^\nu\Psi(p_1,z)\,,\nonumber\\
S^{k}_{hLL}&=&\sqrt{2\kappa^2}\int dz\sqrt{g}\,e^{-\phi}z^4\,\epsilon^{TT}_{\mu\nu}h(k,z)K^{\mu\nu}(q,q',\epsilon,\epsilon',z)\,.\nonumber\\
\label{vertices5}
\eea
We  have set $h_{\mu\nu}=\epsilon^{TT}_{\mu\nu}h(k,z)$ (where $\epsilon^{TT}_{\mu\nu}$ is transverse and traceless polarization tensor), $q^2=-Q^2$, $q^{\prime 2}=-Q^{\prime 2}$  for space-like momenta, and defined

\bea
&&K^{\mu\nu}(q,q',\epsilon,\epsilon',z)\equiv  B_1^{\mu\nu}\mathcal{V}(Q,z)\mathcal{V}(Q',z)
-B_0^{\mu\nu}\partial_z\mathcal{V}(Q,z)\partial_z\mathcal{V}(Q',z)\,,\nonumber\\
&&B_{0}^{\mu\nu}(\epsilon,\epsilon')\equiv \epsilon^\mu \epsilon'^\nu\,,\nonumber\\
&&B_{1}^{\mu\nu}(q,q',\epsilon,\epsilon')\equiv
\epsilon\cdot \epsilon' \,q^\mu q'^\nu-q\cdot \epsilon'\, \epsilon^\mu q'^\nu
-q^\prime \cdot \epsilon\, q^\mu \epsilon^{\prime \nu}+q\cdot q^\prime\, \epsilon^\mu \epsilon^{\prime \nu}\,.
\label{BK}
\eea
with $B_{1,0}=\eta_{\mu\nu}B_{1,0}^{\mu\nu}$, $K=\eta_{\mu\nu}K^{\mu\nu}$, and the non-normalizable wave function for the virtual photon $\mathcal{V}(Q,z)$ given in (\ref{Gauge}).

\subsection{Scattering amplitude}

The t-channel Compton exchange of a spin-2 glueball of mass $m_n$ in AdS reads

\be
\label{Amph}
&&i{\cal A}^{h}_{Lp\rightarrow  Lp} (s,t)=\sum_n i\tilde{{\cal A}}^{h}_{Lp\rightarrow L p} (m_n,s,t)\nonumber\\
&&i\tilde{{\cal A}}^{h}_{L p\rightarrow L p} (m_n,s,t)=(-i)V_{hLL}^{\mu\nu(TT)}(q,q',k,m_n)\times \tilde{G}^{TT}_{\mu\nu\alpha\beta}(m_n,t)\times
 (-i)V_{h\bar\Psi\Psi}^{\alpha\beta(TT)}(p_1,p_2,k,m_n)\,,	\nonumber\\
\ee
with the bulk vertices ($k=p_2-p_1=q-q'$)

\be
V_{hLL}^{\mu\nu(TT)}(q,q',k,m_n)\equiv&& \left(\frac{\delta S_{hLL}^k}{\delta (\epsilon^{TT}_{\mu\nu}h(k,z))}\right)\,J_{h}(m_n,z)=\sqrt{2\kappa^2}\times\frac{1}{2}\int dz\sqrt{g}\,e^{-\phi}z^4K^{\mu\nu}(q,q',\epsilon,\epsilon',z)J_{h}(m_n,z)\,,\nonumber\\
V_{h\bar\Psi\Psi}^{\alpha\beta(TT)}(p_1,p_2,k,m_n)\equiv &&\left(\frac{\delta S_{h\bar\Psi\Psi}^k}{\delta (\epsilon^{TT}_{\alpha\beta}h(k,z))}\right)\,J_{h}(m_n,z)=-\sqrt{2\kappa^2}\times\int dz\sqrt{g}\,e^{-\phi}z\bar\Psi(p_2,z)\gamma^\alpha p^\beta\Psi(p_1,z)J_{h}(m_n,z)\,,\nonumber\\ \label{vh}\nonumber\\
\ee
with $p=({p_1+p_2})/{2}$.
The bulk-to-bulk transverse and traceless graviton propagator $G_{\mu\nu\alpha\beta}=G_{\mu\nu\alpha\beta}^{TT}$ for the $2^{++}$ glueball is~\cite{Raju:2011mp,DHoker:1999bve}

\be
 \label{Gh}
&&G_{\mu\nu\alpha\beta}^{TT}(m_n,t,z,z')=J_{h}(m_n,z)\tilde{G}_{\mu\nu\alpha\beta}^{TT}(m_n,t)J_{h}(m_n,z')\,,\nonumber\\
&&\tilde{G}_{\mu\nu\alpha\beta}^{TT}(m_n,t)=
{1 \over 2} \left({\cal T}_{\mu\alpha} {\cal T}_{\nu\beta} + {\cal T}_{\mu\beta} {\cal T}_{\nu\alpha} -
\frac 23 {\cal T}_{\mu\nu} {\cal T}_{\alpha\beta}\right)\frac{i} {t-m_n^2+i\epsilon}\,,\nonumber
\ee
\end{widetext}
with

\be
 \label{wfSW}
{\cal T}_{\mu\nu} =&& -\eta_{\mu\nu} + k_{\mu} k_{\nu}/m_n^2 \,,\nonumber\\
J_h(m_n,z)\equiv&&\psi_{n}(z)=c_n\,z^{4}L_{n}^{\Delta(j)-2}(2\xi)\nonumber\\
\eea
and 
\be
c_n=\Bigg(\frac{2^{4}\tilde{\kappa}_{N}^{6}\Gamma(n+1)}{\Gamma(n+3)}\Bigg)^{\frac 12}\,,
\ee
normalized according to

\be
\int dz\,\sqrt{g}e^{-\phi}\,\abs{g^{xx}}\,\psi_n(z)\psi_m(z)=\delta_{nm}\,.
\ee

For $z'\rightarrow 0$, we can simplify  (\ref{Amph}) as  ($t=-K^2$), 

\begin{widetext}
\be
\label{nAmph}
&&i{\cal A}^{h}_{Lp\rightarrow L p} (s,t)\approx(-i)\mathcal{V}^{\mu\nu(TT)}_{hLL}(q_1,q_2,k_z)\times \bigg(\frac{i}{2}\eta_{\mu\alpha}\eta_{\nu\beta}\bigg)\times(-i)\mathcal{V}^{\alpha\beta(TT)}_{h\bar\Psi\Psi}(p_1,p_2,k_z)\,,	\nonumber\\
\ee
with
\be
&&\mathcal{V}^{\mu\nu(TT)}_{hLL}(q_1,q_2,k_z)=\sqrt{2\kappa^2}\times\frac{1}{2}\int dz\sqrt{g}\,e^{-\phi}z^4K^{\mu\nu}(q,q',\epsilon,\epsilon',z)\frac{z^4}{4}\,,\nonumber\\
&&\mathcal{V}^{\alpha\beta(TT)}_{h\bar\Psi\Psi}(p_1,p_2,k_z)=-\sqrt{2\kappa^2}\times\int dz\,\sqrt{g}\,e^{-\phi}z\,\bar\Psi(p_2,z)\gamma^\mu p^\nu\,\Psi(p_1,z)\mathcal{H}(K,z)\,,
\nonumber\\
\ee
where

\bea
\mathcal{H}(K,z)
&&=\sum_n \frac{\sqrt{2}\kappa F_n\psi_n(z')}{K^2+m_n^2}\nonumber\\
&&=4z^{4}\Gamma(a_K +2)U\Big(a_K+2,3;2\xi\Big)
=\Gamma(a_K+2)U\Big(a_K,-1;2\xi\Big)\nonumber\\
&&=\frac{\Gamma(a_K+2)}{\Gamma(a_K)}
\int_{0}^{1}dx\,x^{a_K-1}(1-x){\rm exp}\Big(-\frac{x}{1-x}(2\xi)\Big)\,,
\label{BBSWj2}
\eea
with $a_K={a}/{2}={K^2}/{8\tilde{\kappa}_N^2}$,
\be
F_n=\frac{1}{\sqrt{2}\kappa}\bigg(-\frac{1}{z^{\prime 3}}\partial_{z^\prime}\psi_n(z^\prime)\bigg)_{z^\prime=\epsilon}=-\frac{4}{\sqrt{2}\kappa}c_n L_{n}^{2}(0)\,,\nonumber\\
\ee
We have used the transformation $U(m,n;y)=y^{1-n}U(1+m-n,2-n,y)$ in the second line of (\ref{BBSWj2}).

\subsection{High energy limit}

In the high energy limit $\sqrt{\lambda}/\tilde{\tau}\rightarrow 0$
with $\tilde{\tau}\equiv\log\tilde{s}=\log[s/\tilde{\kappa}_N^2]$,  the  single Pomeron (or spin-j gluballs) contribution to the Compton Scattering amplitude has been evaluated in
\cite{Mamo:2019mka}, with the result

\bea\label{pomeron3}
{\cal A}^{tot}_{L p\rightarrow L p} (s,t)\simeq e^{j_0\tilde{\tau}} \left[(\sqrt{\lambda}/\pi)+ i\right] ( \sqrt{\lambda}/ 2 \pi )^{1/2}\; \tilde{\xi}  \; \frac{e^ {-\sqrt\lambda  \tilde{\xi}^2 / 2\tilde{\tau}}}{\tilde{\tau}^{3/2}}\left(1 + {\cal O}\bigg(\frac{\sqrt{\lambda}}{\tilde{\tau}}\bigg) \right)
\times  G_{5}(j_0,s,t,Q)
\eea
with $\tilde{\xi}-\pi/2=\gamma=0.55772.....$ is Euler-Mascheroni constant, and

\bea
\label{IJ}
G_{5}(j_0,s,t,Q)=&&\Big(\frac{\tilde{\kappa}_N}{\tilde{\kappa}_V}\Big)^{4-\Delta(j)+j-2}\nonumber\\
&&\times\frac{1}{s^2}\bigg(\frac{1}{2}\tilde{\kappa}_V^{4-\Delta(j)+j-2}\Gamma(\Delta(j)-2)\Big(\mathcal{V}^T_{hLL}(j,Q,Q')\times B_{1}^{\alpha\beta}-\mathcal{V}^L_{hLL}(j,Q,Q')\times B_{0}^{\alpha\beta}\Big)\nonumber\\
&&\times\frac{\sqrt{2\kappa^2}}{g_5^2}\times \tilde{\kappa}_N^{j-2+\Delta(j)}A(j,K)
 \bar u(p_2)\gamma_\alpha p_\beta u(p_1)\bigg)\bigg\vert_{j\rightarrow j_0,\,\Delta(j)\rightarrow 2}
\eea
with, $Qz=\xi$,
\bea
 \label{vvJPSI1j}
\mathcal{V}^T_{hLL}(j,Q,Q')=&&\frac{\sqrt{2\kappa^2}}{2}\int_{0}^{\infty} dz\sqrt{g}e^{-z^2\tilde{\kappa}_V^2}\,z^{4+2(j-2)}\times \mathcal{V}(Q,z)\times\mathcal{V}(Q',z)\times C(j) \times z^{\Delta(j)-(j-2)}\nonumber\\
=&& Q^{4-(j+\Delta(j))}\times\frac{\sqrt{2\kappa^2}}{2}\int_{0}^{\infty} d\xi\,e^{-\xi^2\frac{\tilde{\kappa}_V^2}{Q^2}}\,\xi^{4+2(j-2)}\times \mathcal{V}(\xi)\times\mathcal{V}(\xi Q'/Q)\times C(j) \times \xi^{\Delta(j)-(j-2)}\,,\nonumber\\
\mathcal{V}^L_{hLL}(j,Q,Q')=&&\frac{\sqrt{2\kappa^2}}{2}\int_{0}^{\infty} dz\sqrt{g}e^{-z^2\tilde{\kappa}_V^2}\,z^{4+2(j-2)}\times \partial_z\mathcal{V}(Q,z)\times\partial_z\mathcal{V}(Q',z)\times C(j) \times z^{\Delta(j)-(j-2)}\nonumber\\
=&& Q^{6-(j+\Delta(j))}\times\frac{\sqrt{2\kappa^2}}{2}\int_{0}^{\infty} d\xi\,e^{-\xi^2\frac{\tilde{\kappa}_V^2}{Q^2}}\,\xi^{4+2(j-2)}\times\partial_{\xi}\mathcal{V}(\xi)\times\partial_{\xi}\mathcal{V}(\xi Q'/Q)\times C(j) \times \xi^{\Delta(j)-(j-2)}\,,\nonumber\\
\eea
and

\be\label{Aj3}
&&A(j,K)=\frac{\tilde{\kappa}_N^{-(j-2)-\Delta(j)}}{2}\,\frac{\Gamma(c)\Gamma(1-\tilde{b}+c)}{\Gamma(1-\tilde{b}+c+\tilde{a})}\nonumber\\&&\times\bigg(\bigg(\frac{\tilde{n}_R}{\tilde{\kappa}_N^{\tau-1}}\bigg)^2\,_2F_1(\tilde{a},c+1,1-\tilde{b}+c+\tilde{a},-1)
+\bigg(\frac{\tilde{n}_L}{\tilde{\kappa}_N^{\tau}}\bigg)^2\frac{c(1-\tilde{b}+c)}{1-\tilde{b}+c+\tilde{a}}\,_2F_1(\tilde{a}+1,c+1,2-\tilde{b}+c+\tilde{a},-1)\bigg)\,.\nonumber\\
\ee

The parameters are fixed as
\bea
&&1-\tilde{b}+c=(\tau-1)+\frac{j-2}{2}+\frac{\Delta(j)}{2}\nonumber\\
&&1-\tilde{b}+c+\tilde{a}=(\tau+1)+\frac{j-2}{2}+a_K\nonumber\\
&&c=(\tau+1)+\frac{j-2}{2}-\frac{\Delta(j)}{2}\nonumber\\
&&\tilde{n}_R=\tilde{n}_L \tilde{\kappa}_N^{-1}\sqrt{\tau-1}\qquad
\tilde{n}_L=\tilde{\kappa}_N^{\tau}\sqrt{{2}/{\Gamma(\tau)}}\nonumber\\
\eea
and
\bea
\label{PARA}
&&C(j)= \tilde{\kappa}_V^{2\Delta(j)-4} \times\frac 1{\Delta(j)}
\frac{2^{\Delta(j)-2}\Gamma(a_K+\frac{\Delta(j)}{2})}{\Gamma(\Delta(j)-2)}\qquad\nonumber\\
&&\Delta(j)=2+\sqrt{2\sqrt{\lambda}(j-j_0)}\qquad{\rm and}\qquad a_K=\frac a2=\frac{K^2}{8{\tilde\kappa}^2}\qquad{\rm and}\qquad j_0=2-\frac{2}{\sqrt{\lambda}}\,.
\eea

We can rewrite $G_{5}(j_0,s,t,Q,Q')$ of (\ref{IJ}) more compactly as
\bea
\label{IJ2}
&&G_{5}(j_0,s,t,Q,Q')=\frac{2\kappa^2}{g_5^2}\times\Big(\frac{Q}{\tilde{\kappa}}\Big)^{2-(j+\Delta(j))}\nonumber\\
&&\times\frac{1}{s^2}\mathcal{F}(j,K)\bigg(I_{\xi}^T(j,Q,Q')\times B_{1}^{\alpha\beta}p_{\alpha}p_{\beta}-I_{\xi}^L(j,Q,Q')\times B_{0}^{\alpha\beta}p_{\alpha}p_{\beta}Q^2\bigg)\bigg\vert_{j\rightarrow j_0,\,\Delta(j)\rightarrow 2}\nonumber\\ 
\eea
where we have set $\tilde{\kappa}_V=\tilde{\kappa}_N=\tilde{\kappa}$, and defined the dimensionless functions
\bea
\mathcal{F}(j,K) &\equiv & \tilde{\kappa}^{j+2-\Delta(j)}\times\Gamma(\Delta(j)-2)\times C(j,K)\times A(j,K)\,,\nonumber\\ 
I_{\xi}^T(j,Q,Q') &\equiv & \frac{1}{2}\int_{0}^{\infty} d\xi\,e^{-\xi^2\frac{\tilde{\kappa}_V^2}{Q^2}}\,\xi^{\Delta(j)+j+2}\times \mathcal{V}(\xi)\times \mathcal{V}(\xi Q'/Q)\,,\nonumber\\
&\approx &\frac{1}{2}\times 2^{\Delta(j)+j+2}\,\frac{Q'}{Q} \,G_{2,2}^{2,2}\left(\frac{Q^2}{Q'^2}\bigg |
\begin{array}{c}
 \frac{1}{2},\frac{3}{2} \\
 \frac{1}{2} (j+\Delta(j) +4),\frac{1}{2} (j+\Delta(j) +6) \\
\end{array}
\right)\,,\nonumber\\
I_{\xi}^L(j,Q,Q') &\equiv & \frac{1}{2}\int_{0}^{\infty} d\xi\,e^{-\xi^2\frac{\tilde{\kappa}_V^2}{Q^2}}\,\xi^{\Delta(j)+j+2}\times\partial_{\xi}\mathcal{V}(\xi)\times\partial_{\xi}\mathcal{V}(\xi Q'/Q) \nonumber\\
&\approx & \frac{1}{2}\times 2^{\Delta (j)+j+2}\,\frac{Q'^2}{Q^2}
\,G_{2,2}^{2,2}\left(\frac{Q^2}{Q'^2}\bigg |
\begin{array}{c}
 1,1 \\
 \frac{1}{2} (j+\Delta(j) +5),\frac{1}{2} (j+\Delta(j) +5) \\
\end{array}
\right)\,,
\eea
using the identities (\ref{LIM}).

We can also rewrite the amplitude (\ref{pomeron3}) as
\bea\label{pomeron32}
{\cal A}^{tot}_{L  p\rightarrow L p} (s,t)&\simeq & \frac{2\kappa^2}{g_5^2}\times\Big(\frac{Q}{\tilde{\kappa}}\Big)^{2+2/\sqrt{\lambda}}\times\Big(\frac{s}{\tilde{\kappa}^2}\Big)^{-2/\sqrt{\lambda}}\times\frac{e^ {-\sqrt\lambda  \tilde{\xi}^2 / 2\log[s/\tilde{\kappa}^2]}}{(\log[s/\tilde{\kappa}^2])^{3/2}}\nonumber\\
&\times &\left[(\sqrt{\lambda}/\pi)+ i\right]\times (\sqrt{\lambda}/ 2 \pi )^{1/2}\; \tilde{\xi}  \;\left(1 + {\cal O}\bigg(\frac{\sqrt{\lambda}}{\log[s/\tilde{\kappa}^2]}\bigg) \right)\times \tilde{G}_{5}(j_0,t,Q,Q')\nonumber\\
\eea
where we have explicitly used $\tilde{\tau}=\log[s/\tilde{\kappa}^2]$, $j_0=2-\frac{2}{\sqrt{\lambda}}$, $\Delta(j_0)=2$,  $\tilde{\xi}-\pi/2=\gamma=0.55772.....$ is Euler-Mascheroni constant, and defined 
\be
\tilde{G}_{5}(j_0,s,t,Q,Q')\equiv \frac{1}{Q^4}\times\frac{g_5^2}{2\kappa^2}\times\Big(\frac{\tilde{\kappa}}{Q}\Big)^{2-(j_0+\Delta(j_0))}\times s^2\times G_5(j_0,s,t,Q,Q')
\ee
For small-x, we have $s\simeq {Q^2}/{x}$, we can rewrite the amplitude (\ref{pomeron32}) in terms of $x$ as
\bea\label{pomeron33}
{\cal A}^{tot}_{L p\rightarrow L p} (x,Q,t)&\simeq & \frac{1}{2}\times\frac{2\kappa^2}{g_5^2}\times\Big(\frac{Q}{\tilde{\kappa}}\Big)^{2-2/\sqrt{\lambda}}\times\Big(\frac{1}{x}\Big)^{1-2/\sqrt{\lambda}}\times\frac{e^ {-\sqrt\lambda  \tilde{\xi}^2 /2(\log[Q^2/\tilde{\kappa}^2]+\log[1/x])}}{(\log[Q^2/\tilde{\kappa}^2]+\log[1/x])^{3/2}}\nonumber\\
&\times &\left[(\sqrt{\lambda}/\pi)+ i\right]\times (\sqrt{\lambda}/ 2 \pi )^{1/2}\; \tilde{\xi}  \;\left(1 + {\cal O}\bigg(\frac{\sqrt{\lambda}}{\log[Q^2/\tilde{\kappa}^2]+\log[1/x]}\bigg) \right)\times \tilde{\tilde{G}}_{5}(j_0,x,t,Q,Q')\nonumber\\
\eea
where we have defined $\tilde{\tilde{G}}_{5}(j_0,x,t,Q,Q')\equiv 2x\,\tilde{G}_{5}(j_0,s,t,Q,Q')$ with ($\epsilon \cdot q=0$)
\bea
\label{IJ2}
&&\tilde{\tilde{G}}_{5}(j_0,x,t,Q,Q')=\mathcal{F}(j,K)\bigg(I_{\xi}^T(j,Q,Q')\times \Big(\frac{1}{2x}\epsilon^2-\frac{2x}{Q^2}(\epsilon\cdot p)^2\Big)-I_{\xi}^L(j,Q,Q')\times\frac{2x}{Q^2}(\epsilon\cdot p)^2\bigg)\bigg\vert_{j\rightarrow j_0,\,\Delta(j)\rightarrow 2}\,,\nonumber\\ 
\eea


\section{Operator Product Expansion}

The parton model emerges in QCD through a leading twist contribution to the structure functions. 
The twist expansion follows from the operator product expansion (OPE) of the JJ currents. We now
illustrate this expansion to leading order for the charged current contributions in T-ordered product in (\ref{2}). 
Specifically, we have as $x\rightarrow 0$

\be
\label{8}
&&T^*\left(J^{W^-}_\mu(x)J^{W^+}_\nu(0)\right)\approx \nonumber\\
&&e_W^2\left(\overline q(x)T^-T^+\gamma_\mu\frac 12 (1-\gamma_5)\,S(x)\,\gamma_\nu\frac 12 (1-\gamma_5) q(0)+
\overline q(0)T^+T^-\gamma_\nu\frac 12 (1-\gamma_5)\,S(-x)\,\gamma_\mu\frac 12 (1-\gamma_5) q(x)\right)\nonumber\\
\ee
with $S(x)=2i\gamma\cdot x/(2\pi x^2)^2$. With the help of the identity

\be
\label{9}
\gamma_\mu\gamma\cdot x\gamma_\nu=\left(S_{\mu\nu\alpha\beta}+i\epsilon_{\mu\nu\alpha\beta}\gamma_5\right)x^\alpha\gamma^\beta
\ee
with the symmetric tensor

\be
\label{10}
S_{\mu\nu\alpha\beta}=\eta_{\mu\alpha}\eta_{\nu\beta}+\eta_{\mu\beta}\eta{\nu\alpha}-\eta_{\mu\nu}\eta_{\alpha\beta}
\ee
in (\ref{8}), the short distance contribution to the T-ordered product in (\ref{2}) is

\be
\label{11}
T_{\mu\nu}^{-+}\approx 2e_W^2\frac {q^\alpha}{q^2}\left<P\right|\bigg(S_{\mu\nu\alpha\beta}\overline q(0)\tau^3\gamma^\beta(1-\gamma_5) q(0)
-i\epsilon_{\mu\nu\alpha\beta}\overline q(0)\gamma^\beta(1-\gamma_5) q(0)\bigg)\left|P\right>
\ee
For unpolarized scattering, a comparison of (\ref{11}) to (\ref{2}) suggests that the parity odd structure function $F_3$ 
can be identified with the antisymmetric tensor contribution, 

\be
\label{12}
F_3(x,Q^2) P^\beta\approx (2e_W)^2{\rm Im}\bigg(\frac 1x \left<P\right|\overline q \gamma^\beta(1-\gamma_5) q\left|P\right>\bigg)
\ee
(\ref{12}) is suggestive of a $j=1$ exchange through the left singlet current, since $s^{j=1}\leftrightarrow 1/x$.

Although the twist-2 contribution to the OPE is real, it provides the relevant starting point for the Reggeization 
by summing over higher spin-j states in holography~\cite{HATTAX}. For that, we first note that the holographic dual 
of the singlet current  form factor is

\begin{eqnarray}
\label{XP1}
\left<P_X\right|\overline q \gamma^\beta(1-\gamma_5) q\left|P\right>=&&
\left[e_N\overline{u}_N(P_X)\gamma^\beta (1-\gamma^5)u_N(P)\right]\nonumber\\
&&\times \int dz \sqrt{g}e^{-{\phi}}\,{{\cal V}(Q,z)}\,\frac{z}{R}\,
\bigg[\frac 12\bigg(\frac{z^{2}}{R^2}\bigg)^2\, \tilde f_L^{n_x}(z)\tilde f_L^{0}(z)+\frac 12\bigg(\frac{z^{2}}{R^2}\bigg)^2\, \tilde f_R^{n_x}(z)\tilde f_R^{0}(z)\bigg]
\end{eqnarray}
The bulk-to-boundary propagator ${\cal V}(Q,z^\prime)$ relates to the bulk-to-bulk propagator $G(Q, z, z^\prime)$ through

\be
\label{XP2}
\lim_{z\to 0}\frac 2{z^{\prime 2}}\, G_1(Q,z^\prime, z)={\cal V}(Q,z)
\ee
Using (\ref{XP2}) into (\ref{XP1}) gives

\begin{eqnarray}
\label{XP4}
\left<P_X\right|\overline q \gamma^\beta(1-\gamma_5) q\left|P\right>=&&
\left[e_N\overline{u}_N(P_X)\gamma^\beta (1-\gamma^5)u_N(P)\right]\,\nonumber\\
&&\times \lim_{z^\prime\to 0}\frac 2{z^{\prime 2}} \int dz\sqrt{g}e^{-{\phi}}\,G_1(Q,z^\prime, z)\,\nonumber\\
&&\times \frac{z}{R}\,
\bigg[\frac 12\bigg(\frac{z^{2}}{R^2}\bigg)^2\, \tilde f_L^{n_x}(z)\tilde f_L^{0}(z)+\frac 12\bigg(\frac{z^{2}}{R^2}\bigg)^2\, \tilde f_R^{n_x}(z)\tilde f_R^{0}(z)\bigg]
\end{eqnarray}

\subsection{Hard wall}

For the hard wall model with $\phi=0$, the bulk-to-bulk propagator can be readily constructed

\be
\label{XP3}
G_1(Q,z^\prime, z)=zz^\prime\bigg(I_0(Qz_0)K_1(Qz_>)+K_0(Qz_0)I_1(Qz_>)\bigg)\frac{I_1(Qz_<)}{I_0(Qz_0)}
\ee
and (\ref{XP2}) explicitly checked. However, for the Reggeization it is more useful to recall 
 that the bulk-to-bulk propagator for the $U_L(1)$ vector field, obeys the Green$^\prime$s equation in warped space ($Q^2=-q^2$)

\be
\label{XP5} 
\bigg(\bigg(\Delta_{j=1}-z^{ 2}Q^2+m_{j=1}^2\bigg)=\bigg(-z^{2}(\partial_{z}^2+Q^2)+z\partial_{z}\bigg)\bigg)G_{j=1}(Q, z^\prime, z)=\frac{\delta(z-z^\prime)}{\sqrt{g}}
\ee
with $m^2_{j=1}=-3$. Using the open-string Regge trajectory 

\be
\label{OPEN}
j=1+\alpha^\prime (m^2_j-m_1^2)\qquad  {\rm with}\qquad \alpha^\prime=l_s^2=1/\sqrt\lambda
\ee
 (\ref{XP5}) generalizes to spin-j

\be
\label{XP6} 
\bigg(\Delta_{j}-z^{ 2}Q^2+m_j^2\bigg)G_{j}(Q, z^\prime, z)=\frac{\delta(z-z^\prime)}{\sqrt{g}}
\ee
with the recursive relation for the warped Laplacian-like

\be
\label{XP7}
\Delta_j=z^{1-j}\,\Delta_1\,z^{j-1}
\ee
(\ref{XP6}) can be formally inverted

\be
\label{XP8}
{\sqrt{g}}\,G_{j}(Q, z^\prime, z)=\frac 1{(\Delta_{j}-z^{ 2}Q^2+m_j^2)}\,   {\delta(z-z^\prime)}=z^{2-j}\,\frac 1{(\Delta_{2}-z^{ 2}Q^2+m_j^2)}\, z^{j-2}\,{\delta(z-z^\prime)}
\ee
Changing to the conformal variable $z^2=e^{-\rho}$, noting that $\Delta_2=-4\partial_\rho^2+4$ and using the plane-wave identity

\be
\label{XP9}
z^\prime\delta(z-z^\prime)=\bigg(\frac{z^\prime}z\bigg)^{j-2}\int\frac{d\nu}{\pi}\,e^{i\nu(\rho-\rho^\prime)}
\ee
we can recast (\ref{XP8})  in the form 

\be
\label{XP10}
{\sqrt{g}}\,G_{j}(0, z^\prime, z)=\sqrt{g^\prime}(g^{\prime xx})^j(zz^\prime)^{2-j}
\int\frac{d\nu}\pi\,\frac 1{4\nu^2+4+m_j^2}\,e^{i\nu(\rho-\rho^\prime)}
\ee
for $Q=0$. The Reggeized form of the spin-j and twist-2  extension of (\ref{XP4})  is

\begin{eqnarray}
\label{XP11}
&&\sum_j\,\frac 1{x^j}\,\left<P\right|\overline q \gamma^\beta\,\partial^{j-1}\,(1-\gamma_5) q\left|P\right>= \int_{\mathbb C}\frac{dj}{4i}\frac{1-e^{-i\pi j}}{{\rm sin}\pi j}\frac 1{x^j}\,
\left[e_N\overline{u}_N(P)\gamma^\beta\,\partial^{j-1}\, (1-\gamma^5)u_N(P)\right]\nonumber\\&&\times\lim_{z^\prime\to 0}\frac 2{z^{\prime 2}} \int dz
\sqrt{g^\prime}(g^{\prime xx})^j(zz^\prime)^{2-j}
\int\frac{d\nu}\pi\,\frac 1{4\nu^2+1+\sqrt{\lambda}(j-1)}\,e^{i\nu(\rho-\rho^\prime)}\,\frac{z}{R}\,\bigg[\frac 12\psi_L^2(z)+\frac 12\psi_R^2(z)\bigg] \,\nonumber\\
\end{eqnarray}
using the open string Regge trajectory (\ref{OPEN}) in the forward limit ($Q=0$). Here $\psi_L(z)$ is  the lowest left-chirality bulk fermionic wavefunction for the hard wall.
The contour $\mathbb C$ is to the left-most of the poles $j=1,3, ...$ and to the right of the pole
$j_0=1-1/\sqrt\lambda$. Undoing the contour integration-${\mathbb C}$ by closing to the left and picking the single pole $j_0$, and then performing the $\nu$-integration yield

\begin{eqnarray}
\label{XP12}
&&{\rm Im}\,\sum^{\rm odd}_j\,\frac 1{x^j}\,\left<P\right|\overline q \gamma^\beta\,\partial^{j-1}\,(1-\gamma_5) q\left|P\right>=\frac 1{x^{j_0}}\,
\left[e_N\overline{u}_N(P)\gamma^\beta\,\partial^{j_0-1}\, (1-\gamma^5)u_N(P)\right]\nonumber\\
&&\times\frac{\pi}{2\sqrt\lambda}\lim_{z^\prime\to 0}\frac 2{z^{\prime 2}} \int dz
\sqrt{g^\prime}(g^{\prime xx})^{j_L}(zz^\prime)^{2-j_0}\,
\frac{e^{-(\rho-\rho^\prime)/4D\chi}}{\sqrt{\pi D\chi}}\,\frac{z}{R}\,\bigg[\frac 12\psi_L^2(z)+\frac 12\psi_R^2(z)\bigg] \,\nonumber\\
\end{eqnarray}
The Gribov time is $\chi ={\rm ln}(1/x)$ and the diffusion constant of the Reggeon is $D=4/\sqrt\lambda$. (\ref{XP12}) fixes the odd structure function in 
(\ref{12}) in the forward direction using this semi-quantitative OPE argument,

\be
F_3(0,x)\approx \frac 1{x^{j_0}}\approx \frac 1{x^{1-1/\sqrt\lambda}}
\ee

\subsection{Soft wall}

For the soft-wall model, the Reggeized current form factor is given by

\begin{eqnarray}
\label{XPSW}
&&\sum^{\rm odd}_j\,\frac 1{x^j}\,\left<P\right|\overline q \gamma^\beta\,\partial^{j-1}\,(1-\gamma_5) q\left|P\right> = \nonumber\\
&&-x^{j_{0}}\int_{-\infty}^{j_0}\frac{dj}{\pi}
\left(\frac{1 +e^{-i\pi}}{{\rm sin}\,\pi j}\right)x^{j-j_{0}}\,\text{Im}\left[\,2\times\frac{2}{3}\times\tilde{\kappa}^{(j-1)+\Delta(j)+1}\times\frac{\Gamma(\Delta(j)-2+a)}{\Gamma(\Delta(j)-2)}\times\frac{1}{g_5}\times\mathcal{V}^{\beta}_{L\bar\Psi\Psi}(p_1=p_2=p,k_z=0)\right]\nonumber\\
&=&-x^{j_{0}}\int_{-\infty}^{j_0}\frac{dj}{\pi}
\left(\frac{1 +e^{-i\pi}}{{\rm sin}\,\pi j}\right)x^{j-j_{0}}\,\text{Im}\left[\,2\times\frac{2}{3}\times\tilde{\kappa}^{(j-1)+\Delta(j)+1}\times\frac{\Gamma(\Delta(j)-2+a)}{\Gamma(\Delta(j)-2)}\times\frac{1}{g_5}\times g_5F_1^{(LN)}(j,K=0)\times\bar u(p)\gamma^\beta u(p)\right] \nonumber\\
\end{eqnarray}
where $\mathcal{V}^{\beta}_{L\bar\Psi\Psi}(p_1,p_2,k_z)$ is given by 
(\ref{pvertices}), and $F_1^{(LN)}(j,K)$ is given by (\ref{Fj1}). Note that the bracket

\be
\bigg[2\times\frac{2}{3}\times\tilde{\kappa}^{(j-1)+\Delta(j)+1}\times\frac{\Gamma(\Delta(j)-2+a)}{\Gamma(\Delta(j)-2)}\times\frac{1}{g_5}\times g_5F_1^{(LN)}(j,K)\bigg]
\ee 
is the spin-j form factor which reduces to the spin-1 form factor for  $j=1$, for the current operator $2\times\tilde J_L^\beta(0)=\overline q \gamma^\beta(1-\gamma_5) q$ sourced by $\frac{1}{2}\times \frac{3}{2}\times L_\beta^0(K,z\rightarrow 0)$ at the boundary. Also note that the momentum transfer is $k_z\equiv q_z$
and that $-k^2=K^2\equiv Q^2$ with $a={K^2}/{4\tilde\kappa^2}\equiv {Q^2}/{4\tilde\kappa^2}$. The momentum of the in-coming nucleon is
$p_1 = p$, and the momentum of the out-going nucleon is $p_2=p$,  with  $k=p_2-p_1\equiv q$. 

Following the reasoning in  Appendix \ref{DRE}, we can evaluate the integral in (\ref{XPSW}) with the result

\begin{eqnarray}
\label{SPSW}
&&{\rm Im} \sum^{\rm odd}_j\,\frac 1{x^j}\,\left<P\right|\overline q \gamma^\beta\,\partial^{j-1}\,(1-\gamma_5) q\left|P\right>\simeq \nonumber\\
&&e^{j_0\tau_x} \left[0\times(\sqrt{\lambda}/\pi)+ i/i\right] ( \sqrt{\lambda}/ 2 \pi )^{1/2}\; \tilde{\xi}  \; \frac{e^ {-\sqrt\lambda  \tilde{\xi}^2 / 2\tau_x}}{\tau_x^{3/2}}\left(1 + {\cal O}\bigg(\frac{\sqrt{\lambda}}{\tau_x}\bigg) \right)
\times  \mathcal{G}_{5}(j_0,x,Q=0)\nonumber\\
\end{eqnarray}
with 
\bea
&&\mathcal{G}_{5}(j_0,x,Q=0)=\nonumber\\
&&\bigg(\Gamma(\Delta(j)-2)\times 2\times\frac{2}{3}\times\tilde{\kappa}^{(j-1)+\Delta(j)+1}\times\frac{\Gamma(\Delta(j)-2+a)}{\Gamma(\Delta(j)-2)}\times\frac{1}{g_5}\times g_5F_1^{(LN)}(j,Q)
\times 2P^{\beta}\bigg)\bigg\vert_{j\rightarrow j_0,\,\Delta(j)\rightarrow 2, Q\rightarrow 0}\nonumber\\
\eea
Again,  $j_0=1-\frac{1}{\sqrt{\lambda}}$, $\tau_x=\log [1/x]$, $\bar u(p)\gamma^\beta u(p)=2p^{\beta}=2P^{\beta}$, and $\tilde{\xi}-\pi/2=\gamma=0.55772.....$ is the Euler-Mascheroni constant. Finally, comparing (\ref{SPSW}) to (\ref{12}), we find
\be
F_3(0,x)\approx \frac 1{x^{1-1/\sqrt\lambda}}\times ( \sqrt{\lambda}/ 2 \pi )^{1/2}\; \tilde{\xi}  \; \frac{e^ {-\sqrt\lambda  \tilde{\xi}^2 / 2\tau_x}}{\tau_x^{3/2}}\left(1 + {\cal O}\bigg(\frac{\sqrt{\lambda}}{\tau_x}\bigg) \right)
\times \tilde{\mathcal{G}}_{5}(j_0,x,0) 
\ee
with 
\bea
&&\tilde{\mathcal{G}}_{5}(j_0,x,0)=\nonumber\\
&&\bigg(\Gamma(\Delta(j)-2)\times 2\times\frac{2}{3}\times\tilde{\kappa}^{(j-1)+\Delta(j)+1}\times\frac{\Gamma(\Delta(j)-2+a)}{\Gamma(\Delta(j)-2)}\times\frac{1}{g_5}\times g_5F_1^{(LN)}(j,Q)
\bigg)\bigg\vert_{j\rightarrow j_0,\,\Delta(j)\rightarrow 2, Q\rightarrow 0}\,.\nonumber\\
\eea

\section{Trace of Gamma matrices }

Note that the Dirac traces do not depend on the specific form of the $\gamma^0,\gamma^1,
\gamma^2,\gamma^3$ matrices but are completely determined by the
Clifford algebra
\be
\{\gamma^\mu,\gamma^\nu\}\ \equiv\ \gamma^\mu\gamma^\nu\,+\,\gamma^\nu\gamma^\mu\
=\ 2\eta^{\mu\nu}\,,
\ee
and some useful identities for carrying some of the  Dirac  traces of gamma matrices above,
are given by (note that $\gamma^5=i\gamma^0\gamma^1\gamma^2\gamma^3$ and it satisfies $\gamma^5\gamma^\mu=-\gamma^\mu\gamma^5$)

\be
\tr(\gamma^\mu\gamma^\nu)= 4\eta^{\mu\nu}\,,
\ee

\be
\tr(\gamma^\mu\gamma^\nu\gamma^5)=0\,,
\ee

\be
\tr(\gamma^\alpha\gamma^\mu\gamma^\beta\gamma^\nu)=4\eta^{\alpha\mu}\eta^{\beta\nu}\ -\ 4\eta^{\alpha\beta}\eta^{\mu\nu}\
  +\ 4\eta^{\alpha\nu}\eta^{\mu\beta}\,,
\ee

\be
\tr(\gamma^\alpha\gamma^\mu\gamma^\beta\gamma^\nu\gamma^5)=-4i\epsilon^{\alpha\mu\beta\nu}\,,
\ee

\be
\tr(\gamma^\alpha\gamma^\mu\gamma^{\tilde{\nu}}\gamma^\beta\gamma^{\tilde{\mu}}\gamma^\nu)&= 4\eta^{\alpha\mu}\times\Bigl( \eta^{{\tilde{\nu}}\beta}\eta^{{\tilde{\mu}}\nu}\,
		-\,\eta^{{\tilde{\nu}}{\tilde{\mu}}}\eta^{\beta\nu}\,+\,\eta^{{\tilde{\nu}}\nu}\eta^{\beta{\tilde{\mu}}}\Bigr)\cr
&\qquad-\ 4\eta^{\alpha{\tilde{\nu}}}\times\Bigl( \eta^{\mu\beta}\eta^{{\tilde{\mu}}\nu}\,
		-\,\eta^{\mu{\tilde{\mu}}}\eta^{\beta\nu}\,+\,\eta^{\mu\nu}\eta^{\beta{\tilde{\mu}}}\Bigr)\cr
&\qquad+\ 4\eta^{\alpha\beta}\times\Bigl( \eta^{\mu{\tilde{\nu}}}\eta^{{\tilde{\mu}}\nu}\,
		-\,\eta^{\mu{\tilde{\mu}}}\eta^{{\tilde{\nu}}\nu}\,+\,\eta^{\mu\nu}\eta^{{\tilde{\nu}}{\tilde{\mu}}}\Bigr)\cr
&\qquad-\ 4\eta^{\alpha{\tilde{\mu}}}\times\Bigl( \eta^{\mu{\tilde{\nu}}}\eta^{\beta\nu}\,
		-\,\eta^{\mu\beta}\eta^{{\tilde{\nu}}\nu}\,+\,\eta^{\mu\nu}\eta^{{\tilde{\nu}}\beta}\Bigr)\cr
&\qquad+\ 4\eta^{\alpha\nu}\times\Bigl( \eta^{\mu{\tilde{\nu}}}\eta^{\beta{\tilde{\mu}}}\,
		-\,\eta^{\mu\beta}\eta^{{\tilde{\nu}}{\tilde{\mu}}}\,+\,\eta^{\mu{\tilde{\mu}}}\eta^{{\tilde{\nu}}\beta}\Bigr)\,,\nonumber\\
\ee

\begin{eqnarray}
\tr(\gamma^{\alpha}\gamma^{\mu}\gamma^{{\tilde{\nu}}}\gamma^{\beta}\gamma^{{\tilde{\mu}}}\gamma^{\nu}\gamma^{5})=&-&4i(\eta^{\alpha\mu}\epsilon^{{\tilde{\nu}}\beta{\tilde{\mu}}\nu}-\eta^{\alpha{\tilde{\nu}}}\epsilon^{\mu\beta{\tilde{\mu}}\nu}+\eta^{{\tilde{\nu}}\mu}\epsilon^{\alpha\beta{\tilde{\mu}}\nu}-\eta^{{\tilde{\mu}}\nu}\epsilon^{\beta\alpha\mu{\tilde{\nu}}}\nonumber\\
&+&\eta^{\beta\nu}\epsilon^{{\tilde{\mu}}\alpha\mu{\tilde{\nu}}}-\eta^{\beta{\tilde{\mu}}}\epsilon^{\nu\alpha\mu{\tilde{\nu}}})\,,
\end{eqnarray}
and
\bea
\tr(\gamma^{\nu_1}\cdots\gamma^{\nu_n}\gamma^5)\ &=&\ 0\quad
\forall\ {\rm odd}\ n\,,\nonumber\\
\tr(\gamma^{\nu_1}\cdots\gamma^{\nu_n})\ &=&\ 0\quad
\forall\ {\rm odd}\ n\,.
\eea

\end{widetext}


 \vfil

\end{document}